\documentclass[useAMS,usenatbib]{mnras}
\pdfoutput=1

\usepackage{amsmath}
\usepackage{txfonts}
\usepackage{graphicx}
\usepackage{url}
\usepackage{color}

\newcommand{\pd}{\partial}
\newcommand{\ud}{\ensuremath{\mathrm{d}}}

\newcommand{\altaffilmark}[1]{$^{ #1}$}

\title[Supernova Simulations from a 3D Progenitor Model]{Supernova Simulations from a 3D Progenitor Model --
Impact of Perturbations and Evolution of Explosion Properties}

\author[B.~M\"uller et al.]{Bernhard~M\"uller\altaffilmark{1,2,6}\thanks{E-mail: b.mueller@qub.ac.uk},
Tobias~Melson\altaffilmark{3},
  Alexander~Heger\altaffilmark{2,4,5,6},
  Hans-Thomas Janka\altaffilmark{3}%}
\\
\altaffilmark{1}{Astrophysics Research Centre, School
of Mathematics and Physics, Queen's University
Belfast, Belfast, BT7~1NN, United Kingdom;
  \href{mailto:b.mueller@qub.ac.uk}{b.mueller@qub.ac.uk}}
\\
\altaffilmark{2}{Monash Centre for Astrophysics, School of
  Physics and Astronomy, Monash University, Victoria
  3800, Australia}
\\
\altaffilmark{3}{Max-Planck-Institut f\"ur Astrophysik,
  Karl-Schwarzschild-Str. 1, 85748 Garching, Germany}
\\
\altaffilmark{4}{School of Physics \& Astronomy,
  University of Minnesota, Minneapolis, MN 55455, U.S.A.}
\\
\altaffilmark{5}{Center for Nuclear Astrophysics, Department of
  Physics and Astronomy, Shanghai Jiao-Tong University, Shanghai
  200240, P. R. China.}
\\
\altaffilmark{6}{Joint Institute for Nuclear Astrophysics, 1 Cyclotron
  Laboratory, National Superconducting Cyclotron Laboratory},\\ Michigan
  State University, East Lansing, MI 48824-1321, U.S.A.
}

\begin{document}

\label{firstpage}
\pagerange{\pageref{firstpage}--\pageref{lastpage}}
\pagerange{\pageref{firstpage}--22}

\maketitle

\begin{abstract}
 We study the impact of large-scale perturbations from convective
 shell burning on the core-collapse supernova explosion mechanism
 using three-dimensional (3D) multi-group neutrino hydrodynamics
 simulations of an $18 M_\odot$ progenitor.  Seed asphericities in the
 O shell, obtained from a recent 3D model of O shell burning, help
 trigger a neutrino-driven explosion $330\, \mathrm{ms}$ after bounce
 whereas the shock is not revived in a model based on a spherically
 symmetric progenitor for at least another $300 \, \mathrm{ms}$.  We
 tentatively infer a reduction of the critical luminosity for shock
 revival by $\mathord{\sim}20\%$ due to pre-collapse
 perturbations. This indicates that convective seed perturbations play
 an important role in the explosion mechanism in some progenitors.  We
 follow the evolution of the $18 M_\odot$ model into the explosion
 phase for more than $2\, \mathrm{s}$ and find that the cycle of
 accretion and mass ejection is still ongoing at this stage.  With a
 preliminary value of $7.7 \times 10^{50}\, \mathrm{erg}$ for the
 diagnostic explosion energy, a baryonic neutron star mass of $1.85
 M_\odot$, a neutron star kick of $\mathord{\sim}600 \, \mathrm{km} \,
 \mathrm{s}^{-1}$ and a neutron star spin period of $\mathord{\sim}20
 \, \mathrm{ms}$ at the end of the simulation, the explosion and
 remnant properties are slightly atypical, but still lie comfortably
 within the observed distribution.  Although more refined simulations
 and a larger survey of progenitors are still called for, this
 suggests that a solution to the problem of shock revival and
 explosion energies in the ballpark of observations are within reach
 for neutrino-driven explosions in 3D.
\end{abstract}

\begin{keywords}
supernovae: general --  stars:massive -- convection -- hydrodynamics  -- turbulence
\end{keywords}

\section{Introduction}
\label{sec:intro}
After several decades of research, the mechanism powering supernova
explosions of massive stars still remains under investigation.  From
the tortuous history of the field, the neutrino-driven mechanism has
emerged as the most popular -- and likely most prevalent -- explosion
scenario for core-collapse supernovae. In all but the lightest
supernova progenitors \citep{kitaura_06,janka_08,fischer_10,melson_15a}, this
mechanism appears to hinge on the breaking of spherical symmetry by
hydrodynamic instabilities like convection
\citep{herant_94,burrows_95,janka_96} and the standing accretion shock
instability (SASI;
\citealp{blondin_03,foglizzo_07,laming_07,guilet_12}). These
instabilities improve the conditions for neutrino-driven runaway shock
expansion by various means, including the ``turbulent pressure''
provided by non-spherical fluid motions
\citep{burrows_95,murphy_12,couch_14,mueller_15a} and the feedback of
mixing on the post-shock stratification and neutrino heating and
cooling \citep{herant_94,janka_96}. 

The computational tools for capturing these two indispensable
ingredients -- neutrino heating and cooling and three-dimensional (3D)
fluid flow -- with a high degree of confidence have become available
during the last few years (see \citealt{janka_16} and
\citealt{mueller_16b} for an overview of recent results and simulation
methodologies). Several groups have presented 3D core-collapse
supernova simulations employing various algorithms for multi-group
neutrino transport ranging from the very rigorous \textsc{Vertex}
\citep{hanke_13,melson_15a,melson_15b,janka_16} and \textsc{Chimera}
\citep{lentz_15} models to more approximate methods
\citep{takiwaki_12,takiwaki_14,takiwaki_16,mueller_15b,roberts_16}.

The findings from the first generation of 3D multi-group neutrino
hydrodynamics simulations present a challenge: In general, the 3D
models prove more reluctant to achieve shock revival than their
axisymmetric (2D) counterparts and either marginally fail to explode
\citep{hanke_13} or explode later than in 2D
\citep{takiwaki_14,melson_15b,lentz_15}. This behaviour is compatible
with earlier models using a simple light-bulb approximation
\citep{hanke_12,couch_12b} and has been ascribed to the different
turbulent energy cascade in 3D and 2D, which leads to smaller
structures and stronger dissipation in the neutrino heating region in
3D (though there may be compensating effect, see
\citealt{mueller_16b}).

The serious implication is that with the best available physics
treatment, we can currently not yet obtain successful 3D models of
neutrino-driven explosions across a large range of progenitor
masses. Considering the paucity of first-principle 3D simulations and
the fact that some of them apparently come very close to shock
revival, this may reflect an unlucky choice of progenitors or minor
numerical problems (such as resolution issues) rather than a
fundamental problem. Nonetheless, the first generation of 3D
multi-group neutrino hydrodynamics simulations certainly provides
motivation for exploring missing ingredients in the neutrino-driven
mechanism that could lead to more robust and earlier explosions, not
least because they may also help solve the ``energy problem'' in many
successful 2D simulations \citep{janka_12b,nakamura_15,oconnor_16}, in
which the explosion energy grows rather slowly and may fall short of
the typical observed values. Various possible ingredients
for more robust neutrino-driven explosions have
recently been explored, including rapid rotation
\citep{janka_16,takiwaki_16} and a reduction of
neutral-current neutrino-nucleon scattering opacities
\citep{melson_15b,horowitz_17,burrows_17}.

Another idea posits that initial asymmetries in the supernova
progenitors can precipitate shock revival by boosting convection
and/or the SASI \citep{couch_13,mueller_15a}. Such initial
perturbations (mostly solenoidal velocity perturbations) at the
pre-collapse stage could arise naturally in active convective burning
shells, and in that sense the ``perturbation-aided'' neutrino-driven
mechanism does not need to invoke new physics. The efficacy of this
mechanism still remains unclear, however. Simulations using
parameterised initial conditions for the initial perturbations using a
leakage scheme \citep{couch_13,couch_14} or multi-group neutrino
transport \citep{mueller_15a,burrows_17} found a range of effect
sizes from nil to a significant impact on shock revival. This is
mirrored by analytic estimates  \citep{mueller_16c,abdikamalov_16} for the reduction of the critical
neutrino luminosity \citep{burrows_93} for shock revival due to the
injection of extra turbulent kinetic energy into the post-shock region
by infalling perturbations.  There
is consensus that the initial convective Mach number during shell
burning and the spatial scale of the initial perturbations are the
decisive parameters. Although these parameters can be estimated from
spherically symmetric (1D) stellar evolution models
\citep{mueller_16c}, firm conclusions can only be drawn from supernova
simulations based on 3D progenitor models. This has been attempted by
\citet{couch_15}, but the conclusive force of their results is
limited. On the methodological side, the approximations employed for
their 3D model of silicon shell burning alter the dynamics of the
convective flow \citep{mueller_16c}, and their use of a leakage scheme
also significantly affects the dynamics in the supernova core after
collapse.\footnote{For example, the diagnostic explosion energy already
  starts to grow around $20 \, \mathrm{ms}$ after bounce even for the
  case of a 1D initial model, which is not even remotely similar to
  recent multi-group neutrino hydrodynamics simulations.} Furthermore,
they did not find a qualitative difference between the supernova
models employing 1D and 3D initial conditions: Positive diagnostic
explosion energies develop in both cases at the same time; however,
the model with 3D initial conditions shows faster shock propagation
from $100 \, \mathrm{ms}$ after bounce onward. Two key questions thus
remain open: Can pre-collapse perturbations make a qualitative
difference for shock revival, and can they help bring the predicted
explosion properties close to observed values?

In this paper, we take the next step towards answering these questions
extending a cursory preview of our recent work on the
perturbation-aided mechanism in \citet{mueller_16b}.  We present
results from the first successful multi-group neutrino hydrodynamics
simulation of a core-collapse supernova explosion using such a 3D
progenitor model for an $18 M_\odot$ star \citep{mueller_16c} and also
provide a comparison to a 1D initial model and a 3D case with
artificially reduced nuclear burning rates and convective velocities,
respectively.  The goals of our study are twofold: First, we aim to
better quantify the effect of infalling initial perturbations on the
post-shock flow dynamics, the neutrino heating conditions, and the
critical luminosity for shock revival. With a base of comparison of
only three models, we can obviously only make limited progress on this
front, but our analysis is also meant as preparatory work for similar
simulations that will undoubtedly emerge in the near future.  Second,
we seek to assess whether the inclusion of 3D initial conditions is
helpful or even sufficient to bring simulations in line with
observational constraints (energy, nickel mass, neutron star mass,
spin, and kick). To this end, the $18 M_\odot$ explosion model has
been extended to more than $2.35 \, \mathrm{s}$ after core bounce.

Our paper is structured as follows: After reviewing the initial models
and the numerical methods used in our supernova simulations in
Section~\ref{sec:numerics}, we provide a brief overview of the three
different simulations in Section~\ref{sec:comparison}, followed by a
more detailed analysis of the impact of the level of initial
perturbations on the model dynamics.
The explosion dynamics of the $18 M_\odot$ progenitor model of
\citet{mueller_16c} is discussed in Section~\ref{sec:explosion}.  In
Section~\ref{sec:conclusions} we outline the implications and
limitations of our findings and point out open questions left for
future research.

\begin{figure}
  \includegraphics[width=\linewidth]{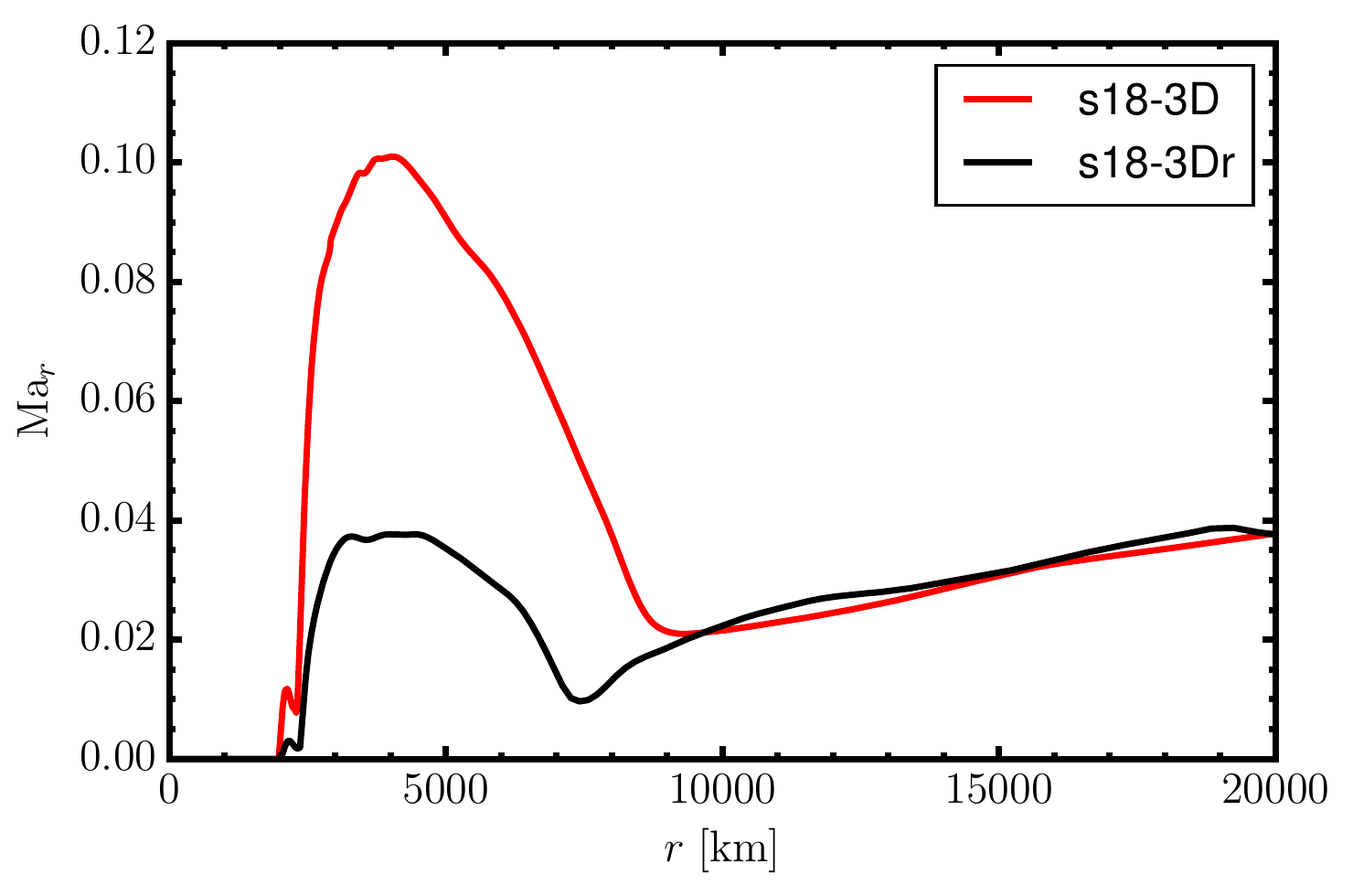}
  \caption{Profiles of the turbulent Mach number $\mathrm{Ma}_r$ of radial velocity
    fluctuations at the onset of collapse for models s18-3Dr (black)
    and s18-3D (red). Due to an artificial reduction of the
    burning rates in s18-3Dr, the maximum convective Mach number
    in the oxygen shell is considerably lower than for s18-3D.
    \label{fig:mach_ini}}
\end{figure}

\begin{figure*}
  \includegraphics[width=0.49 \linewidth]{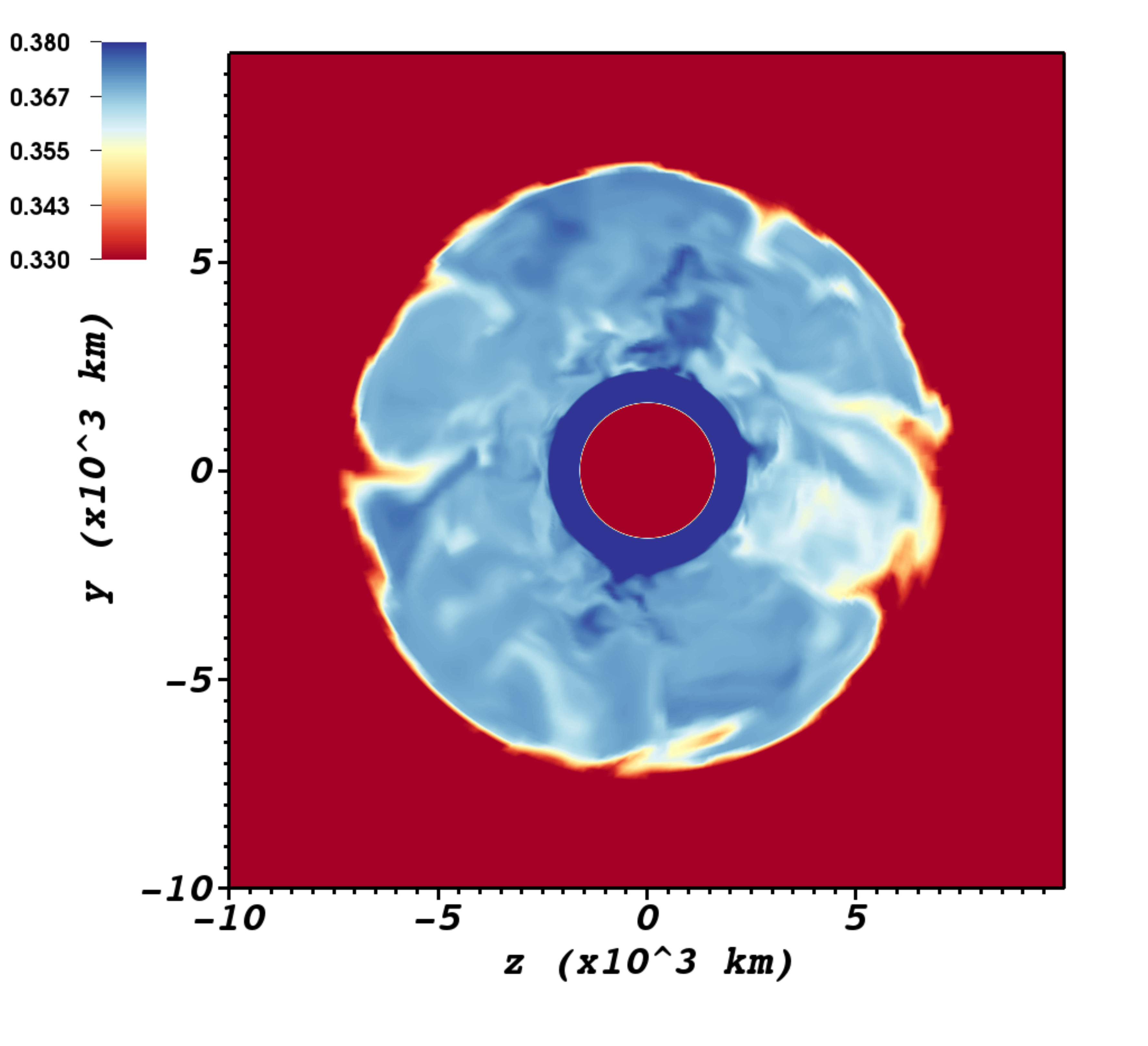}
  \includegraphics[width=0.49 \linewidth]{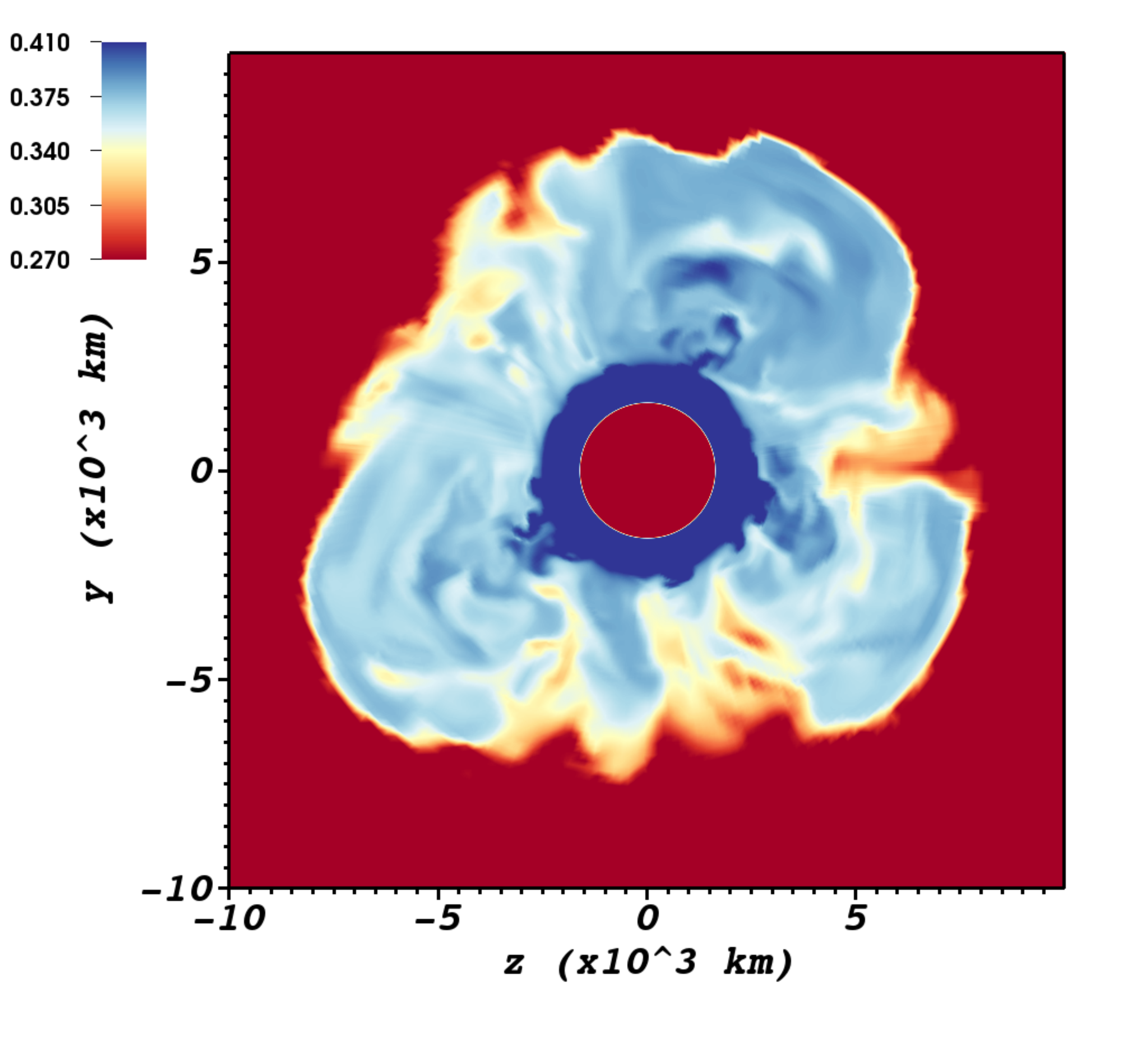}
  \caption{Slices showing the mass fraction
    $X_\mathrm{Si}$ of silicon at the onset of collapse
    in models s18-3Dr (left) and s18-3D (right). Both
    models are characterised by 2--3 silicon rich plumes (darker
    shades of blue. Due to the higher convective velocities,
    the boundary between the oxygen shell and the carbon shell
    is more strongly distorted by interfacial wave breaking
    in model s18-3D.)
\label{fig:x_si_ini}}
\end{figure*}

\section{Setup and Numerical Methods}
\label{sec:numerics}

\subsection{Initial Models}
We compute the collapse and post-bounce evolution of a 1D initial
model (s18-1D) and two 3D initial models (s18-3D and s18-3Dr) of an
$18 M_\odot$ solar metallicity progenitor with a helium core mass of
$5.3 M_\odot$.

s18-3D is a 3D pre-collapse model that has been obtained by simulating the
final 5 minutes of oxygen shell burning in 3D \citep{mueller_16c}.
As
\citet{mueller_16c} simulated only the region between mass coordinates
$m$ 
of $1.68 M_\odot$ and $4.07 M_\odot$ in 3D, data from the
corresponding 1D stellar evolution model is used outside this domain.

In order to better analyse the response of the accretion shock to the
amplitude of the pre-collapse perturbations, we have also constructed
another 3D progenitor model s18-3Dr with reduced convective
velocities. In this model, we assume a fixed temperature profile for
computing the nuclear reaction rates
during the last 5~minutes of the pre-collapse evolution (while taking
  composition changes into account). Otherwise
the setup is identical to s18-3D.

The resulting artificial reduction of the burning rate in s18-3Dr results
in smaller convective Mach numbers in the oxygen shell at collapse.
This is illustrated by radial
profiles of the root-mean-square average Mach number $\mathrm{Ma}_r$
of fluctuations of the radial velocity $v_r$ (Figure~\ref{fig:mach_ini}),
\begin{equation}
  \mathrm{Ma}_r=
  \frac{
    \left \langle (v_r -\langle v_r\rangle)^2 \right\rangle^{1/2}}
  {\langle c_\mathrm{s} \rangle},
\end{equation}
where angled brackets denote spherical Favre (i.e.\ density-weighted) averages.\footnote{
  We also adhere to this convention in the remainder of the paper except for spherical averages
of the density itself, which are always volume-weighted. }
The maximum value of $\mathrm{Ma}_r$ is less than $0.04$
in model s18-3Dr compared to $0.1$ in s18-3D.
On the other hand, the scales
of the convective flow at the pre-collapse stage are similar
for both models with 2--3 prominent silicon-rich plumes
(Figure~\ref{fig:x_si_ini}) and a convective velocity
field dominated by an $\ell=2$ mode.

In the case of the spherically symmetric model s18-1D, spherical symmetry
needs to be broken by hand in the supernova simulation.
To this end, we impose random seed perturbations
of $\delta v_r/v_r=5\times 10^{-3}$ onto the radial velocity field
around $110 \, \mathrm{ms}$ after the onset of collapse.

\subsection{Core-Collapse Supernova Simulations}
We compute the collapse and post-bounce evolution of the three initial
models with the \textsc{CoCoNuT-FMT} code \citep{mueller_15a}. The
hydro module \textsc{CoCoNuT} solves the equations of general
relativistic hydrodynamics in spherical polar coordinates using
piecewise parabolic reconstruction \citep{colella_84}, a hybrid
HLLC/HLLE Riemann solver \citep{mignone_05_a}, and second-order time
integration. It employs the xCFC approximation for the space-time
metric \citep{cordero_09} and currently assumes a spherically
symmetric metric for 3D simulations. To tame the coordinate
singularity at the grid axis, we use a mesh coarsening scheme with
variable resolution in the $\varphi$ direction depending on latitude
as in \citet{mueller_15b}. The interior of the proto-neutron star (at
densities higher than $5\times 10^{11} \, \mathrm{g}\, \mathrm{cm}^{-3}$) is
treated in spherical symmetry, and the effect of proto-neutron star
convection is captured by means of mixing-length theory using a
similar implementation as in \citet{mirizzi_16} and 
Bollig et al.\ (in preparation).

Neutrino transport is treated using the energy-dependent \textsc{FMT}
scheme of \citet{mueller_15a}, which makes various simplifying
assumptions, but still achieves reasonable quantitative agreement with
more sophisticated methods for many of the quantities relevant for the
supernova explosion problem (neutrino luminosities and mean energies,
heating conditions, etc.)  as shown in Appendix~A of
  \citet{mueller_15a}.  Technically, the FMT scheme involves the
  solution of the zeroth moment equation in fully decoupled energy
  groups under the assumption of stationarity. At high and
  intermediate optical depths, the flux factor is obtained from a
  two-stream solution of the Boltzmann equation in such a way as to
  produce the correct diffusion limit. Where the flux factor exceeds
  $1/2$, we use a two-moment closure based on the assumption maximum
  packing. Velocity-dependent terms in the transport equation are
  currently neglected (although gravitational redshift is included),
  and the ray-by-ray method \citep{buras_06a} is used to simplify the
  problem of multi-dimensional transport. Simplificatiions in the
  treatment of the neutrino microphysics include the omission of
  neutrino electron scattering, an approximate treatment of recoil
  energy transfer in neutrino-nucleon scattering by an effective
  absorption opacity, and the omission of absorption and emission
  processes other than nucleon bremsstrahlung (treated by an effective
  single-particle rate) for heavy flavour neutrinos. To better reproduce
the collapse dynamics of the inner iron core, the deleptonisation
scheme of \citet{liebendoerfer_05_b} is used during the collapse phase.
Modifications of neutrino nucleon reactions due to weak magnetism
and nucleon correlations are not taken into account in the present study.
For details we refer the reader to Appendix~A of \citet{mueller_15a}.

Our calculations are performed on a grid of $550 \times 128 \times
256$ zones in the $r$-, $\theta$- and $\varphi$-direction
(corresponding to an angular resolution of $1.4^\circ$) with a
non-equidistant radial grid extending out to $10^5 \, \mathrm{km}$.

We use the equation of state of \citet{lattimer_91} with a bulk
incompressibility modulus of $K=220 \, \mathrm{MeV}$ in the
high-density regime. In the low-density regime, we use an EoS
accounting for photons, electrons, and positrons of arbitrary
degeneracy and an ideal gas contributions from 17 nuclear species.
Nuclear statistical equilibrium is assumed above $5 \times 10^{9}\,
\mathrm{K}$, and nuclear reactions below this temperature are treated
using the approximate ``flashing'' method of \citet{rampp_02}.

\begin{figure}
  \includegraphics[width=\linewidth]{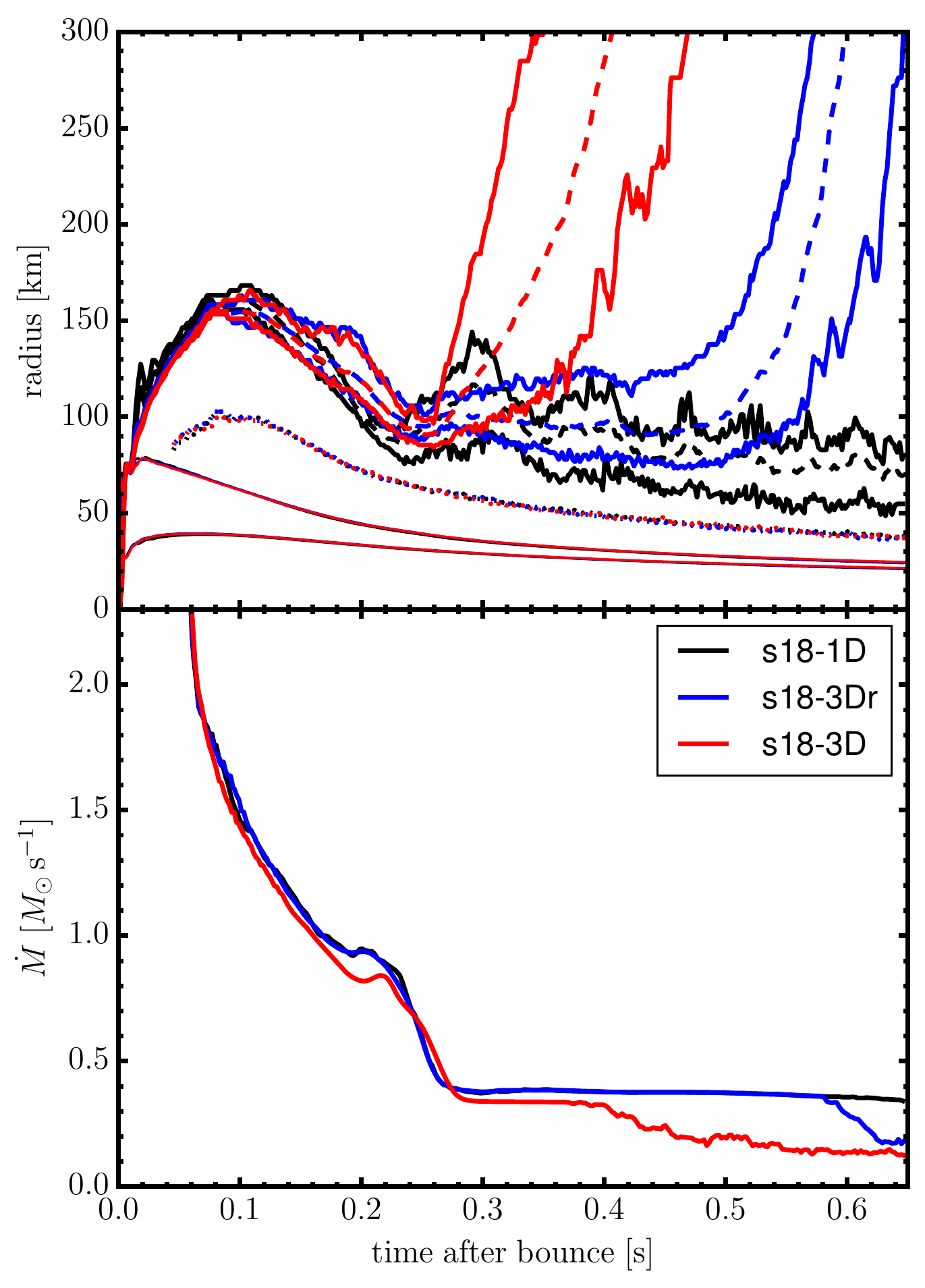}
  \caption{Top panel:
Evolution of the maximum, minimum and average shock
radius (thick solid and dashed curves),
the gain radius (dotted)
and the radii corresponding
to densities of $10^{11}\,\mathrm{g}\,\mathrm{cm}^{-3}$
and $10^{12}\,\mathrm{g}\,\mathrm{cm}^{-3}$ (thin solid lines)
for models s18-1D (black), s18-3Dr (blue),
and s18-3D (red). Bottom panel:
Mass accretion rate $\dot{M}$
for s18-1D, s18-3Dr, and s18-3D, measured
at a radius of $400 \, \mathrm{km}$. 
 \label{fig:shock_comparison}}
\end{figure}

\begin{figure}
  \includegraphics[width=\linewidth]{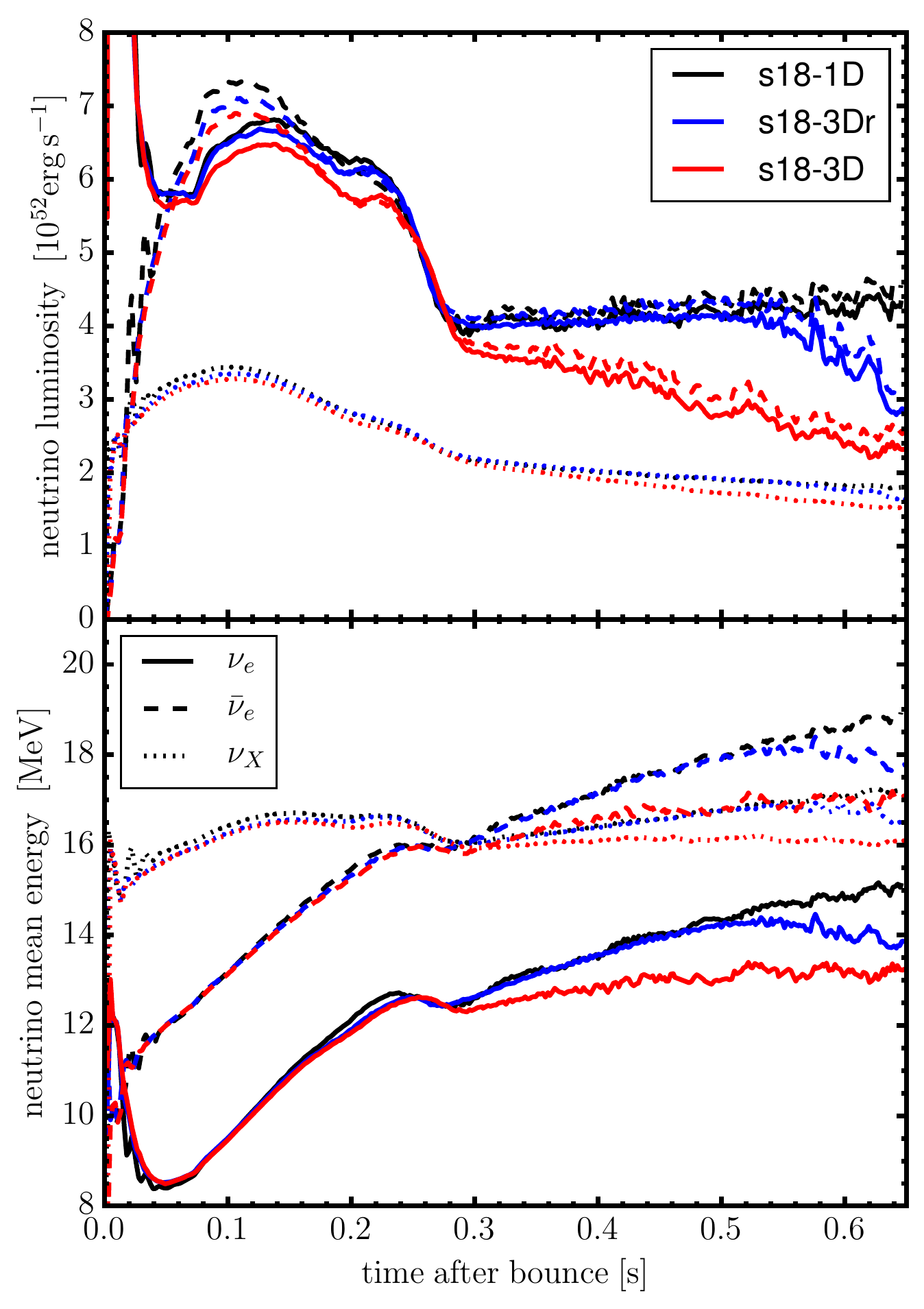}
  \caption{Neutrino luminosities (top panel) and mean energies of
    $\nu_e$ (solid), $\bar{\nu}_e$ (dashed), and heavy flavour
    neutrinos $\nu_X$ (dotted) in models s18-1D (black), s18-3Dr (blue), and
    s18-3D (red). Note that a slightly different development of prompt
    convection leads to differences in the neutrino luminosities at
    early times for $\bar{\nu}_e$, and that the slightly smaller
    accretion rate in s18-3D results in a small reduction of the
    electron flavour luminosity between $0.1 \, \mathrm{s}$ and $0.24
      \, \mathrm{s}$ after bounce. Models s18-3D and s18-3Dr exhibit a
      drop in the electron flavour luminosity and a slower rise of the
      mean energies after shock revival. \label{fig:neutrino}}
\end{figure}

\begin{figure*}
  \centering
  \includegraphics[width=0.32\textwidth]{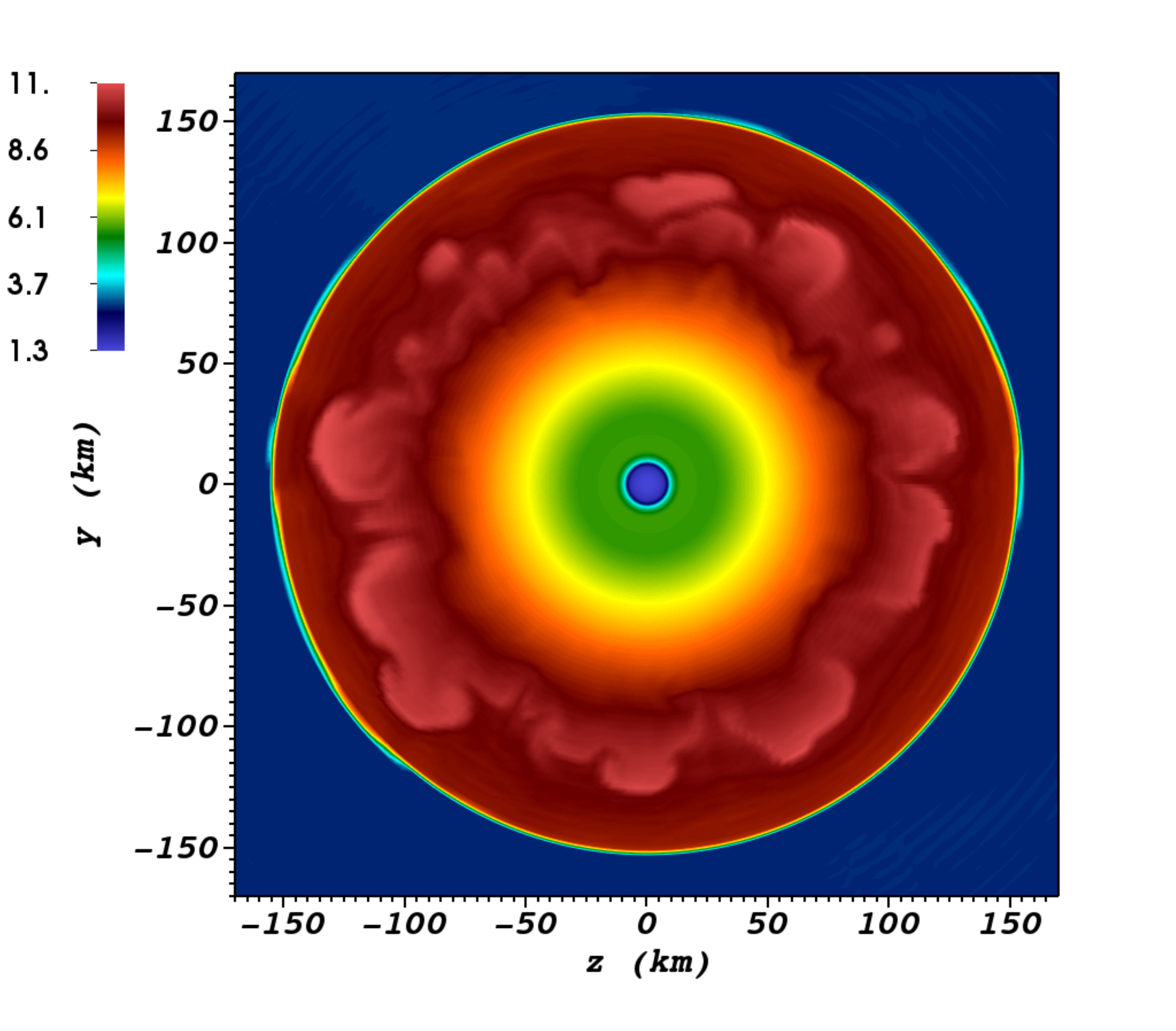}
  \includegraphics[width=0.32\textwidth]{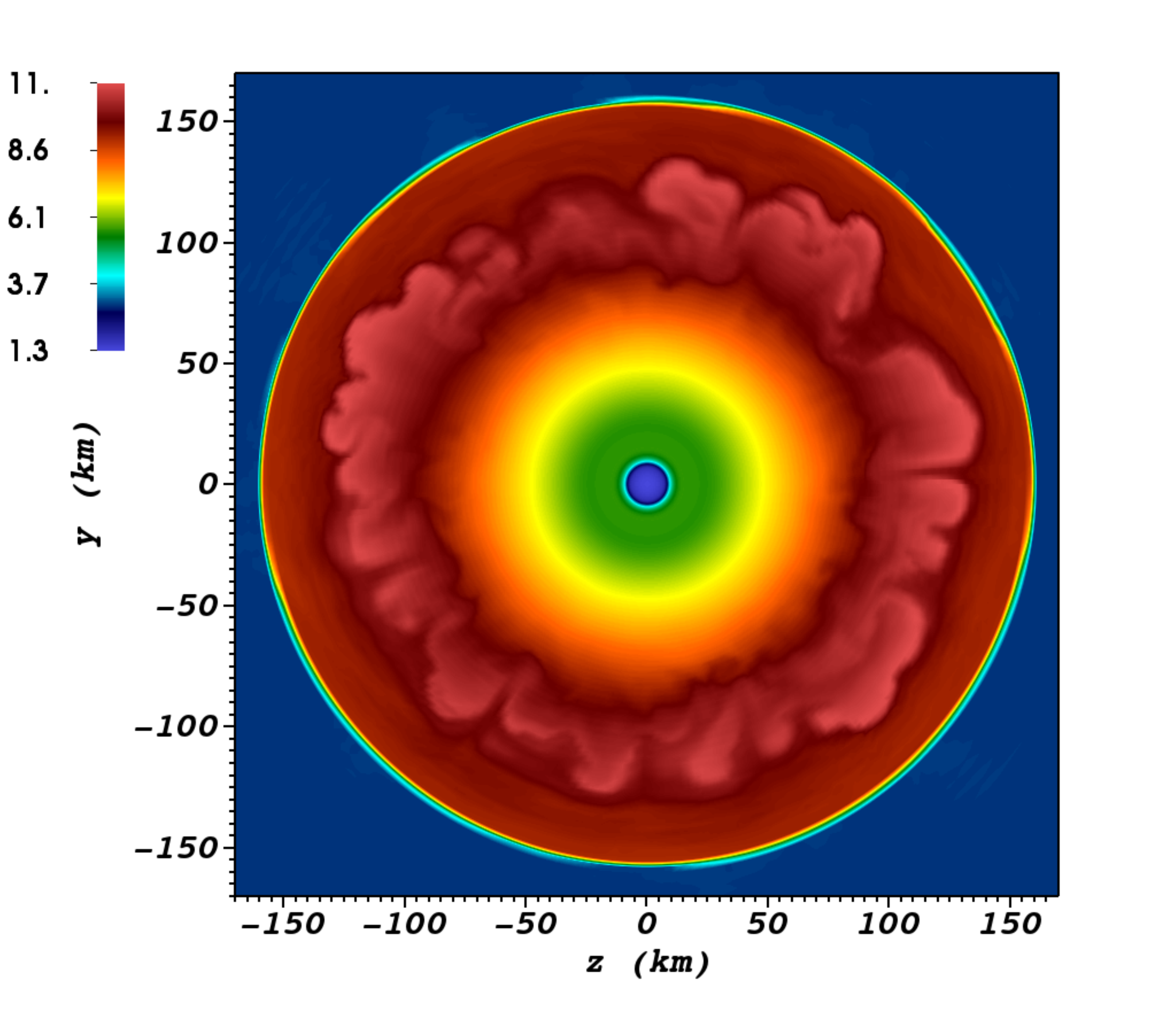}
  \includegraphics[width=0.32\textwidth]{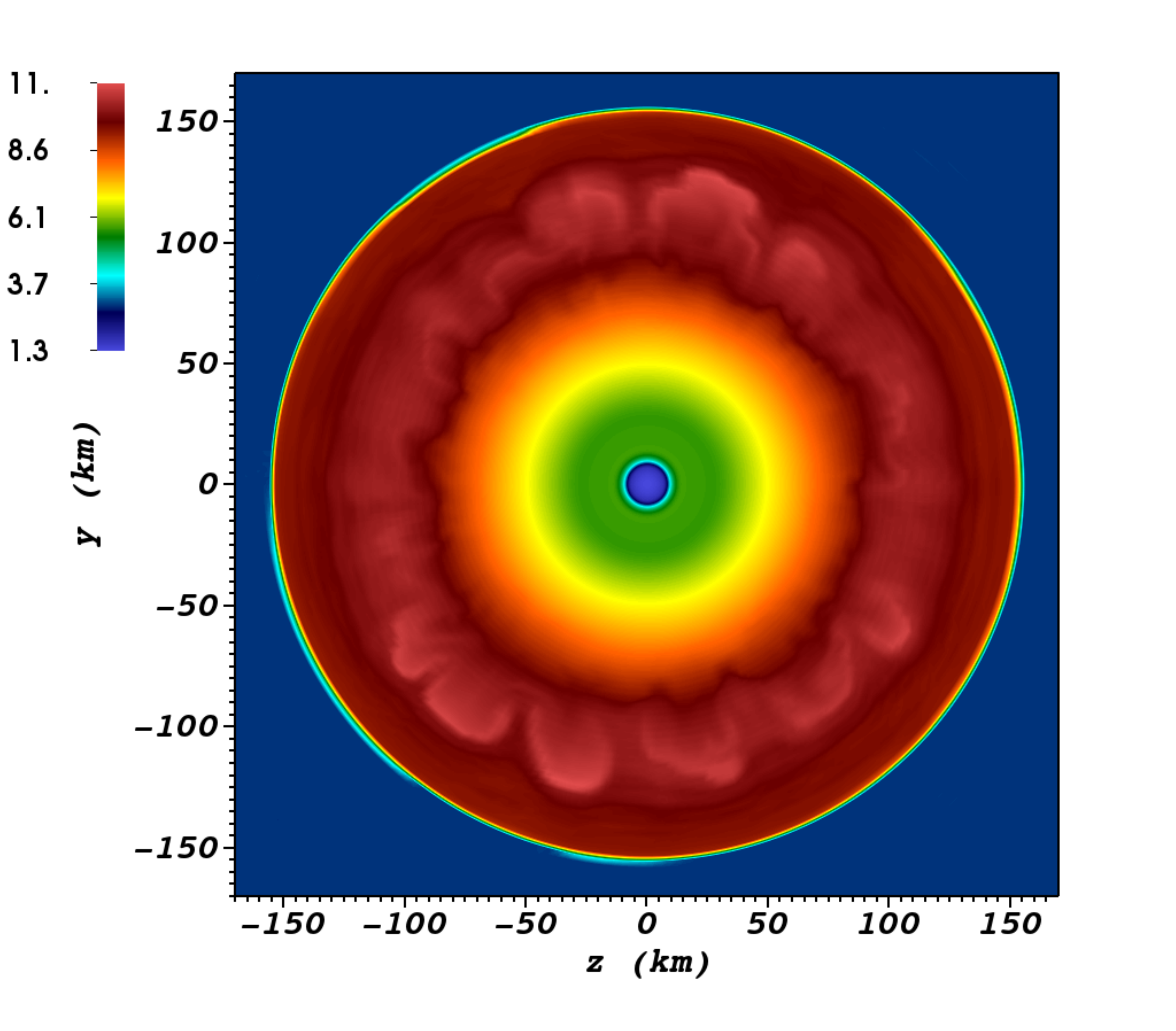}\\
  \includegraphics[width=0.32\textwidth]{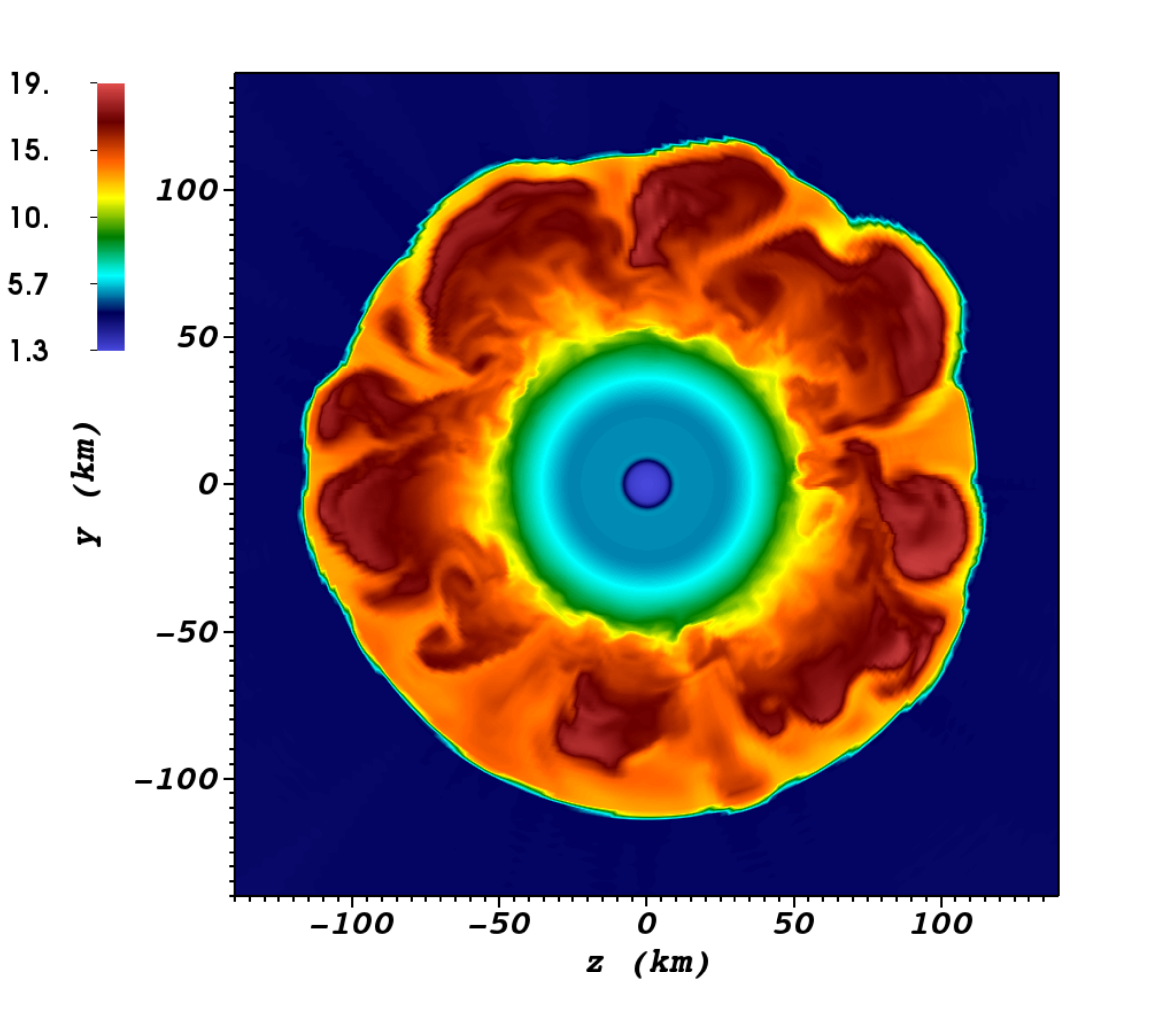}
  \includegraphics[width=0.32\textwidth]{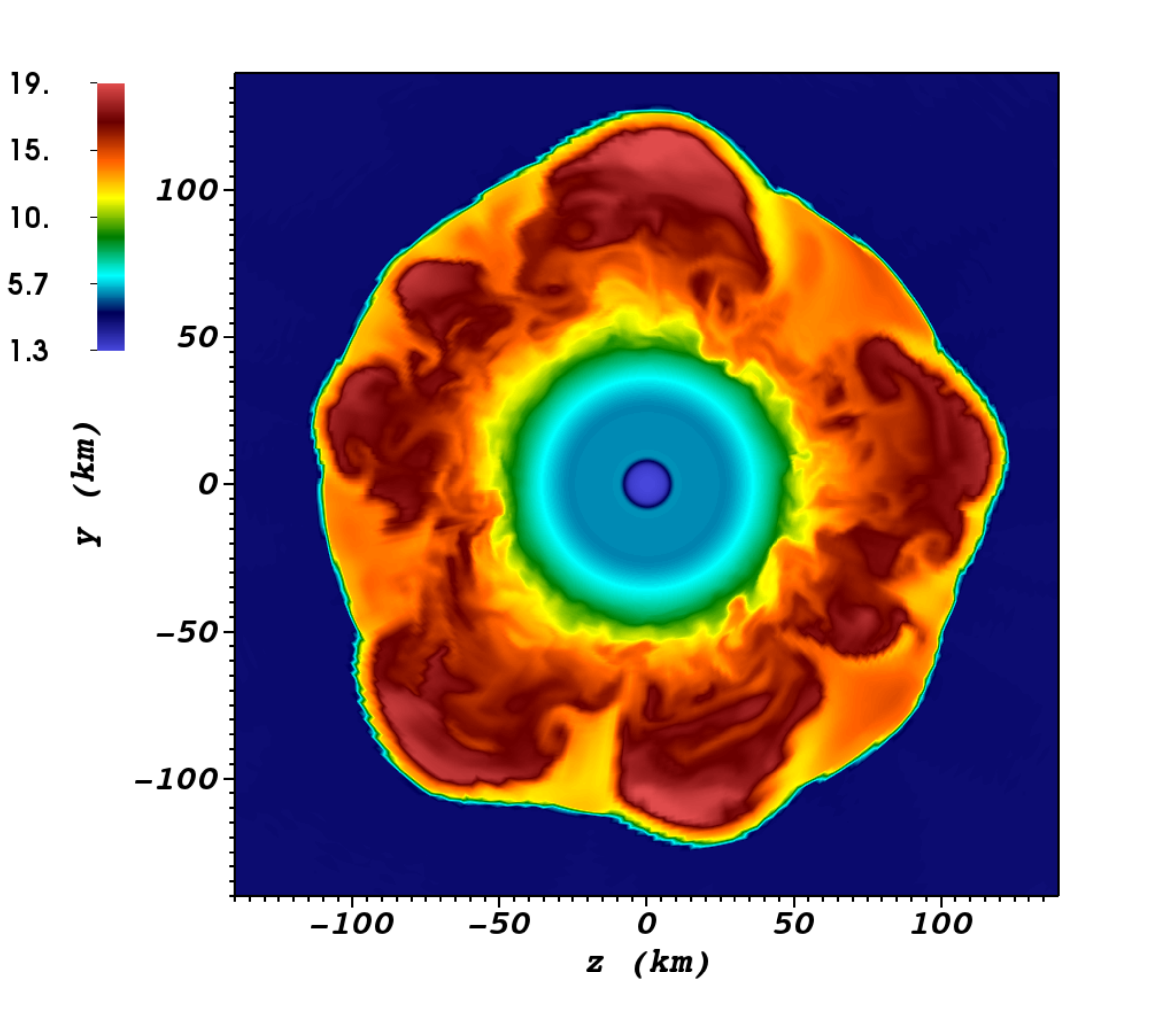}
  \includegraphics[width=0.32\textwidth]{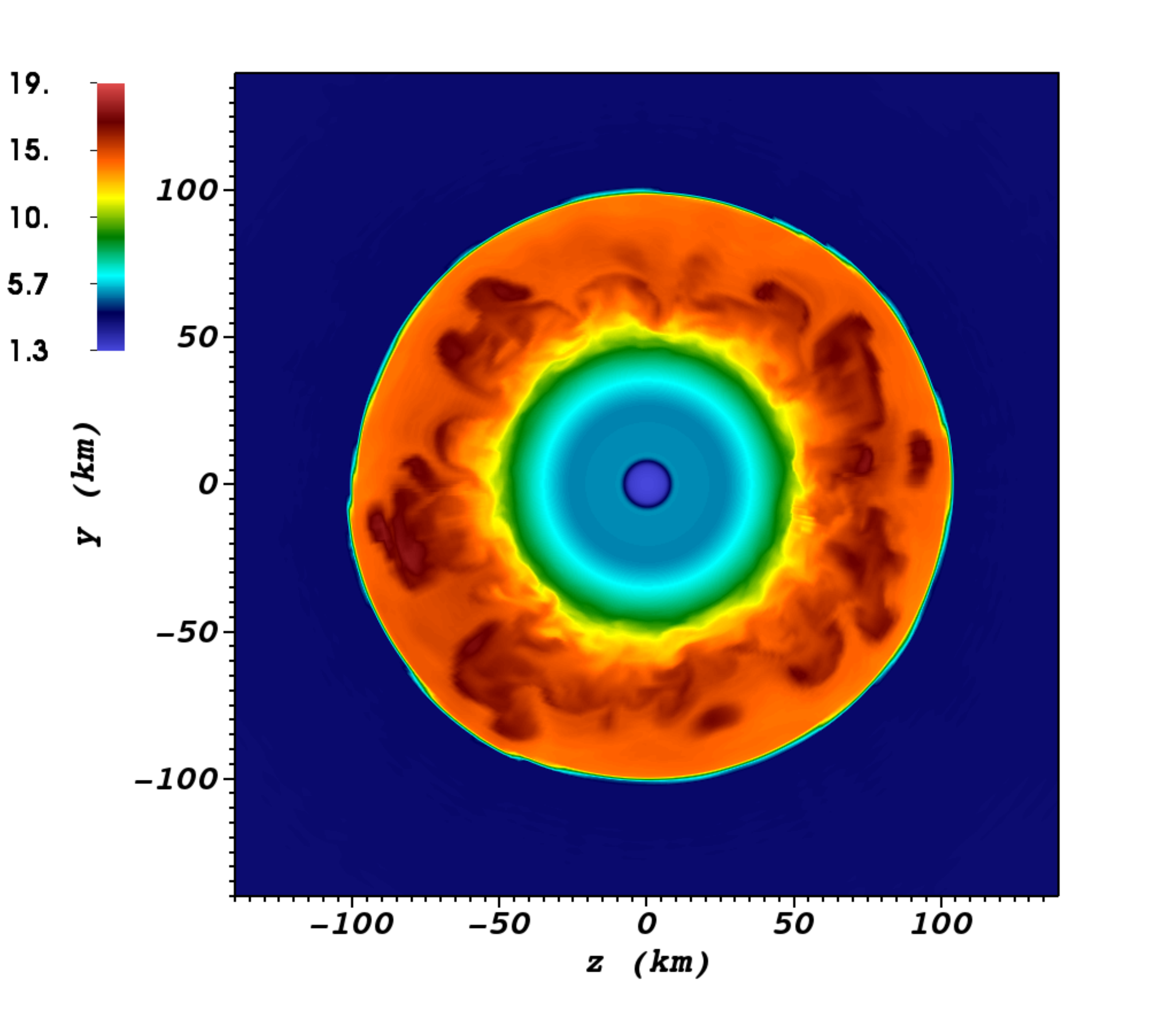}\\
  \includegraphics[width=0.32\textwidth]{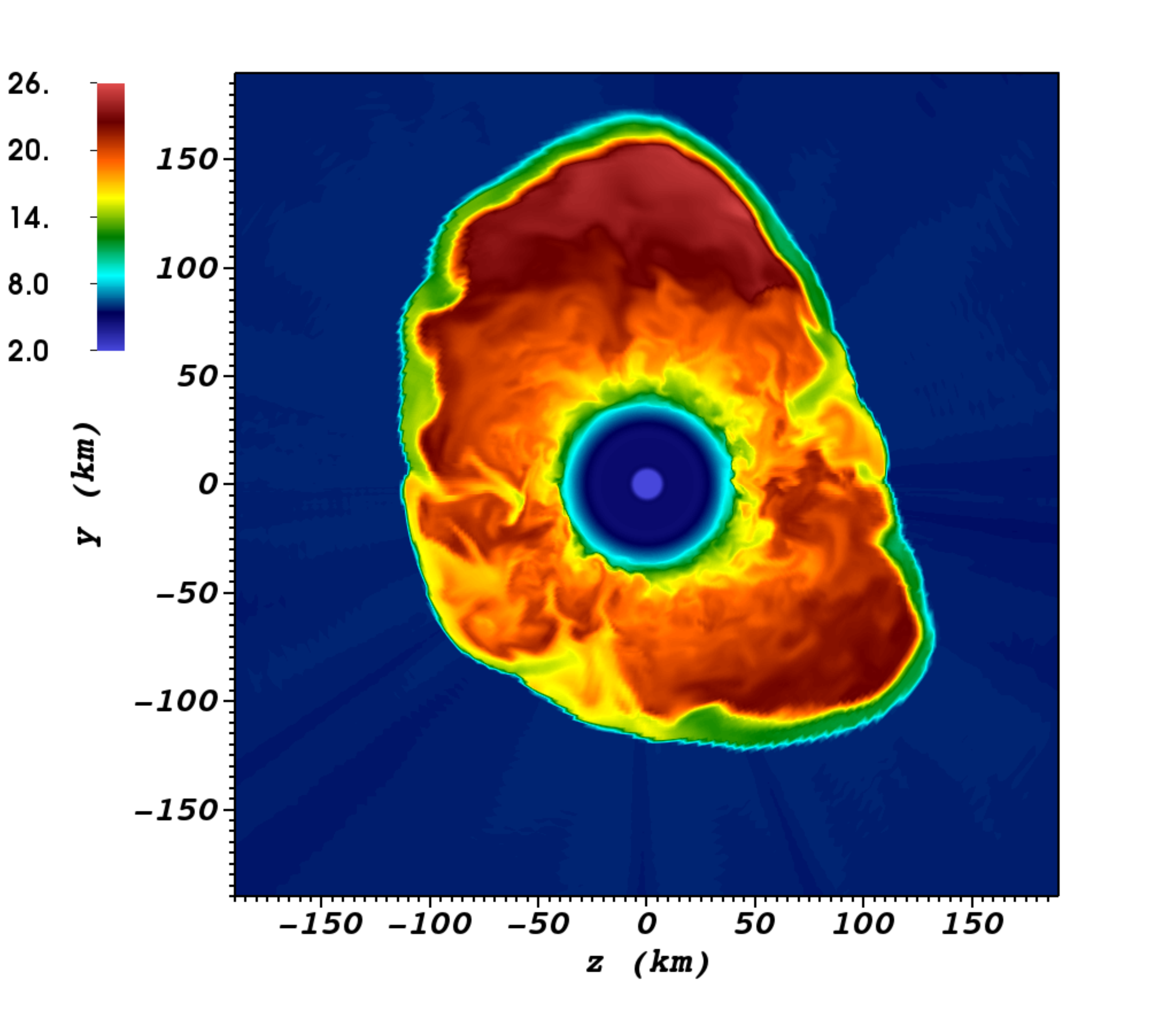}
  \includegraphics[width=0.32\textwidth]{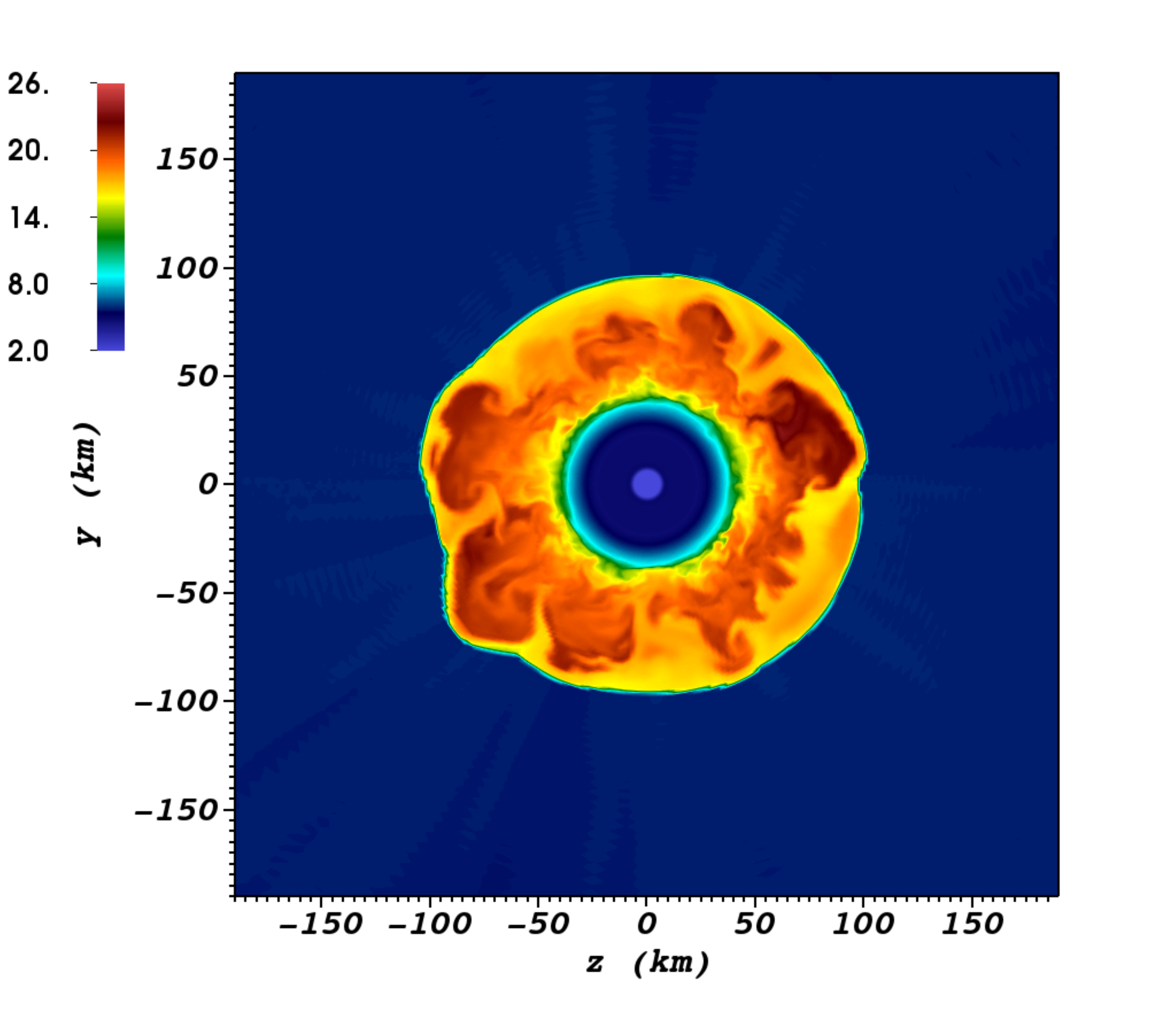}
  \includegraphics[width=0.32\textwidth]{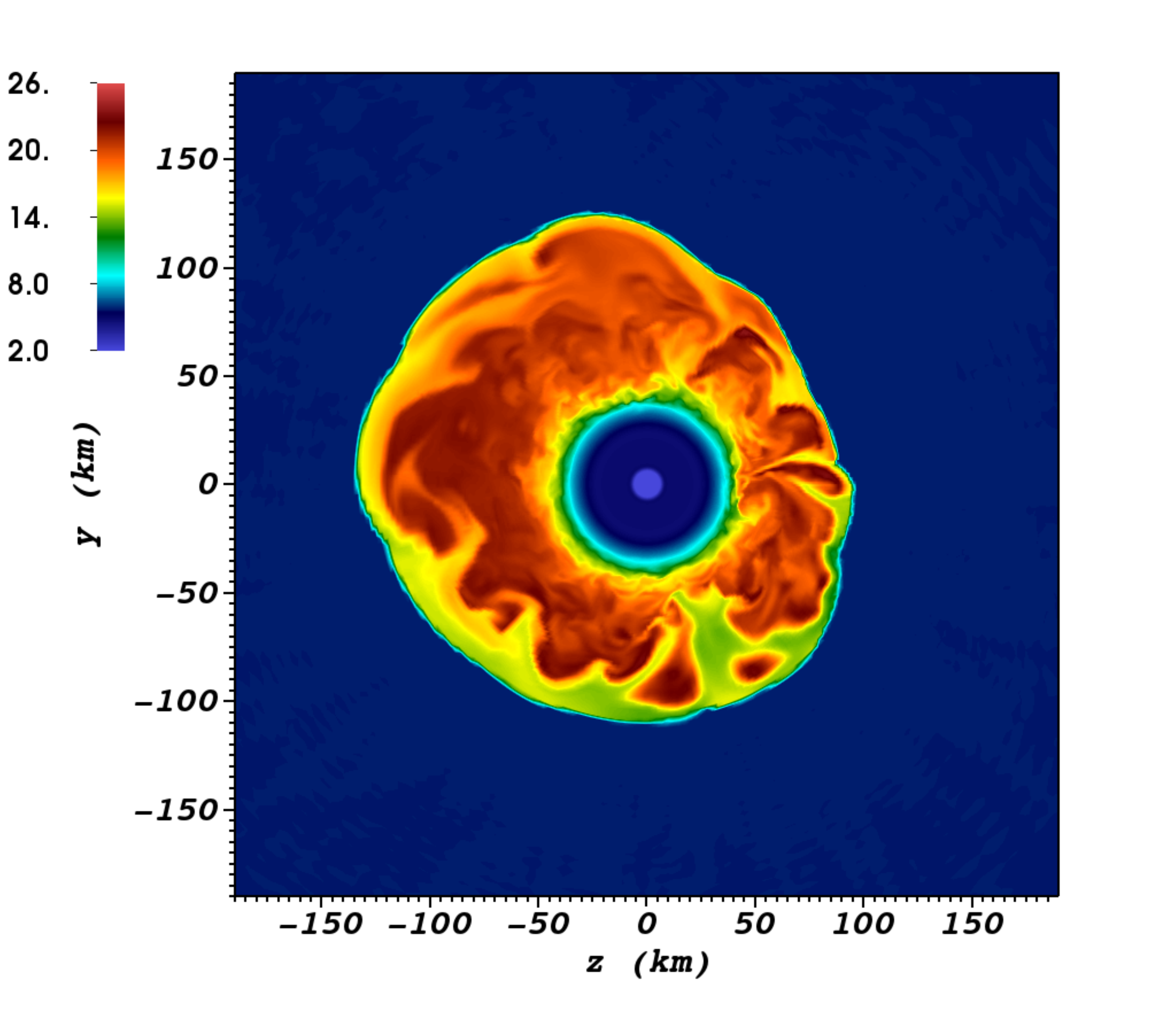}\\
  \includegraphics[width=0.32\textwidth]{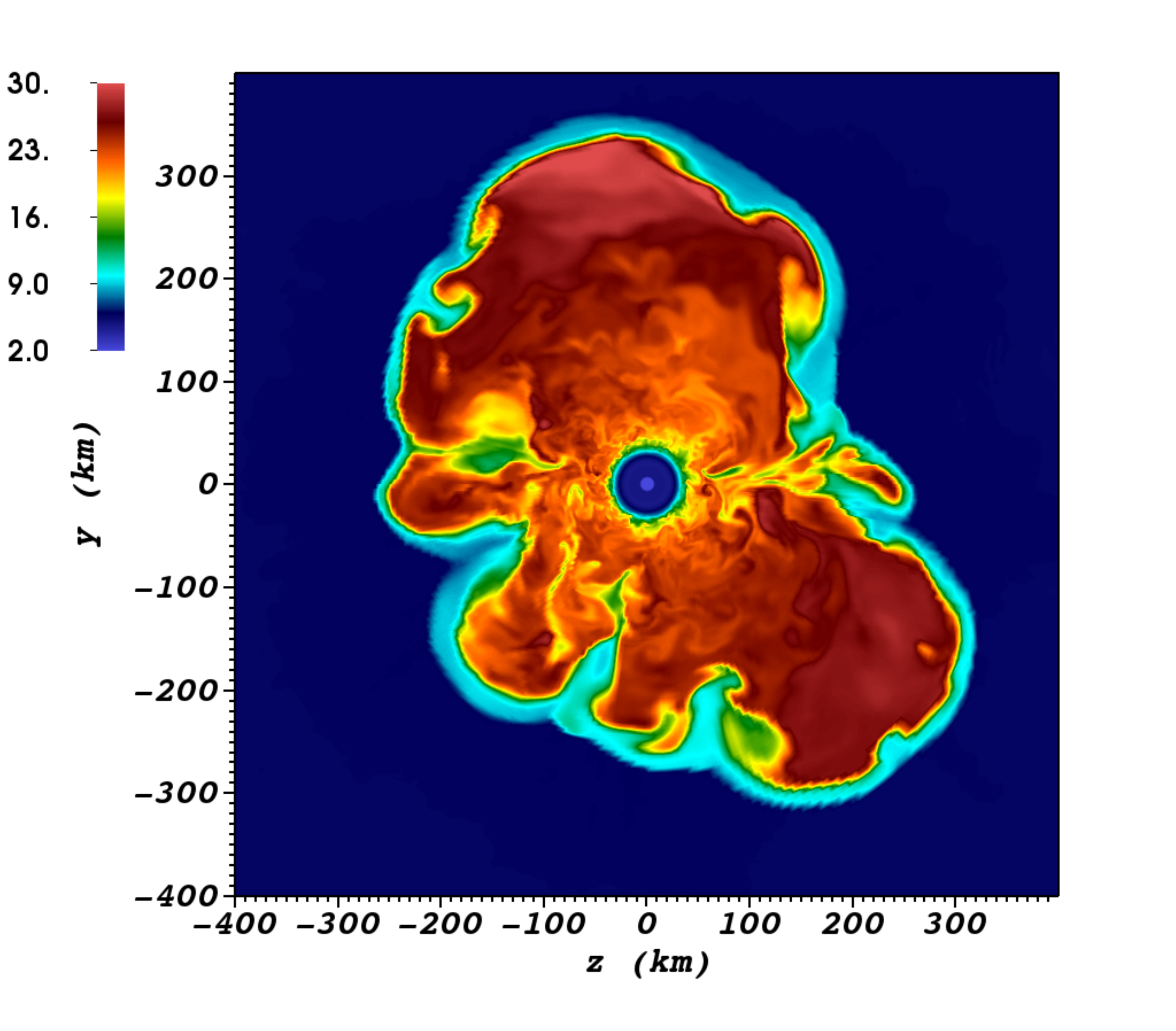}
  \includegraphics[width=0.32\textwidth]{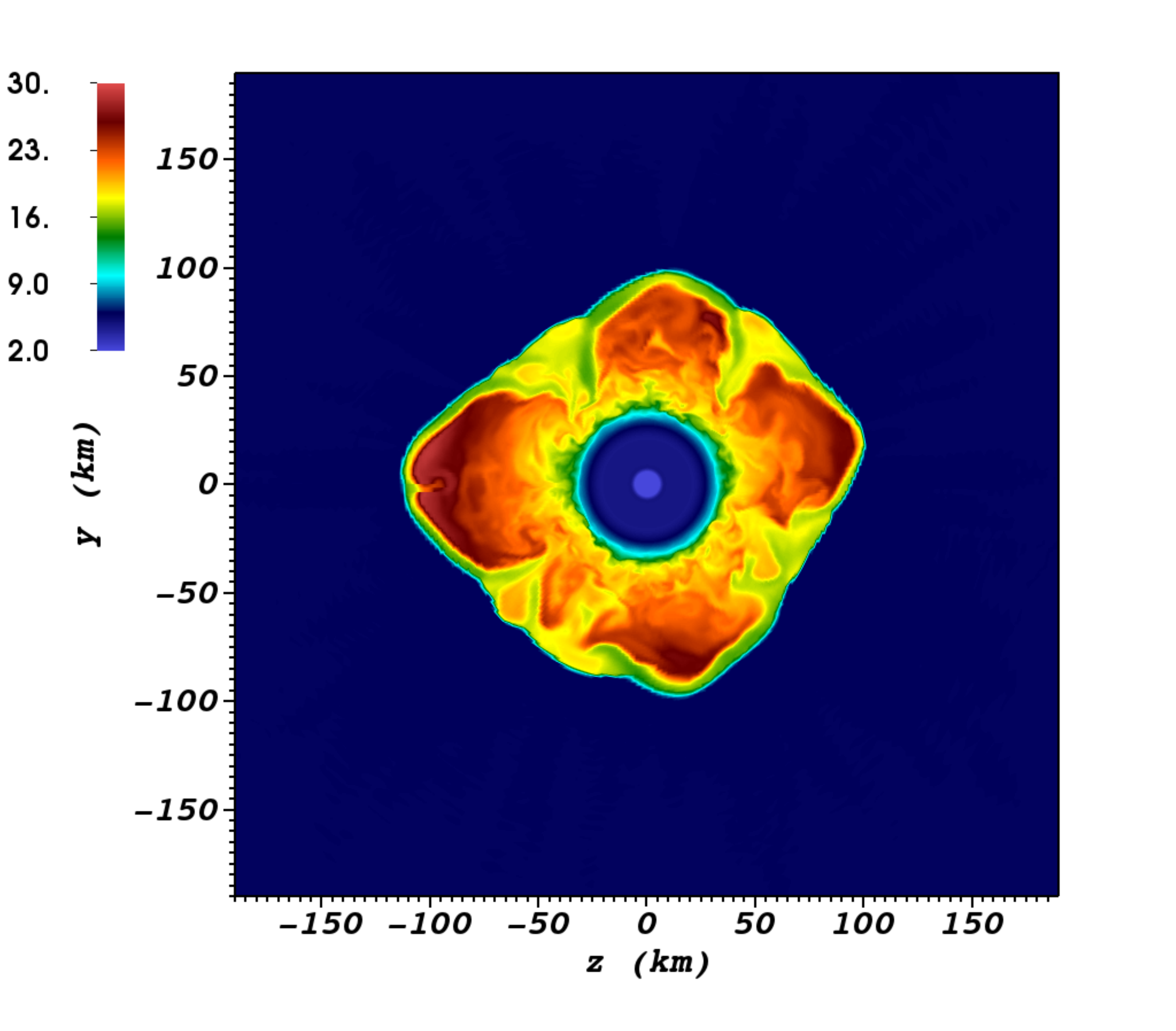}
  \includegraphics[width=0.32\textwidth]{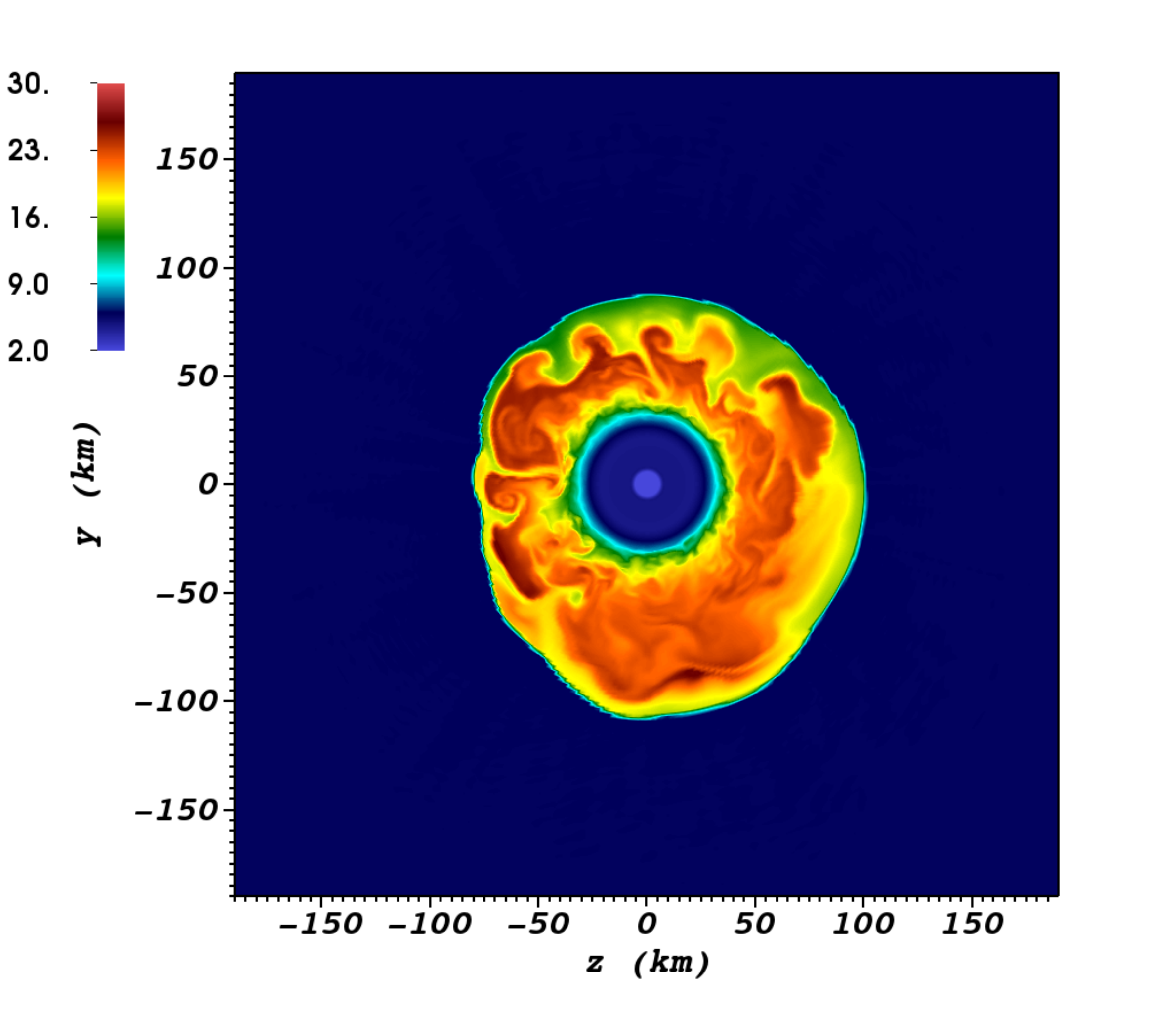}\\
  \caption{Entropy in the $z$-$y$-plane 
at $80\, \mathrm{ms}$,
$200\, \mathrm{ms}$,
$300\, \mathrm{ms}$, and
$400\, \mathrm{ms}$ after bounce (top to bottom) in
models
s18-3D (left column), s18-3Dr (middle column), and
s18-1D (right column).
At $80 \, \mathrm{ms}$ (top row),
all three models shows the development
of neutrino-driven convection without any noticeable
influence of initial perturbations.
At $200 \, \mathrm{ms}$ (second) row, infalling perturbations
start to interact with large-scale SASI oscillations
in s18-3D and s18-3Dr.
At $300 \, \mathrm{ms}$ (third row),
s18-3D is on the way to shock-revival aided by
forced-shock deformation, and infalling perturbations
largely destroy SASI oscillations in s18-3Dr.
At $400 \, \mathrm{ms}$ (bottom row), model s18-3Dr also 
exhibit strong forced shock deformation, while
model s18-1D continues to be dominated by the SASI 
spiral mode.
\label{fig:entropy_slices}}
\end{figure*}

\section{Impact of Initial Perturbations on Shock Revival}
\label{sec:comparison}
\subsection{Qualitative Impact on Shock Evolution and
Hydrodynamics Instabilities}
\label{sec:shock_evolution}
To illustrate the role of initial perturbations in models s18-3D,
s18-3Dr, and s18-1D, we compare the evolution of the shock, gain, and
proto-neutron star radii and the mass accretion rate $\dot{M}$
in Figure~\ref{fig:shock_comparison}.
We also show neutrino luminosities $L_\nu$ and mean energies $E_\nu$ for
all three models in Figure~\ref{fig:neutrino}.
Meridional slices of the entropy
for all three models at selected times are presented in
Figure~\ref{fig:entropy_slices}.

As expected, differences between the 3D models are minute at early
times. A minor peculiarity of model s18-1D is the development of more
violent prompt convection and shock ringing prior to $50 \,
\mathrm{ms}$ after bounce. This behaviour is connected to the
imposition of random seed perturbations in s18-1D on the entire grid,
i.e.\ also in the Fe and Si core, which is not explicitly perturbed in
models s18-3D and s18-3Dr. Moreover, patching the 3D O shell burning
simulation and the core of the 1D stellar evolution model together
results in slight hydrostatic adjustment in model s18-3D, which
slightly reduces the mass accretion rate and the electron flavour
luminosity (top panel of Figure~\ref{fig:neutrino}) compared to s18-1D
and s18-3Dr.  Despite these differences, the shock trajectories in the
three models nonetheless converge again after this transient phase of
prompt convection. $80 \, \mathrm{ms}$ after bounce (top row
Figure~\ref{fig:entropy_slices}), they all show very similar shock
radii and incipient neutrino-driven convection with small-scale plumes
of similar size.

\begin{figure}
  \centering
  \includegraphics[width=\linewidth]{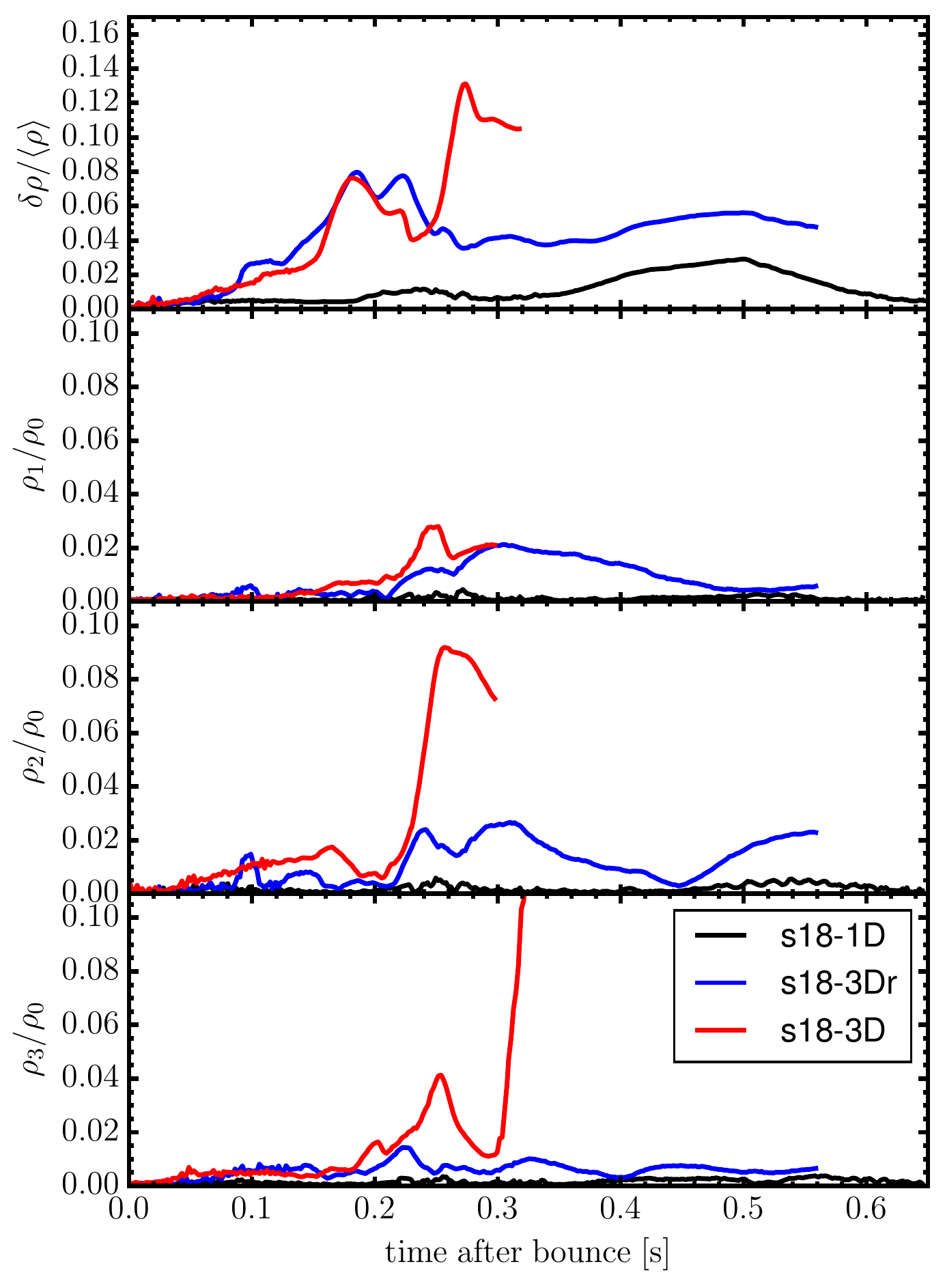}
  \caption{ RMS deviation $\delta\rho/\langle\rho\rangle$ of the density from its
    spherical average and normalised power $\rho_\ell/\rho_0$
    (with $\rho_0=\langle \rho\rangle$) in the
    first multipoles of the pre-shock density perturbations computed
    according to Equation~(\ref{eq:multipoles_rho})
for models s18-1D (black), s18-3Dr (blue) and s18-3D (red). Since the density
    perturbations are evaluated at a radius of $250 \, \mathrm{km}$,
    no data are shown once the maximum shock radius exceeds this value
    in models s18-3Dr and s18-3D. 
\label{fig:dflukt}}
\end{figure}

\begin{figure}
  \centering
  \includegraphics[width=\linewidth]{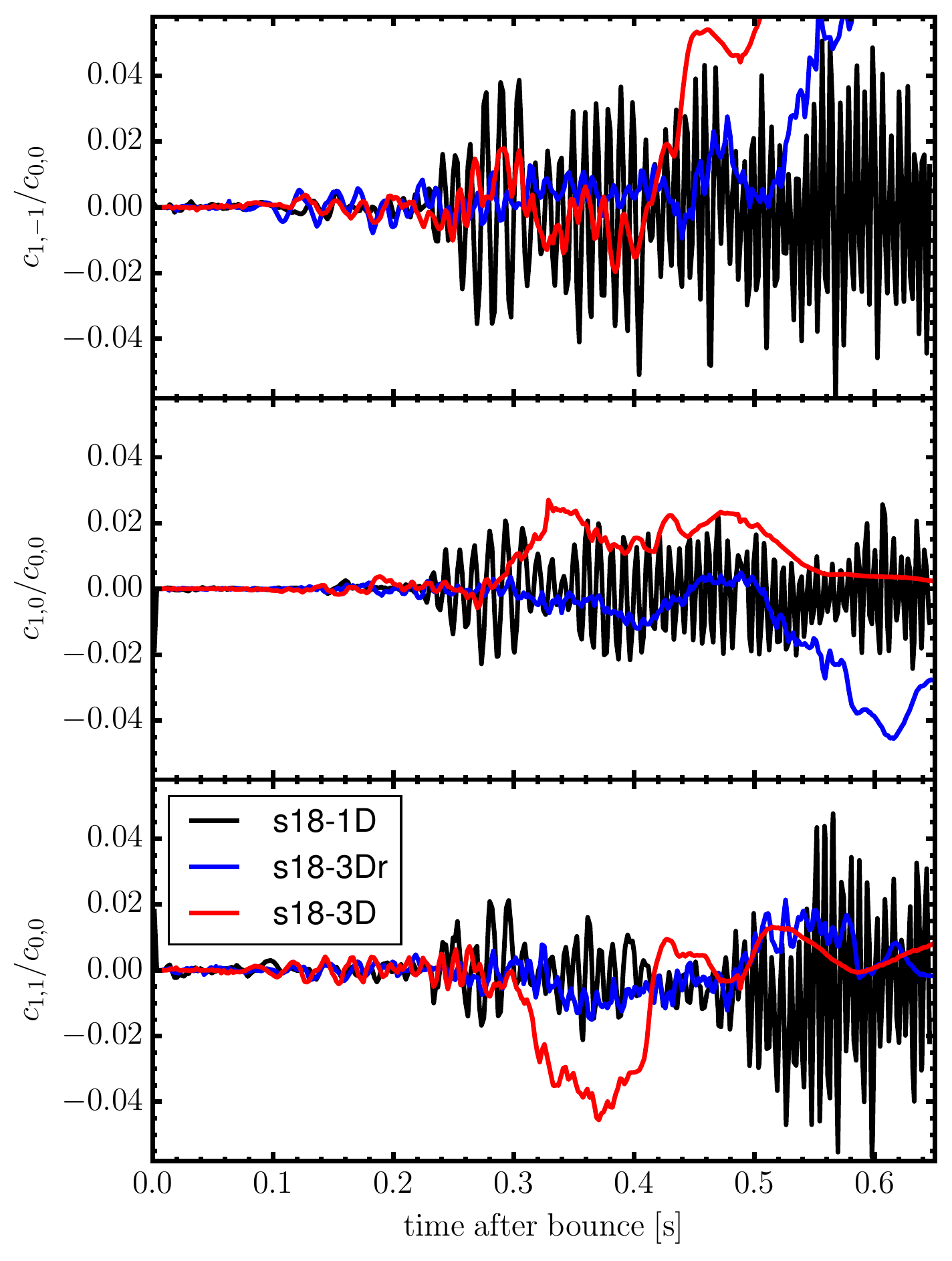}
  \caption{Normalised coefficients $c_{1,m}$ for the dipolar shock
    deformation modes with different $m$ computed according to
    Equation~(\ref{eq:multipoles}) for models s18-1D (black), s18-3Dr
    (blue), and s18-3D (red).  The dipole coefficients illustrate the
    interaction of infalling large-scale perturbations with the SASI,
    which dominates the unperturbed model s18-1D.  Initially (from
    about $80 \, \mathrm{ms}$ to $220 \, \mathrm{ms}$), the infalling
    perturbations are not harmful to the SASI and can even enhance
    quasi-periodic shock oscillations. As the infalling large-scale
    perturbations become stronger, regular SASI oscillations are first
    reduced in amplitude and then give way completely to forced shock
    deformation with small stochastic oscillations on top.
\label{fig:sasi}}
\end{figure}

\begin{figure}
\centering
  \includegraphics[width=\linewidth]{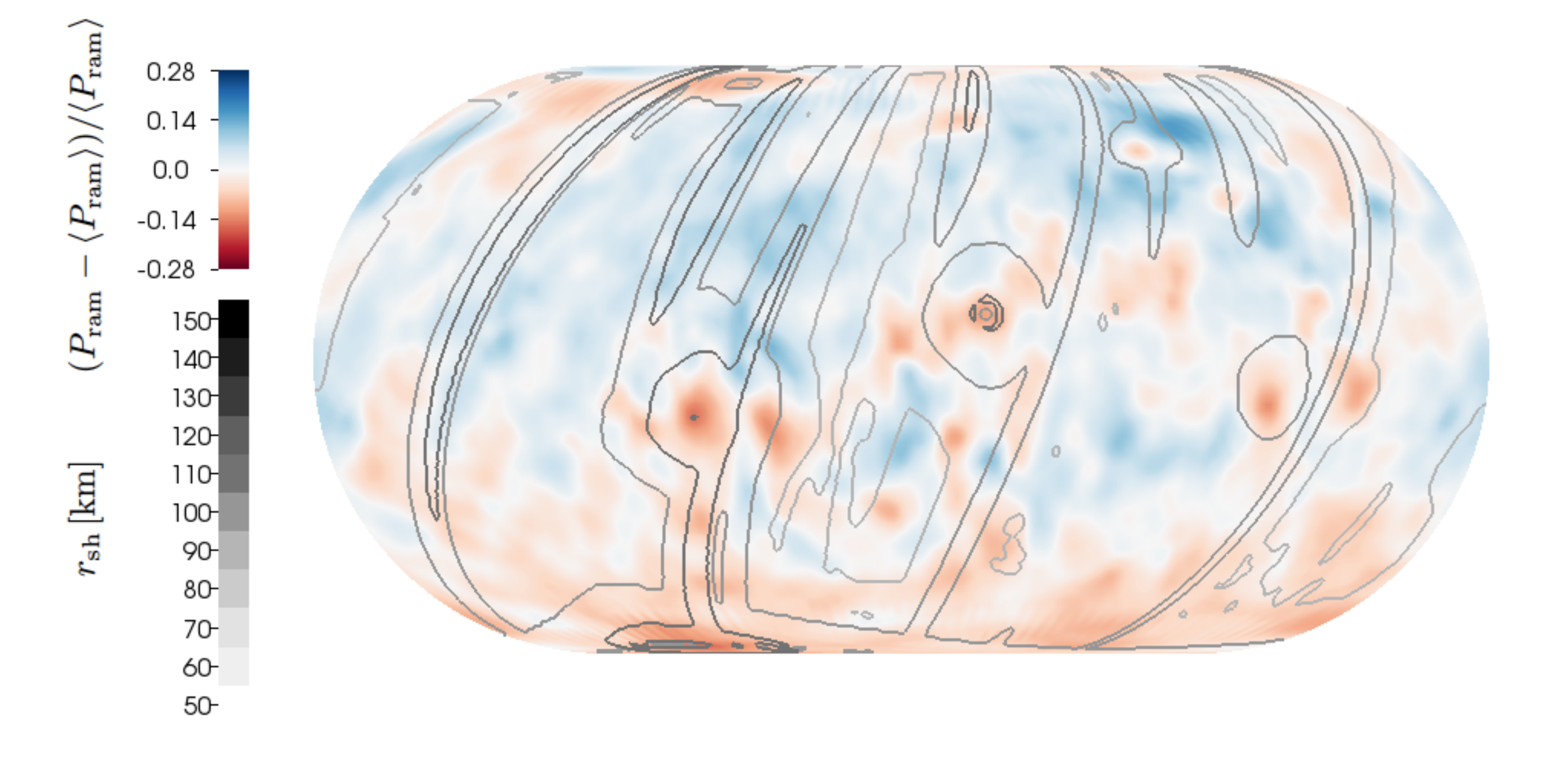}\\
  \includegraphics[width=\linewidth]{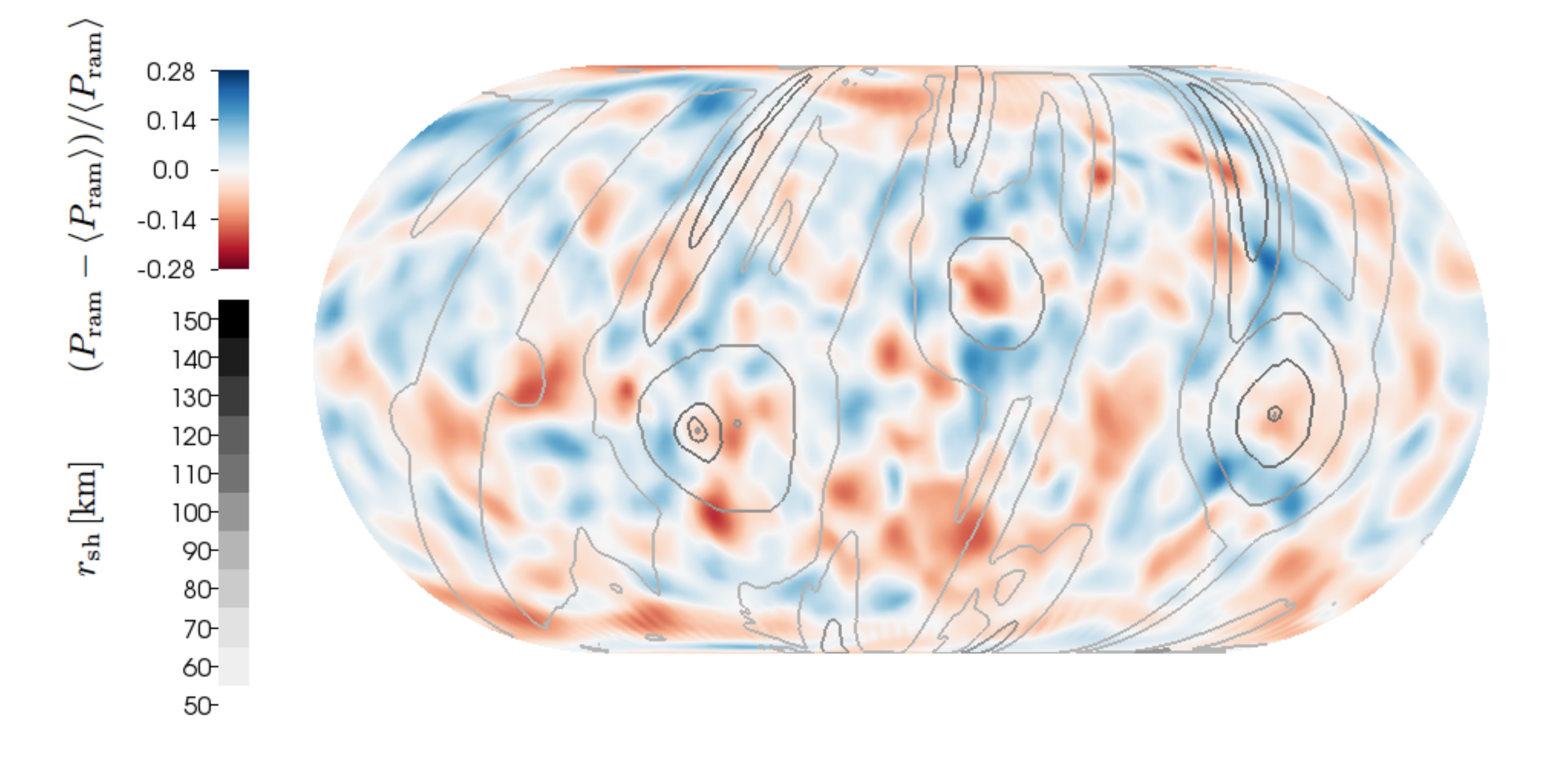}\\
  \includegraphics[width=\linewidth]{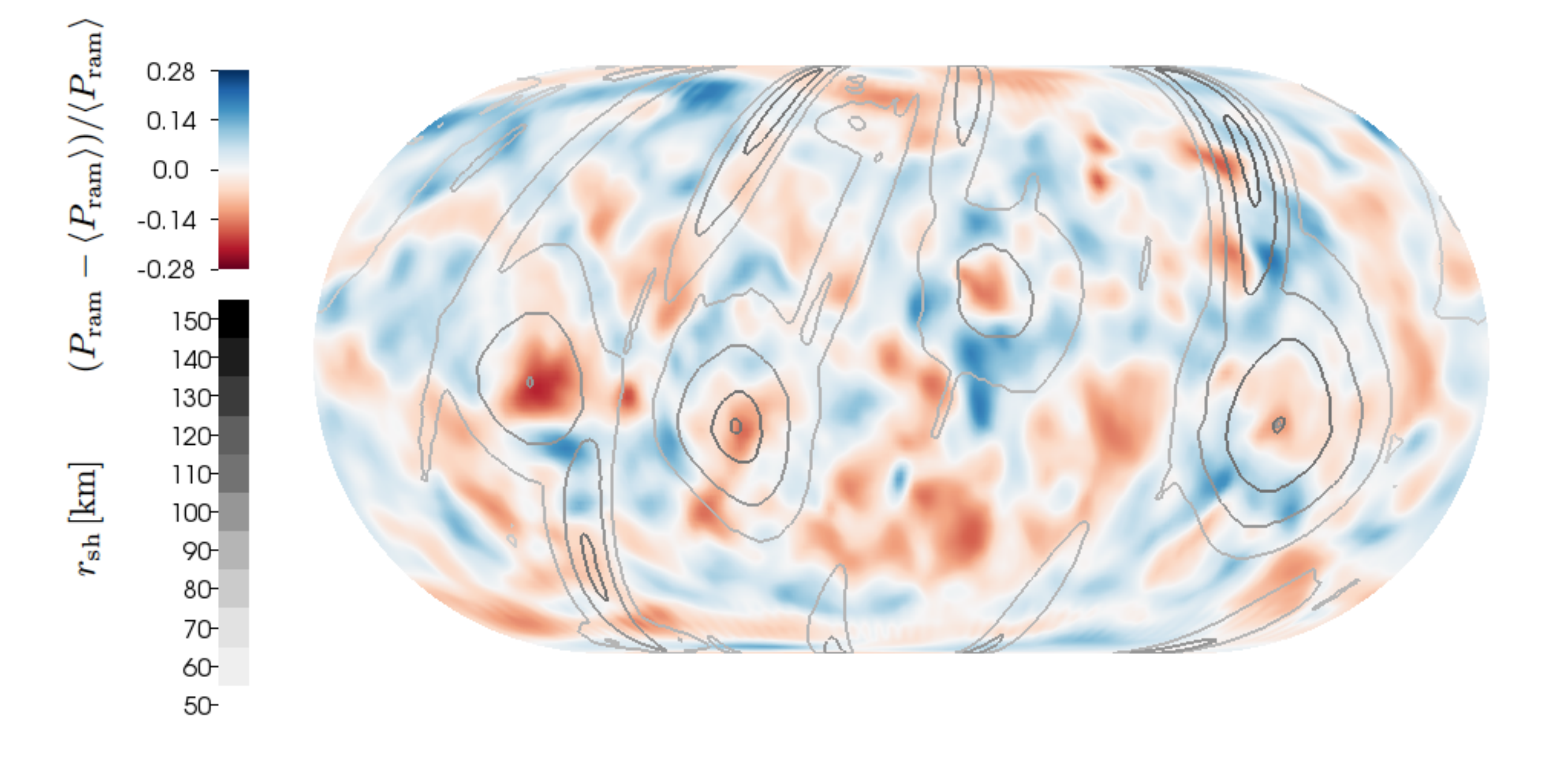}\\
  \includegraphics[width=\linewidth]{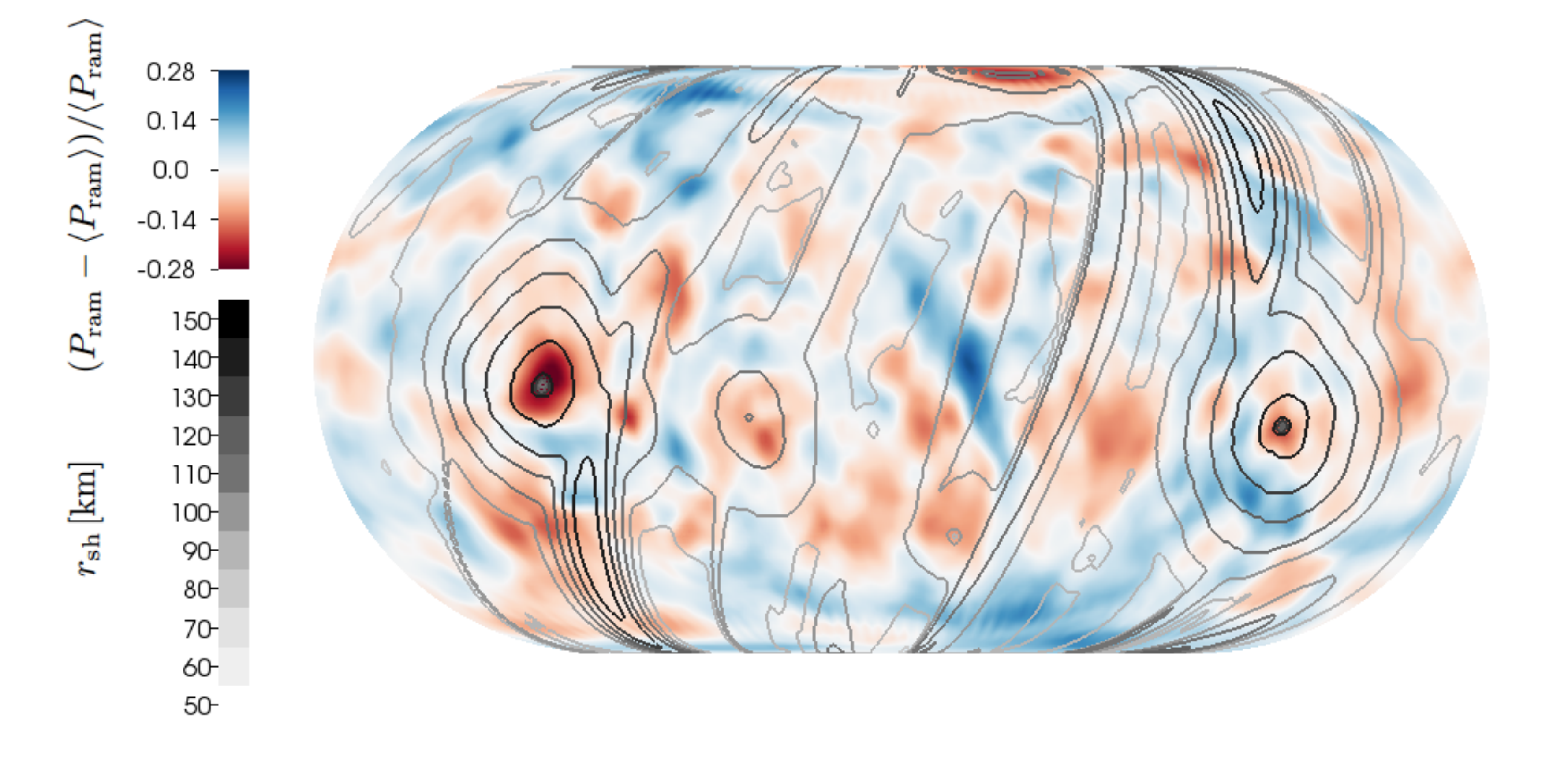}\\
  \caption{Aitoff projections of the
    relative variations in pre-shock ram pressure
    $(P_\mathrm{ram}-\langle P_\mathrm{ram}\rangle)/\langle P_\mathrm{ram} \rangle$
    at a radius of $170 \, \mathrm{km}$
    (encoded on a red-white-blue colour scale)
    and isocontours of the angle-dependent shock position
$r_\mathrm{sh}$
(in units of $\mathrm{km}$)
    in model s18-3Dr
    at post-bounce times of $326\, \mathrm{ms}$,
$408\, \mathrm{ms}$,
$428\, \mathrm{ms}$, and
$482\, \mathrm{ms}$. 
The poles lie on the $z$-axis, and
the central meridian is aligned with
the $x$-$z$-plane.
As model s18-3Dr remains
    in the regime of forced shock deformation for about
    $200 \, \mathrm{ms}$, it nicely illustrates the
    adjustment of the shock geometry to changes in the
    infalling perturbations. The shock
    initially develops three protrusions
    in the equatorial plane and
    one at the South pole due to the low density
    and ram pressure ahead of the shock
    (top). The three protrusions in the
    equatorial plane remain stable despite
    the emergence of stronger underdensities
    ahead of the shock in other spots (second panel),
    although one of these original
    protrusions gradually disappears in favour
    of a strong bulge in the Western hemisphere
    (third and fourth panel).
  \label{fig:dpram_and_rsh}}
\end{figure}

The evolution of the models starts to diverge around $150 \,
\mathrm{ms}$ after bounce with slightly larger shock radii in models
s18-3D and s18-3Dr, This is well before the arrival of the Si/O shell
interface at $200\text{-}250 \,\mathrm{ms}$, but convection
in the O shell can already make itself felt by generating
density perturbations in the 
stable Si shell (via g-mode excitation), which will then
undergo amplification during collapse \citep{lai_00}.
Shortly before the arrival of the Si/O shell interface,
the density perturbations ahead of the shock can already
become sizeable. To quantify the level of pre-shock
density perturbations, we evaluate
the root-mean-square (RMS) deviation of the density $\rho$ from  its spherical average $\langle\rho\rangle$
\begin{equation}
\label{eq:delta_rho}
\delta \rho(r) =\left [\frac{\int (\rho-\langle \rho\rangle )^2\,\ud
    \Omega}{4\pi} \right]^{1/2},
\end{equation}
at a radius  $r$ of $250 \, \mathrm{km}$
(Figure~\ref{fig:dflukt}, top panel).

Prior to the arrival of the Si/O interface, the scale of the infalling
density perturbations remains small, however.  Only once the O shell
reaches the shock do we observe large-scale density perturbations in
s18-3D and s18-3Dr with angular wavenumbers $\ell\approx 2$
corresponding to the dominant convective eddies in the pre-collapse
models. This is illustrated by Figure~\ref{fig:dflukt} 
(bottom panel), which shows the normalised power
$\rho_\ell/\rho_0$ of the density perturbations
for the lowest multipoles. Here, $\rho_\ell$ is computed as
\begin{equation}
\label{eq:multipoles_rho}
\rho_\ell(r)
=
\left(\sum_{m=-\ell}^\ell
\left | \int \rho Y^*_{\ell,m} (\theta,\varphi)
\, \ud \Omega\right|^2\right)^{1/2}
\end{equation}
where  $Y_{\ell,m}$ are real spherical harmonics
(so that $Y_{\ell,m}^\star (\theta,\varphi)=Y_{\ell,m}(\theta, \varphi)$)
with the same normalisation as in \citet{burrows_13}
(see their Equation~2).

The interaction of the infalling perturbations with the shock turns
out to be very subtle upon close examination and evolves through
several distinct regimes as the strength of the perturbations
and the character of the post-shock flow in the unperturbed model
change. Initially  (second row of Figure~\ref{fig:entropy_slices} $200\,
\mathrm{ms}$), the unperturbed model shows low-amplitude
SASI oscillations with buoyancy-driven plumes emerging
in some directions, which is suggestive of a transition regime
between strongly SASI-dominated and convection-dominated flow.
The infalling perturbations that interact
with the oscillating shock are still small during this phase.
Their quantitative impact on the SASI oscillation can be
gauged by decomposing the angle-dependent shock position into
spherical harmonics $c_{\ell,m}$ (Figure~\ref{fig:sasi}),
\begin{equation}
\label{eq:multipoles}
c_{\ell,m}
=
\int
r_\mathrm{sh} (\theta,\varphi)  Y^*_{\ell,m} (\theta,\varphi)
\, \ud \Omega.
\end{equation}
Figure~\ref{fig:sasi} shows the coefficients $c_{1,m}$ for the dipole
mode. In the first regime of ``weak perturbations'' (up
to about $200 \, \mathrm{ms}$ after bounce), the quasi-periodicity of
the SASI oscillations is not destroyed, and their amplitudes are
unaffected or even slightly enhanced by the pre-shock perturbations.

The situation changes as strong, large-scale pre-shock
density perturbations reach the shock in models
s18-3D and s18-3Dr (third and fourth row in Figure~\ref{fig:entropy_slices}).

Compared to s18-1D, which transitions to the more strongly SASI-dominated regime
and develops a pronounced spiral mode
(right column in Figure~\ref{fig:entropy_slices}), quasi-periodic SASI
oscillations in the two simulations starting from 3D progenitor models
are now considerably weaker, though they subsist until $250 \,
\mathrm{ms}$ after bounce in s18-3D and almost until  $500 \,
\mathrm{ms}$ in s18-3Dr.  A likely explanation for this behaviour is
that the infalling perturbations are sufficiently strong in this
regime to break the coherence of the SASI amplification cycle
\citep{guilet_10}.  Both s18-3D and s18-3Dr go through a phase where
the reduced SASI activity also results in smaller maximum and average
shock radii compared to s18-1D.

Later on, models s18-3D and s18-3Dr exhibit strong ``forced shock
deformation'', which was identified by \citet{mueller_15a} as the
principal mechanism whereby infalling perturbations can facilitate
shock revival. In this regime, the models are characterised by large
and long-lived high-entropy bubbles.
Despite the fact that small quasi-periodic dipolar shock oscillations
still persist, the fact that the  bubbles clearly have higher
characteristic angular wavenumbers than the typical
$\ell=1$ and $\ell=2$ modes of the SASI
in model s18-3Dr in particular (bottom row of Figure~\ref{fig:entropy_slices}, middle panel) suggest
that the strong perturbations push the models from 
the SASI-dominated regime into a buoyancy-dominated regime.

 To 
illustrate that the evolution of these bubbles is dictated
by the infalling density perturbations,
we show Aitoff projections of the 
relative perturbations in ram pressure
$(P_\mathrm{ram}-\langle P_\mathrm{ram}\rangle)/\langle P_\mathrm{ram} \rangle$
ahead of the shock and of the angle-dependent
shock radius for model s18-3Dr
in Figure~\ref{fig:dpram_and_rsh}. The variations
in $P_\mathrm{ram}$ mostly reflect
the pre-shock density perturbations; the relative variations
in pre-shock velocity are smaller by more than an order
of magnitude.

Bubbles generally emerge below
strong underdensities in the pre-shock region. The geometry
of the pre-shock density perturbations is not strictly mirrored
by the the shock, however. The shape of the shock only
adjusts to changes in the infalling perturbations with
some time lag, and rather reflects the long-term average
of the forcing by density and ram pressure perturbations.

In s18-3D, the formation of two
high-entropy bubbles very quickly results in shock revival (around $300\,
\mathrm{ms}$ after bounce).  Due to the lower amplitude of the initial
perturbations, model s18-3Dr exhibits stable shock deformation for
about $200 \, \mathrm{ms}$ and maintains somewhat higher shock radii
than s18-1D before the shock eventually starts to re-expand around
$500 \, \mathrm{ms}$ after bounce. The shock is
not revived in s18-1D, and continues to undergo
SASI oscillations until the end of the run
($625 \, \mathrm{ms}$ after bounce). 

A key point emerging from this analysis is that the beneficial
  effect of infalling perturbations (at least in the regime considered
  here) is not primarily to add ``turbulent pressure''
  \emph{homogeneously} throughout the gain region (although this does
  not preclude the existence of an effective phenomenological
  description of the reduction in critical luminosity in a
  quasi-spherical picture along the lines of \citealt{mueller_15a} and
  \citealt{summa_16}). The mechanism of forced shock deformation
  crucially depends on the modification of the large-scale flow
  structure; and one of the key effect seems to be that variations in
  ram pressure facilitate the formation of stable, high-entropy
  bubbles that eventually reach a sufficient scale and density
  contrast to expand continuously due to buoyancy, which is critical
  for runaway shock expansion in multi-D
  \citep{dolence_13,fernandez_14a,fernandez_15}. It is important to
  note, however,  that (as shown by model s18-3Dr) there may be a
  considerable delay from the point when local variations in ram pressure allow
  high-entropy bubbles to hover stably in the accretion flow in an
  almost stationary manner to the point when they become sufficiently
buyoant to expand continuously, which requires overcoming the ram
pressure not only in the direction of the shock protrusions
(Figure~\ref{fig:entropy_slices}, bottom centre)
at a local ram pressure minimum, but over a large solid angle with
somewhat higher average ram pressure than at the local minimum.

The three simulations thus demonstrate a significant impact of the
seed asphericities from O shell burning on the post-bounce evolution
of the $18 M_\odot$ progenitor. Depending on the level of
perturbations, shock revival occurs at $300 \, \mathrm{ms}$, $500 \,
\mathrm{ms}$ (if the O burning rate is artificially reduced), or does
not occur at all for the 1D version of the progenitor. Overall, our
models with 3D initial conditions behave very similar to 2D models
with parameterised perturbations of comparable amplitude and scale:
They confirm that the mechanism of forced shock deformation is viable
in 3D and 2D alike, and initial Mach numbers of $\mathord{\sim} 0.1$
or less are already sufficient for a large effect on shock revival
provided that shell convection is dominated by large-scale $\ell
\approx 2$ modes. To what extent perturbation-aided
explosions are still possible in 3D under
less favourable circumstances (i.e.\ lower $\mathrm{Ma}$ or
higher $\ell$) remains to be determined by more systematic studies.

\begin{figure}
  \centering
  \includegraphics[width=\linewidth]{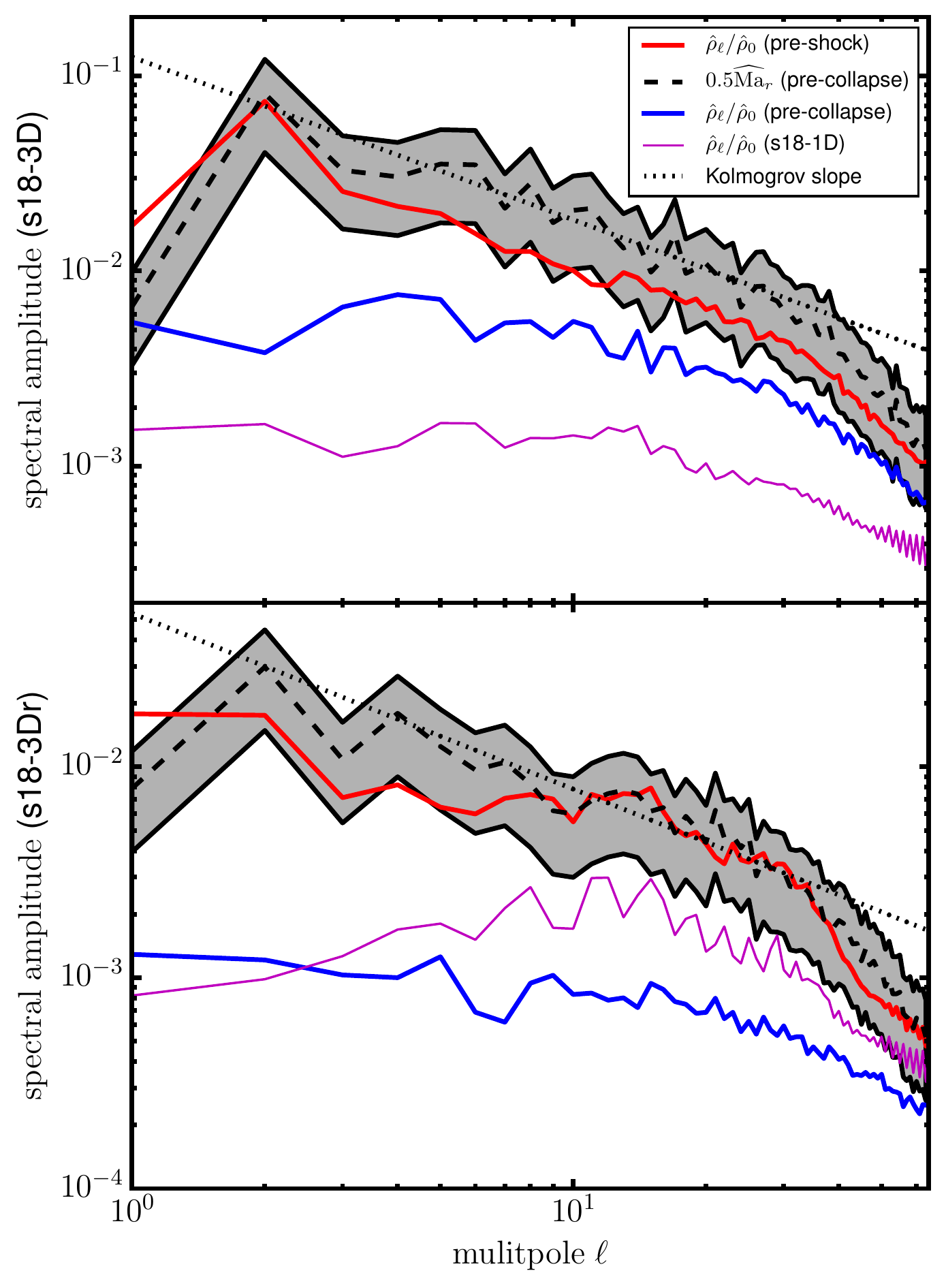}
  \caption{Spectra of pre-shock density perturbations
$\hat{\rho}_\ell/\hat{\rho}_0$
    at a radius of
    $350 \, \mathrm{km}$ (red) for model s18-3D (top, RMS average
between $260 \, \mathrm{ms}$ and $360 \, \mathrm{ms}$
after bounce)
 and s18-3Dr (bottom, RMS average
between $300 \, \mathrm{ms}$ and $400 \, \mathrm{ms}$)
 compared to the spectra of pre-collapse density
    perturbations (blue) and the radial Mach number in the O shell.
      The spectrum
    $\widehat{\mathrm{Ma}}_\ell$ of the radial Mach number is computed according to
    Equation~(\ref{eq:multipoles_ma}). The solid black curves define a
    corridor of $\pm 0.25 \widehat{\mathrm{Ma}}_\ell$ around
$0.5 \widehat{\mathrm{Ma}}_\ell$ (dashed curve); the spectrum of pre-shock density
perturbations mostly falls within this corridor. For comparison,
the spectra of the pre-shock perturbations in model s18-1D
are also included (magenta) for the respective time frame. 
Note that the pre-collapse spectra of density and velocity
perturbations are RMS averages over radii from 
$3352\, \mathrm{km}$
to $4126\, \mathrm{km}$.\label{fig:comp_spec}}
\end{figure}

\begin{figure}
  \centering
  \includegraphics[width=\linewidth]{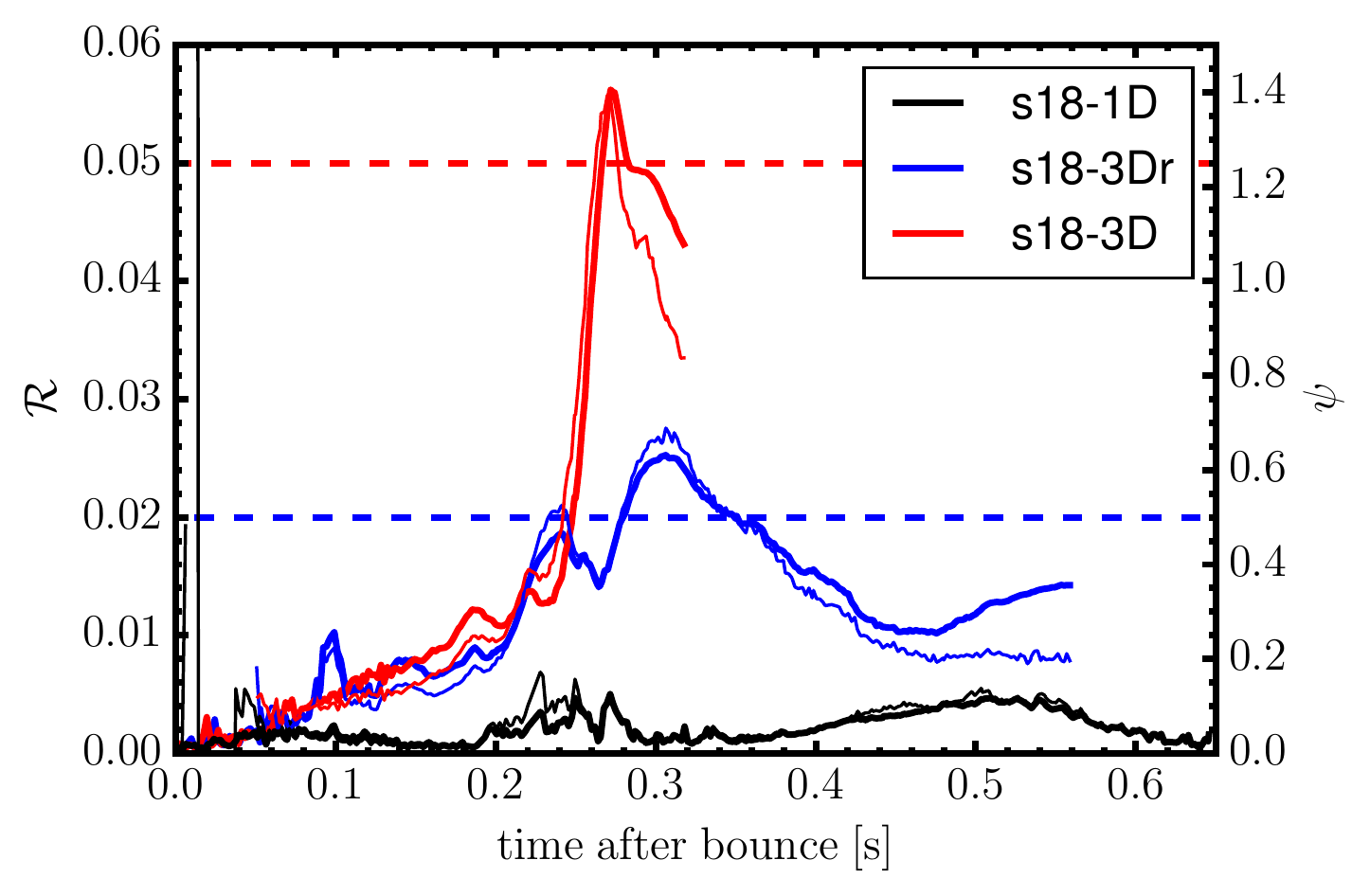}
  \caption{Comparison of the scale-weighted power
 $\mathcal{R}$ in density perturbations
(thick lines) and the parameter $\psi$ (thin)
that quantifies the strength of the external forcing of 
non-spherical motions by infalling perturbations
relative to the neutrino heating as a driver
of convection. Models s18-1D, s18-3Dr, and s18-3D
are shown in black, blue, and red, respectively. 
Dashed lines indicate the level of $\mathcal{R}$
expected based on the tpyical convective
Mach number and angular wave number
in the progenitor (Equation~\ref{eq:pert_prog}).
Note that while 
 the infall of the Si/O shell interface coincides
with an increase (which is quite sharp in s18-3D) of $\mathcal{R}$ to about the
expected value, there are considerable fluctuations
around this value.
\label{fig:pwr_drho}}
\end{figure}

\subsection{Spectrum of Pre-Shock Perturbations}
\citet{mueller_15a} outlined the mechanism of perturbation-aided
neutrino-driven explosions qualitatively, invoking a causal chain from
near-solenoidal initial perturbations, which are converted into density
perturbations via during the infall
\citep{lai_00,takahashi_14}, and increase shock deformation and
the violence of the turbulent post-shock flow
\citep{mueller_15a,couch_13}. Using analytic estimates
\citep{mueller_16c} and \citet{abdikamalov_16} also made attempts to
quantify the resultant effect on the critical neutrino luminosity for
shock revival in terms of the initial Mach number and eddy scale or
the spectrum of pre-shock perturbations. Specifically,
\citet{mueller_15a} estimated that the pre-shock density perturbations
are related to the initial convective Mach number
$\mathrm{Ma}_\mathrm{conv}$ in the convective
burning shell of the progenitor
as $\delta \rho/\rho \sim
\mathrm{Ma}_\mathrm{conv}$, and \citet{mueller_16c} suggested that the
additional work by buoyancy on the infalling perturbations downstream
of the shock results in a reduction of the critical neutrino
luminosity $\Delta L_\mathrm{crit}/L_\mathrm{crit} \propto
\mathrm{Ma}_\mathrm{conv}/\ell_\mathrm{conv}$ in terms of
$\mathrm{Ma}_\mathrm{conv}$ and the dominant angular wavenumber $\ell_\mathrm{conv}$
of convective shell burning.

Previous simulations of perturbation-aided shock revival have only
provided limited insights on how initial perturbations in the
progenitor translate quantitatively into pre-shock perturbations and
then into more violent turbulent motions. The new models allow us to
better test the validity of the proposed analytic approaches to the
perturbation aided mechanism.

Regarding the conversion of initial velocity perturbations into
density perturbations via advective-acoustic coupling, our simulations
are in line with the simple estimate $\delta\rho/\rho \sim
\mathrm{Ma}_\mathrm{conv}$ of \citet{mueller_15a} for the
pre-shock density perturbations: Once the O shell reaches
the shock, we 
find -- roughly as predicted -- $\delta \rho/\rho \approx 0.1$ for
s18-3D 
and $\delta \rho/\rho \approx 0.04$ for s18-3D
(top panel of Figure~\ref{fig:dflukt}). $\delta \rho$ is only a
very gross measure for the pre-shock density perturbations, however.
It is also susceptible to short-wavelength perturbations
that grow from random noise in model s18-1D (where we encounter values of
$\delta \rho/\rho$ of up $0.03$) and could also
arise for purely numerical reasons (e.g.\ odd-even modes).
The relatively large values of $\delta \rho/\rho$ in models
s18-3D and s18-3Dr before the arrival of
the Si/O interface at the shock are also due to such
high-wavenumber modes.

Since small-scale perturbations do not have a significant impact on the
shock \citep{mueller_15a}, it is more appropriate to consider the
spectrum $\rho_\ell$ of pre-shock perturbations instead of $\delta
\rho$. Our simulations suggest that the spectra of the pre-shock
density perturbations and the initial velocity perturbations
in the O shell are remarkably similar, at least in a time-average sense.
Figure~\ref{fig:comp_spec} shows
normalised time-averaged spectra 
$\hat{\rho}_\ell/\hat{\rho}_0$ of the pre-shock density perturbations
with
\begin{equation}
  \hat{\rho}_\ell(r)=\left(\frac{\int_{t_1}^{t_2} \rho_\ell^2(r)\,\ud t}{t_2-t_1}\right)^{1/2},
\end{equation}
for relevant phases before shock revival in models s18-3D and s18-3Dr to
illustrate this. We compare this to the
spectrum $\widehat{\mathrm{Ma}}_{r,\ell}$ of the Mach number
of radial motions in the middle of the O shell, which we 
define as
\begin{equation}
\label{eq:multipoles_ma}
\widehat{\mathrm{Ma}}_{r,\ell} = \left(\sum_{m=-\ell}^\ell \left |\int Y^*_{\ell,m}
(\theta,\varphi) v_r /c_\mathrm{s} \, \ud
\Omega\right|^2\right)^{1/2},
\end{equation}
where we have normalised the velocity by the local sound speed.
To obtain a smoother spectrum, we compute RMS averages of
$\widehat{\mathrm{Ma}}_{r,\ell}$ for the region between $3352\, \mathrm{km}$
and $4126\, \mathrm{km}$ in the progenitor (although
we find $\widehat{\mathrm{Ma}}_{r,\ell}$ to vary little in the interior of the O shell).
To illustrate the generation of density perturbations
from vorticity perturbations due to
advectice-acoustic coupling
during collapse, we also show the RMS averaged
spectrum of initial density in this region in
Figure~\ref{fig:comp_spec}. Spectra of the
pre-shock density perturbations in model s18-1D
are also included for comparison.

With only a few exceptions, the pre-shock density perturbations  fall
nicely within the range,
\begin{equation}
  \label{eq:spectrum}
  \hat{\rho}_\ell/\hat{\rho}_0 =(0.25\text-0.75)  \widehat{\mathrm{Ma}}_{r,\ell},
\end{equation}
for both s18-3D and s18-3Dr
over a wide range of angular wavenumbers $\ell$.
On average, the spectrum of
pre-shock density perturbations
is well described by $\hat{\rho}_\ell/\hat{\rho}_0 \approx  \widehat{\mathrm{Ma}}_{r,\ell} /2$.

The similarity of the initial velocity spectrum and the spectrum of
pre-shock density perturbations is in line with the simple explanation
of density perturbations by ``differential infall'' in
\citet{mueller_15a}, which suggests $\hat{\rho}_\ell /\hat{\rho}_0 \sim  \widehat{\mathrm{Ma}}_{r,\ell} /2$ for all
eddy scales. Interestingly, the simple relation between $\hat{\rho}_\ell$
and $ \widehat{\mathrm{Ma}}_{r,\ell} /2$ has not been established so far by studies of linear
perturbations in Bondi accretion in a general context
\citep{kovalenko_98,foglizzo_01,foglizzo_02} or specifically in
core-collapse supernovae \citep{lai_00,takahashi_14}.  Instead, these
works suggest that stronger density perturbations are created for
higher wavenumbers, e.g.\ $\rho_\ell \propto \ell(\ell+1)$
\citep{lai_00} and $\rho_\ell \propto \ell$ \citep{takahashi_14}.  The
crucial problem here is that some assumptions of these linear studies
need to be adjusted in order to model the growth of perturbations
during core collapse accurately: \citet{lai_00} and
\citet{takahashi_14} do not account for the subdominant role of
compressive velocity perturbations at the pre-collapse stage
\citep{mueller_15a}.  They also assume stationary Bondi flow as a
background solution and neither account for the small initial infall
velocity (much slower than free-fall) of perturbed convective shells
nor for the density gradient in the progenitor. The problem of
linear perturbation growth in collapsing supernova
cores therefore needs to be revisited to provide a theoretical basis
for Equation~(\ref{eq:spectrum}).

Figure~\ref{fig:comp_spec} also clearly shows the excess
power in pre-shock density perturbations at small $\ell$ in s18-3D and s18-3Dr compared
to s18-1D. While s18-1D develops non-negligible pre-shock density
perturbations at late times, these lie predominantly in
the range $\ell =10\text{-}20$ and are dynamically less
significant than the strong large-scale perturbations in
the two other simulations.
For further analysis, it is convenient to introduce
a dimensionless number to characterise the power of
perturbations
in the relevant region of small $\ell$ by a single dimensionless
number to replace the problematic metric $\delta \rho/\rho$.
To this end, we define the scale-weighted power\footnote{Note that this
expression naturally generalises the identity $\delta \rho/\rho=[\sum_{\ell=1}^\infty (\rho_\ell/\rho_0)]^{1/2}$
 between the
RMS value of the density perturbations in real space and the
power in spectral space (which follows from Parseval's identity
for spherical harmonics and also motivates the use of RMS values
for the perturbations in real and spectral space elsewhere). 
}
in perturbations $\mathcal{R}$ as
\begin{equation}
\label{eq:r}
  \mathcal{R}(r)
  =\left(\sum_{\ell=1}^\infty \frac{(\hat{\rho}_\ell/\hat{\rho}_0)^2}{ \ell^2}\right)^{1/2}.
\end{equation}
Because of the similarity of
$\rho_\ell$ and $\widehat{\mathrm{Ma}}_\ell$, $\mathcal{R}$ is also tightly related
to the typical Mach number and angular wavenumber of the convective
shell in question, i.e.\ we have
\begin{equation}
\label{eq:pert_prog}
  \mathcal{R}
\approx
\frac{\mathrm{Ma}_\mathrm{conv}}{2\ell_\mathrm{conv}}
\end{equation}
as shown in Figure~\ref{fig:pwr_drho}.

\begin{figure}
  \includegraphics[width=\linewidth]{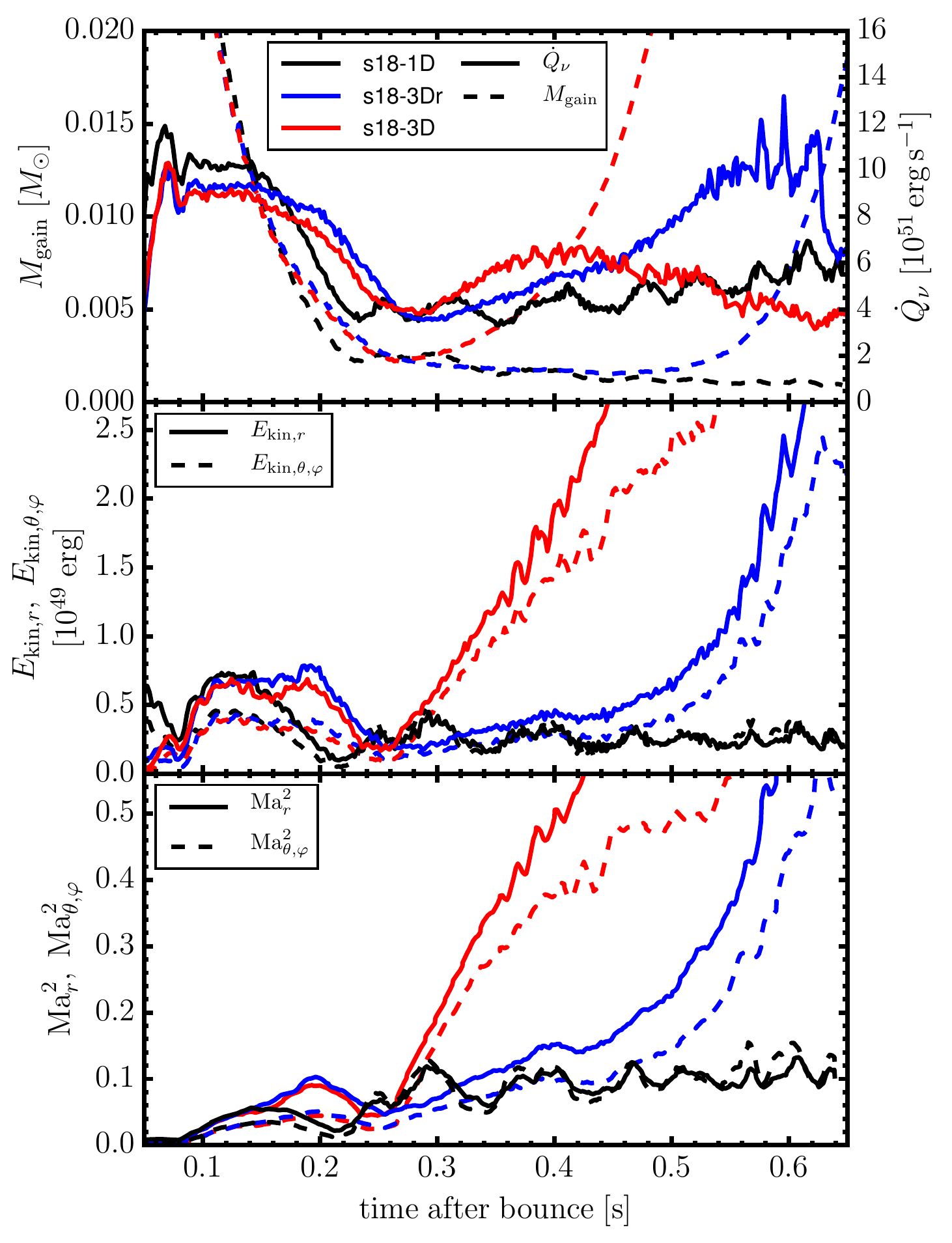}
  \caption{Properties of the gain region for models
s18-1D (black curves), s18-3Dr (blue), and
s18-3D (red). The top panel
shows the mass $M_\mathrm{gain}$ (dashed)
of the gain region and the volume-integrated
neutrino heating rate $\dot{Q}_\nu$ (solid). The second panel shows the kinetic
energy 
contained in the radial ($E_\mathrm{kin,r}$, solid) and non-radial
($E_\mathrm{kin,\theta,\phi}$, dashed) components of turbulent motions in
the gain region. The mean-square Mach numbers corresponding
to these energies are plotted in the bottom panel.
\label{fig:gain}}
\end{figure}

\begin{figure}
  \includegraphics[width=\linewidth]{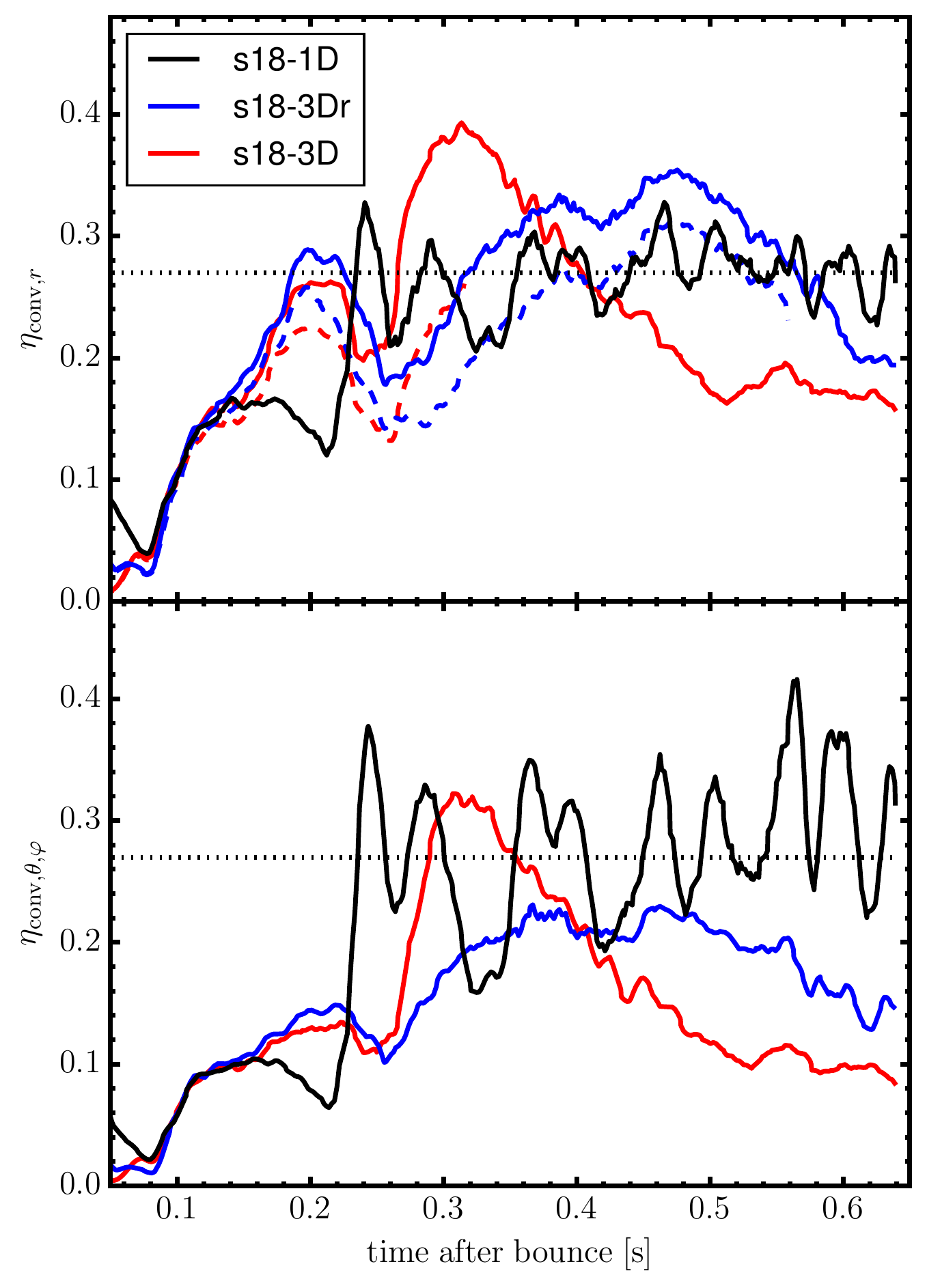}
  \caption{Efficiency factors (solid curves)
$\eta_{\mathrm{conv},r}$ (top panel)
and $\eta_{\mathrm{conv},\theta,\varphi}$ (bottom panel)
for the conversion of neutrino heating into
radial and non-radial turbulent motions
computed according to Equations~(\ref{eq:eff1}--\ref{eq:eff2})
for models s18-1D (black), s18-3Dr (blue), and s18-3D (red).
Note 
that models s18-3Dr and s18-3D exhibit a significantly
higher efficiency $\eta_{\mathrm{conv},r}$ above
the long-term average of $\mathord{\approx} 0.27$
in model s18-1D (marked by dotted lines) between
the infall of the Si/O interface and shock revival.
Dashed curves in the top panel show how the inclusion
of a correction term for the external forcing of
large-scale turbulent motions according to
Equation~(\ref{eq:epert}) reduces the
maximum efficiency factor $\eta_{\mathrm{conv},r}$ to values
comparable to model s18-1D.
\label{fig:qconv}}
\end{figure}

\subsection{Turbulent Kinetic Energy in the Gain Region} 
\label{sec:ekin}

Previous studies of perturbation-aided shock
revival \citep{couch_13,mueller_15a,couch_15} have traced the beneficial
impact of the pre-collapse perturbations to the increased violence
of non-spherical flows in the post-shock region, which 
increases the mass $M_\mathrm{gain}$ in the gain region and hence
the volume-integrated neutrino heating rate
$\dot{Q}_\nu$ (since $
\dot{Q}_\nu \propto M_\mathrm{gain}$ to first order, see \citealp{janka_12}).
Our simulations conform to this picture. Figure~\ref{fig:gain} compares
$M_\mathrm{gain}$ and $\dot{Q}_\nu$ for the three different models as well
as the kinetic energy contained in turbulent motions.
The turbulent kinetic energies in radial
($E_{\mathrm{kin},r}$) and non-radial
($E_{\mathrm{kin},\theta,\varphi}$)
motions are computed separately as integrals over the gain region $V_\mathrm{gain}$,
\begin{eqnarray}
  \label{eq:ekinr}
  E_{\mathrm{kin},r}&=& \frac{1}{2} \int_{V_\mathrm{gain}} \rho (v_r-\langle v_r\rangle)^2\,\ud V
  \\
  \label{eq:ekinlat}
  E_{\mathrm{kin},\theta,\varphi}&=& \frac{1}{2} \int_{V_\mathrm{gain}} \rho (v_\theta^2+v_\varphi^2)\,\ud V,
\end{eqnarray}
where $v_r$, $v_\theta$, and $v_\varphi$ denote the velocity components and
$\langle v_r \rangle$ is the spherical Favre (i.e.\ density-weighted) average of the radial velocity.\footnote{Because
  of the small velocities, the Newtonian expression for the kinetic energy can be used. Special
  and general relativistic corrections to the volume element are implicitly included
  in Equations~(\ref{eq:ekinr}) and (\ref{eq:ekinlat}) as well as in all other
  volume integrals used in this study.}
We also calculate effective Mach numbers
$\mathrm{Ma}_r$ and $\mathrm{Ma}_{\theta,\varphi}$ for the
radial and transverse components of the turbulent velocity fluctuations as
the relevant parameters characterising the importance of non-radial
fluid flow in the gain region. These are defined
as
\begin{equation}
  \mathrm{Ma}_{r}^2
  =\frac{1}{c_\mathrm{s}^2(\langle r_\mathrm{sh}\rangle)}\frac{2E_{\mathrm{kin},r}}{M_\mathrm{gain}}
  \approx \frac{3\langle r_\mathrm{sh}\rangle}{GM }\frac{2E_{\mathrm{kin},r}}{M_\mathrm{gain}},
\end{equation}
and analogously for $\mathrm{Ma}_{\theta,\varphi}$. For convenience we
use the post-shock value of the sound speed $c_\mathrm{s}$ at the average shock radius
$\langle r_\mathrm{sh}\rangle$ for computing effective Mach numbers
and approximate $c_\mathrm{s}^2 (\langle r_\mathrm{sh}\rangle )
\approx GM/(3r_\mathrm{sh})$ as in \citet{mueller_15a}.\footnote{There
    is, in fact, some motivation for using this approximation; it is
    more closely connected to the idea that the ratio of turbulent
    stresses and the pre-shock ram pressure is relevant for the
    interaction of non-radial fluid motions with the shock.} We
  generally find the expected hierarchy between the three models,
  i.e.\ $E_{\mathrm{kin},r}$, $E_{\mathrm{kin},\theta,\varphi}$,
  $\dot{Q}_\nu$, and $M_\mathrm{gain}$ are higher for stronger initial
  perturbations during the accretion phase with the exception of brief
  intervals where s18-1D shows larger shock radii than the two other
  models due to modulations in SASI activity.

\citet{mueller_15a} pointed out, however, that this hierarchy alone
only furnishes indirect evidence that the excitation
of non-radial fluid motions in the presence of infalling
perturbations is more efficient than in the presence of
neutrino heating as a driver of convection (or of
an advective-acoustic amplification cycle for SASI modes)
alone. For this it is useful to consider efficiency
factors $\eta_{\mathrm{conv}}$ for the conversion of neutrino
heating into turbulent kinetic energy,
\begin{eqnarray}
  \label{eq:eff1}
  \eta_{\mathrm{conv},r}
  &=&
  \frac{E_{\mathrm{kin},r}/M_\mathrm{gain} }{[(\langle r_\mathrm{sh}\rangle-r_\mathrm{gain})(\dot{Q}_\nu/M_\mathrm{gain} )]^{2/3}},
  \\
  \label{eq:eff2}
  \eta_{\mathrm{conv},{\theta,\varphi}}
  &=&
  \frac{E_{\mathrm{kin},{\theta,\varphi}}/M_\mathrm{gain} }{[(\langle r_\mathrm{sh}\rangle-r_\mathrm{gain})(\dot{Q}_\nu/M_\mathrm{gain} )]^{2/3}}.
\end{eqnarray}
In the non-perturbed case in 2D, \citet{mueller_15a} found
$\eta_{\mathrm{conv},\theta,\varphi}\approx 0.5$ once the
non-spherical instabilities had reached saturation, which can be
understood as the result of a balance between buoyant driving and
turbulent dissipation that obtains in the gain region
\citep{murphy_12}. As already observed in \citet{mueller_16b},
the efficiency factor in non-perturbed
3D models is considerably smaller
($\eta_\mathrm{conv}\sim 0.2 \ldots 0.35$; Figure~\ref{fig:qconv})
than in the 2D case.
This is likely due to a combination of several factors:
The forward turbulent cascade in 3D results in smaller
eddy structures \citep{hanke_12,murphy_12,couch_12b} and implies
a smaller effective dissipation length. It is also conceivable
that turbulent 3D flow more efficiently transports heat for a given
average convective velocity, so that marginal instability against
convective overturn can be maintained by less violent convective
motions than in 2D. The results of  \citet{murphy_12}, who
found a higher ratio
of the convective heat flux (or convective luminosity in their
terminology) to the volume-integrated neutrino heating rate points
in this direction.

Interestingly, \citet{mueller_15a} detected no
clear impact of the perturbation amplitudes on
$\eta_{\mathrm{conv},{\theta,\varphi}}$; all their perturbed 2D models
showed small fluctuations around
$\eta_{\mathrm{conv},\theta,\varphi}=0.5$ similar to the unperturbed
baseline model.

This counterintuitive finding can be resolved by considering the
contributions to the turbulent kinetic energy from radial and
transverse velocity fluctuations separately. Figure~\ref{fig:gain}
shows that the infalling perturbations appear to selectively boost
radial velocity fluctuations, whereas model s18-1D shows
$E_{\mathrm{kin},r}\approx E_{\mathrm{kin},\theta,\varphi}$ for fully
developed SASI. This conclusion is reinforced by plots of the
efficiency factors $\eta_{\mathrm{conv},r}$ and
$\eta_{\mathrm{conv},\theta,\varphi}$ in Figure~\ref{fig:qconv}.
Prior to and around shock revival, the perturbations increase
$\eta_{\mathrm{conv},r}$ considerably above the typical value of
$\eta_{\mathrm{conv},r}\approx 0.27$ in s18-1D after the accretion of
the Si/O shell interface. This is consistent with the change in
geometry from an $\ell=1$ spiral SASI in s18-1D with rapid mass
motions around the proto-neutron star to $2\text{-}4$ stable
high-entropy bubbles that are supported by (radial) buoyancy
forces. Since it is the radial velocity fluctuations that regulate
turbulent heat transport within the gain region and determine the
momentum carried by convective bubbles colliding with the shock, it is
natural to consider the increase in $\eta_{\mathrm{conv},r}$ as the
relevant factor for achieving better conditions in models with strong
initial perturbations.

Having identified $\eta_{\mathrm{conv},r}$ as a suitable metric that
reflects the stronger activity of turbulent motions in the gain region
due to the infalling perturbations (while factoring out feedback
effects due to the larger shock radius and increased neutrino heating),
it is tempting to ask whether our models can be used to validate
analytic predictions \citep{mueller_16c,abdikamalov_16} for this
phenomenon.  At the current stage, this seems premature, however.  A
comparison with the theory of \citet{abdikamalov_16} is not yet
meaningful: Although their study is based on an elaborate framework
for the interaction of perturbations with the shock in the linear
approximation \citep{ribner_53}, it considers only the interaction of
vorticity and entropy perturbations with the shock and disregards the
action of buoyancy on the shocked perturbations, which is likely the
dominant effect for the injection of extra turbulent kinetic energy
\citep{mueller_16c}. The approach of \citet{mueller_16c} simplifies
the interaction of the perturbations with the shock, but in principle admits
a comparison with our numerical results. Their prediction
can be encapsulated in a correction term in the equation
for the turbulent kinetic energy due to the work of buoyancy 
on the shocked perturbations. Phrased in our variables, they
predict
\begin{equation}
\label{eq:epert}
\frac{E_{\mathrm{kin},r}}{M_\mathrm{gain}} = \eta_{\mathrm{conv},r} \left[
  \frac{(\langle r_\mathrm{sh} \rangle -r_\mathrm{gain}) \dot{Q}_\nu}{M_\mathrm{gain}}
  \right]^{2/3} (1+\psi)^{2/3}.
\end{equation}
Here the correction term $\psi$ is determined by
the ratio of buoyant energy generation $\dot{Q}_\mathrm{pot}$
from the shocked perturbations and the neutrino heating
rate $\dot{Q}_\nu$, and by the postulated ratio of 
turbulent dissipation lengths in the non-perturbed and strongly
perturbed regime,
\begin{equation}
\label{eq:psi}
\psi=
\frac{\pi \langle r_\mathrm{sh}\rangle \dot{Q}_\mathrm{pot}/(\ell M_\mathrm{gain})}
{(r_\mathrm{sh}-r_\mathrm{gain}) \dot{Q}_\nu/M_\mathrm{gain}}.
\end{equation}
$\dot{Q}_\mathrm{pot}$  depends on the density perturbations,
the accretion rate $\dot{M}$ and the difference in gravitational potential
between the shock and gain radius,
\begin{equation}
\dot{Q}_\mathrm{pot}
=
  \frac{\dot{M }\delta \rho}{\rho}
  \left(\frac{GM}{\langle r_\mathrm{sh}\rangle}-\frac{GM}{r_\mathrm{gain}}\right).
\end{equation}
Since $\delta\rho/\rho$ is weighted with $\ell^{-1}$ in
Equation~(\ref{eq:psi}), it is natural to relate the
term $\dot{Q}_\mathrm{pot}/\ell$ to $\mathcal{R}$ for a spectrum
of incident perturbations.
\begin{equation}
\label{eq:qpot}
\frac{\dot{Q}_\mathrm{pot}}{\ell}
=
  \dot{M }\mathcal{R}
  \left(\frac{GM}{\langle r_\mathrm{sh}\rangle}-\frac{GM}{r_\mathrm{gain}}\right).
\end{equation}
To test the prediction of \citet{mueller_16c}, we re-compute
$\eta_{\mathrm{conv},r}$ based on Equations~(\ref{eq:epert}),
(\ref{eq:psi}), and (\ref{eq:qpot}), to check whether the correction
factor $(1+\psi)^{2/3}$ allows us to recover a ``universal'' value of
$\eta_\mathrm{conv,r}$. The corrected values of $\eta_\mathrm{conv,r}$
are included in Figure~\ref{fig:qconv}. We also show the evolution
of $\psi$ (Figure~\ref{fig:pwr_drho}) as a measure for the relative
importance of buoyant energy generation by infalling perturbations.

The result is ambiguous. During the
quasi-stationary evolution after the infall of the Si/O interface, the
correction factor brings $\eta_{\mathrm{conv},r}$ in s18-3Dr nicely in
line with s18-1D, and it also reduces the excursion to $0.4$ in s18-3D
around $300\, \mathrm{ms}$ to about the long-term average of
$\eta_{\mathrm{conv},r}=0.27$.  Before and during the infall of the
Si/O shell interface, the non-stationarity of the models precludes a
reasonable comparison. Moreover, it is not clear whether the
transition from the SASI-dominated regime (s18-1D) to a
buoyancy-dominated regime (s18-3D and s18-3Dr) offsets the comparison.
At present, we can only conclude that the factor for the turbulent
kinetic energy proposed by \citet{mueller_16c} is not
implausible. Firm conclusions can only be based on a larger set of
models, and more idealised simulations controlled setup (in the vein
of \citealt{fernandez_14a,fernandez_15}) may be more informative for
this purpose.

\begin{figure}
  \includegraphics[width=\linewidth]{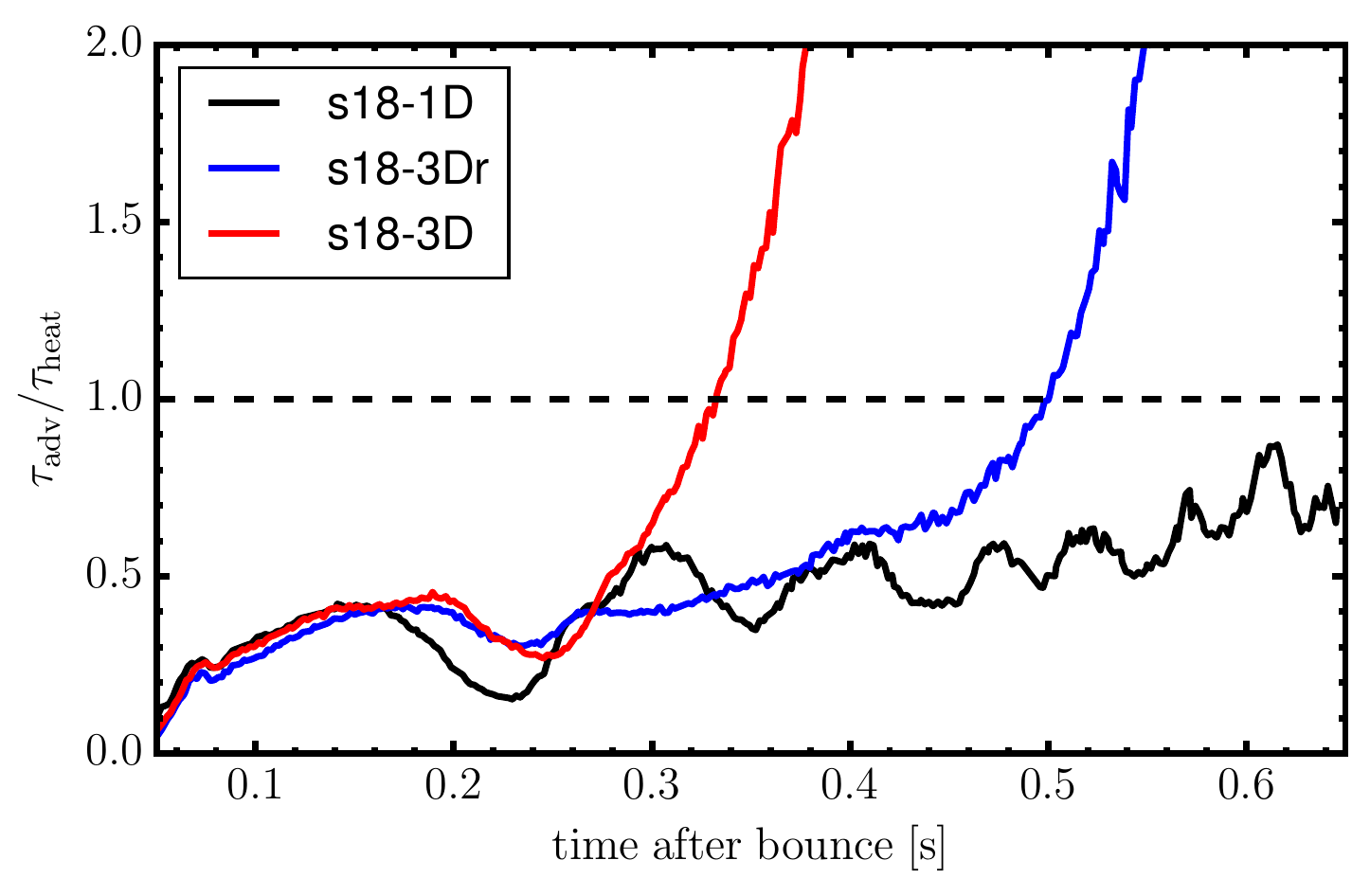}
  \caption{Ratio
$\tau_\mathrm{adv}/\tau_\mathrm{heat}$
of the advection and heating time-scales
$\tau_\mathrm{adv}$ and $\tau_\mathrm{heat}$
for models s18-1D (black), s18-3Dr (blue), s18-3D (red).
The critical value for
runaway shock expansion
$\tau_\mathrm{adv}/\tau_\mathrm{heat}=1$ is denoted
by a dashed line.
Thanks to forced shocked deformation by infalling large-scale
perturbations,
models s18-3D and s18-3Dr undergo shock revival while
the critical ratio is still well below
unity for s18-1D.
\label{fig:ratio}}
\end{figure}

\begin{figure}
  \includegraphics[width=\linewidth]{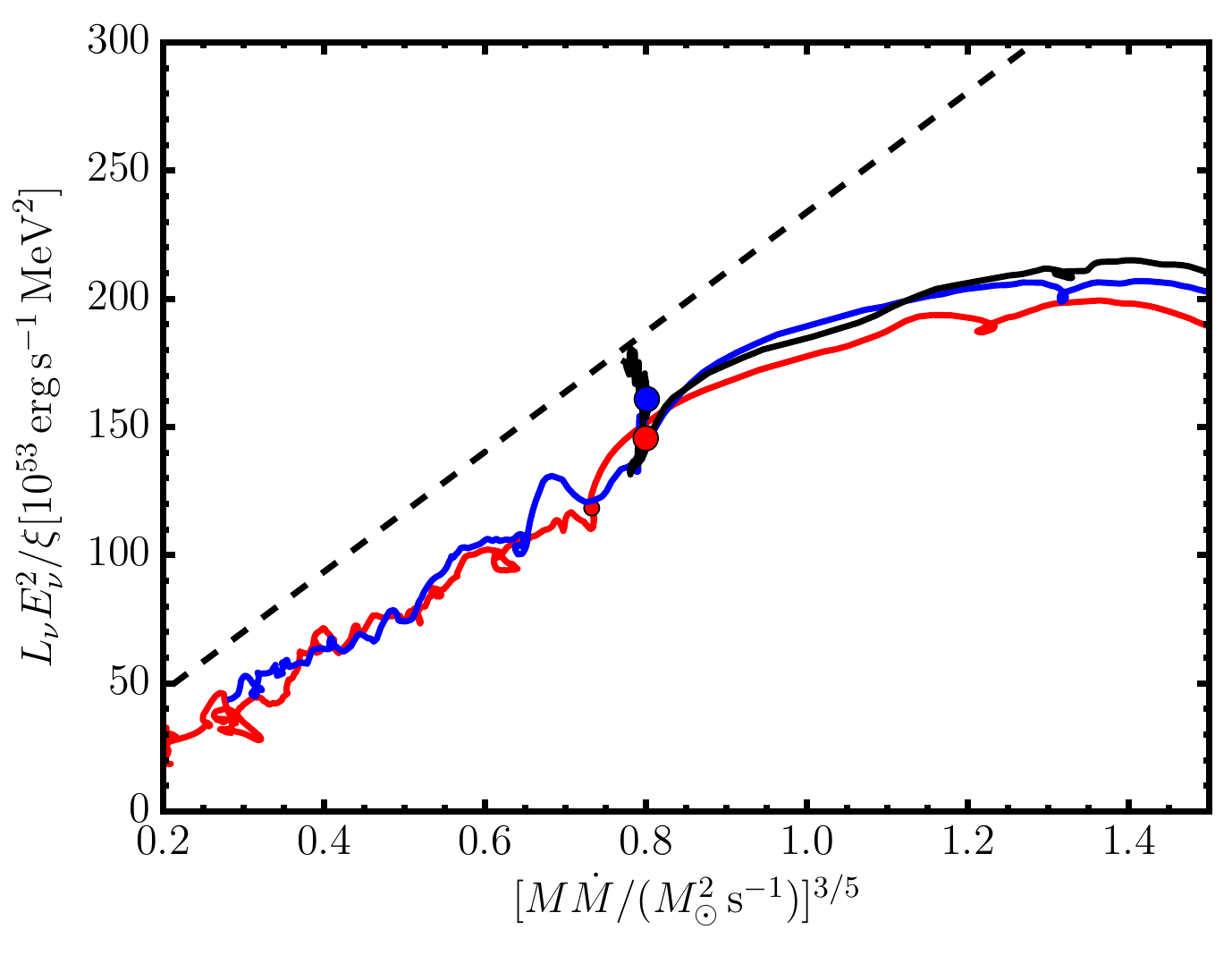}
  \caption{Evolution of models s18-1D (black), s18-3Dr (blue), and
    s18-3D towards the critical curve in the $M \dot{M}$-$L_\nu
    E_\nu^2$-plane.  The dependency of the critical heating functional
    $L_\nu E_\nu^2$ on the gain radius (Equation~\ref{eq:lcrit}) is
    taken into account by means of a correction factor
    $\xi=(r_\mathrm{gain}/100 \, \mathrm{km})^{-2/5}$.  A lower limit for the
    theoretical critical curve (dashed) is obtained by anchoring the
    power law $(L_\nu E_\nu^2)_\mathrm{crit}/\xi \propto (\dot{M}
    M)^{3/5}$ at the endpoint of the trajectory of s18-1D, which has
not undergone shock revival yet at this junction, but is close to 
the threshold. Red and blue dots denote points
on the trajectory of model s18-1D
corresponding to the onset of the explosion in s18-3D and s18-3Dr.
\label{fig:lcrit}}
\end{figure}

\subsection{Impact on Explosion Conditions}
From their different estimates for the additional generation for
turbulent kinetic energy due to infalling perturbations, \citet{mueller_16c} and
\citet{abdikamalov_16} proceeded further to predict the reduction of
the critical neutrino luminosity for explosion in terms of the
parameters of the initial perturbations. Specifically,
\citet{mueller_16c} postulated a reduction of the critical luminosity
by $\mathord{\sim} 0.15 \psi$, which according to their
estimates would amount to
$12\text{-} 24\%$ for model s18-3D. In view of our discussion in
Section~\ref{sec:ekin}, it is obvious that we cannot validate the
quantitative relations and the underlying
physical arguments proposed in the studies. Instead we can only
provide a plausibility check by diagnosing the impact
of the perturbations on the explosion conditions in our models.
Following \citet{mueller_15a}, we quantify the effect of the
perturbations by considering both the critical ratio of the advection
and heating time-scales $\tau_\mathrm{adv}$ and $\tau_\mathrm{heat}$
\citep{janka_98,janka_01,thompson_00,buras_06b,murphy_08b,fernandez_12} and the
trajectories of the models in the generalised $\dot{M}-L_\nu$ plane.

Figure~\ref{fig:ratio} shows the time-scale ratio
$\tau_\mathrm{adv}/\tau_\mathrm{heat}$ for the three different
models. It is evident that the perturbations result in a major
enhancement in heating conditions: Disregarding some fluctuations in
models s18-1D, s18-3D and s18-3Dr show significantly higher values than
s18-1D while the convective O shell is accreted and reach the critical
threshold $\tau_\mathrm{adv}/\tau_\mathrm{heat}=1$ at times of $340 \,
\mathrm{ms}$ and $500 \, \mathrm{ms}$ after bounce respectively.  The
typical values for model s18-1D are only
$\mathord{\sim 0.5}$
at these
times. Incidentally, Figure~\ref{fig:ratio} also demonstrates that the
slightly lower mass accretion rate in s18-3D compared to s18-1D and
s18-3Dr (see discussion in Section~\ref{sec:shock_evolution}) is not
responsible for earlier shock revival in this model; the time-scale
ratio $\tau_\mathrm{adv}/\tau_\mathrm{heat}$ is practically identical
to s18-3Dr until the Si/O interface reaches the shock.

The time-scale criterion is closely related to the concept of the
critical neutrino luminosity \citep{burrows_93} for shock revival
\citep{janka_12}.
Disregarding multi-D effects (and assuming a roughly constant binding
energy at the gain radius), \citet{mueller_15a} showed that it is
tantamount to a critical condition
\begin{equation}
\label{eq:lcrit}
(L_\nu E_\nu^2)_\mathrm{crit}
r_\mathrm{gain}^{2/5}
\propto
(\dot{M} M)^{3/5} ,
\end{equation}
where $L_\nu$ is the total electron flavour luminosity, $E_\nu$ is an
appropriately defined average mean energy for $\nu_e$ and
$\bar{\nu}_e$, and $M$ is the proto-neutron star mass.  Since s18-1D
is close to reaching the critical explosion condition
$\tau_\mathrm{adv}/\tau_\mathrm{heat}=1$, we can anchor the critical
curve defined by Equation~({\ref{eq:lcrit}) at (or slightly above) the
  end point of trajectory of this model in the $\dot{M} M \text{-}
  L_\nu E_\nu^2$-plane (Figure~\ref{fig:lcrit}). 

The explosions in models s18-3D and s18-3Dr occur roughly
 $22\%$ and $16\%$ below the (theoretically
inferred) critical curve.  This reduction of the critical luminosity
is of a similar order as the estimate of $0.15 \psi$ of
\citet{mueller_16c}, though more on the high side.  These effects are
sizeable and comparable to the differences between spherically
symmetric and multi-dimensional models
\citep{murphy_08b,hanke_12,couch_12b,dolence_13}. For s18-3D we must,
however, bear in mind that the explosion is triggered right when the
Si/O shell interface arrives, i.e.\ as soon as the shock experiences
strong forcing by infalling perturbations.  It cannot be excluded that
shock revival could be achieved even earlier if similarly strong
perturbations were already present when
$\tau_\mathrm{adv}/\tau_\mathrm{heat}$ is still down at values of
$\mathord{\sim} 0.25$. On the other hand, the effect of the infalling
perturbations may be magnified during a highly non-stationary phase of
rapidly declining $\dot{M}$ and transient shock expansion.  For
s18-3D, diagnosing the reduction of the critical luminosity is
therefore particularly problematic. Despite these uncertainties, it is
clear, however, that the various diagnostics point to a reduction of
the critical luminosity in the tens-of-percent range due convective
seed perturbations for our particular $18 M_\odot$ progenitor.

For models without strong initial perturbations, \citet{summa_16}
and \citet{janka_16} showed that the effects of a varying binding
energy and of multi-dimensional effects can be absorbed in correction
factors to obtain a \emph{universal} critical luminosity curve that
marks the threshold for shock revival across a large variety of
progenitors. This is a somewhat different perspective than the one
adopted in the present study. Here we essentially seek to \emph{measure}
these correction factors for models of the \emph{same} progenitor and
can ascribe all differences (including differences in the binding
energy of the gain region) to the initial perturbations in the O shell
because the inner boundary conditions for the accretion problem
(proto-neutron star radius and mass, neutrino luminosities) are
extremely similar. 
Whether the concept of a ``universal'' critical
luminosity of \citet{summa_16} and \citet{janka_16} carries over to
models with strong initial perturbations cannot be addressed here yet, as
this would require a larger body of simulations for different
progenitors.

\begin{figure}
  \includegraphics[width=\linewidth]{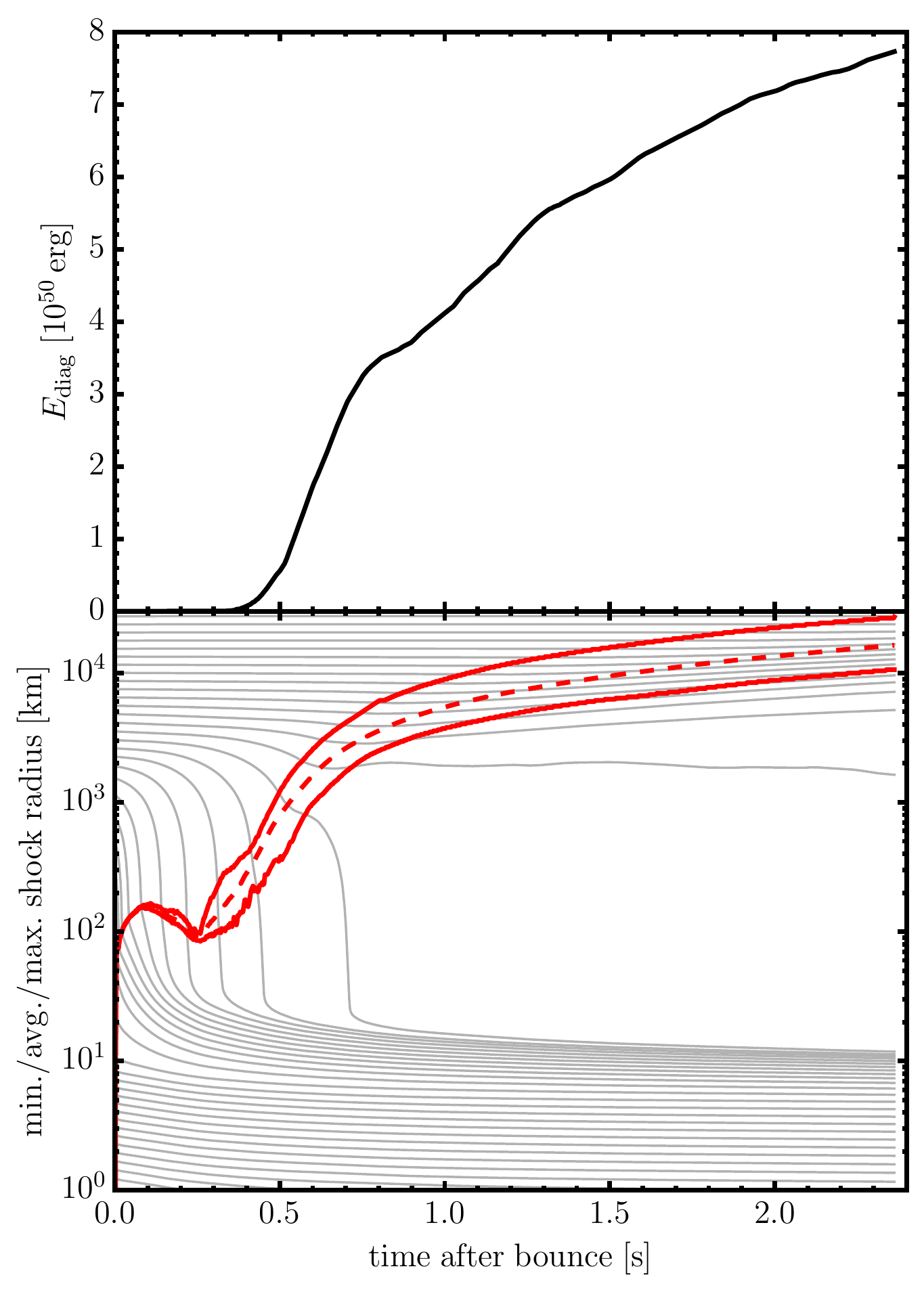}
  \caption{Top panel: Diagnostic explosion energy $E_\mathrm{diag}$ for model
s18-3D. Bottom panel: Maximum, minimum (red solid curves),
and angle-averaged shock radius (red, dashed) and mass
shell trajectories starting from locations
that are uniformly spaced in $\log r$ 
at the
onset of collapse.
 \label{fig:expl}
     }
\end{figure}

\begin{figure*}
  \includegraphics[width=\linewidth]{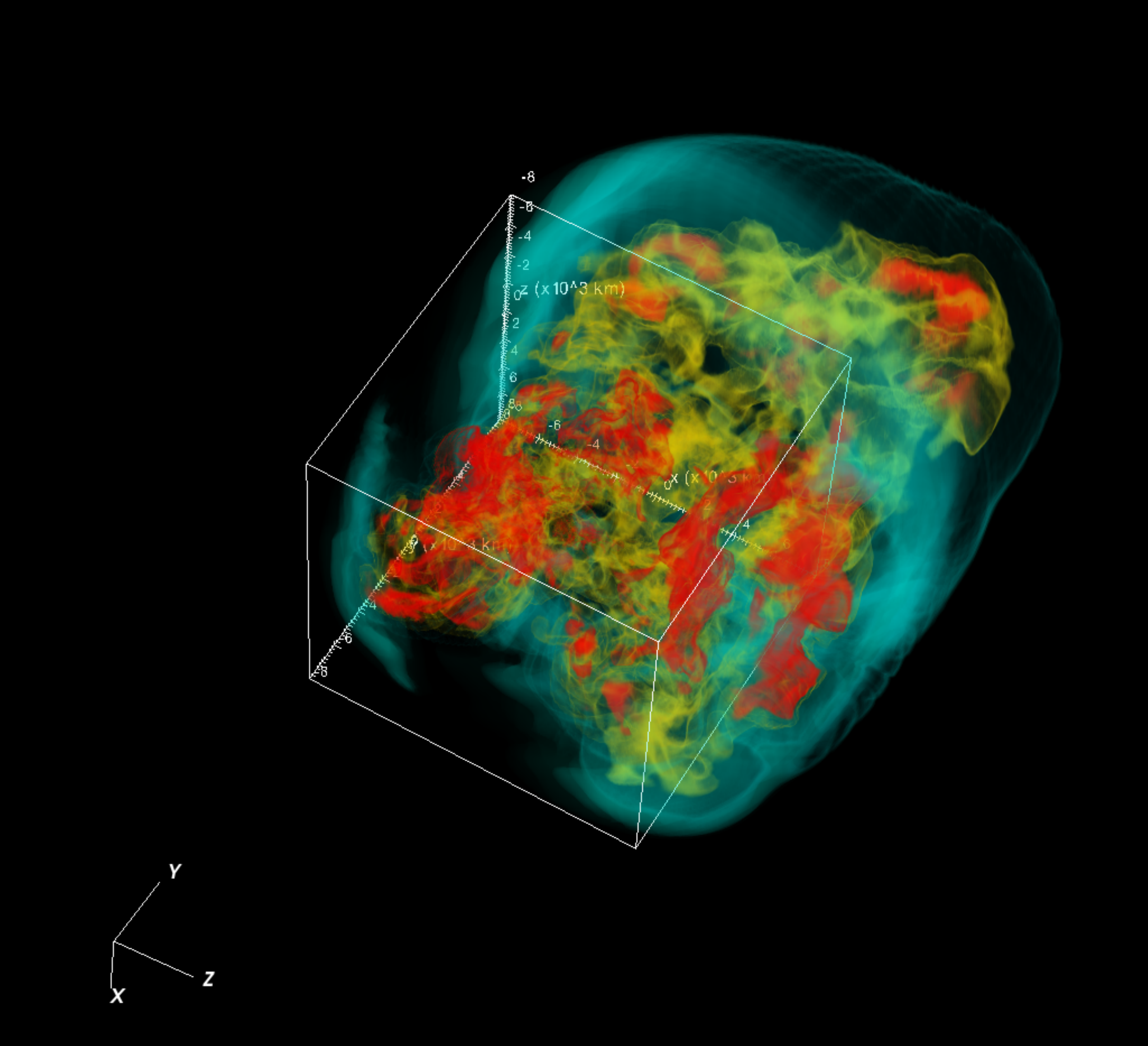}
  \caption{Volume rendering of the entropy in model s18-3D
    $2\, \mathrm{s}$ after bounce. The expanding neutrino-heated bubbles
    (red/yellow)
    retain a pronounced global asymmetry at this stage and drive
    faster shock (translucent cyan surface) expansion in the
    direction of positive $y$ and $x$. The global geometry is still
    very similar to the stage of shock revival shown
    in Figure~\ref{fig:entropy_slices}.\label{fig:3dexpl}
  }
\end{figure*}

\section{Explosion Phase}
\label{sec:explosion}
We have continued model s18-3D to $2.35 \, \mathrm{s}$
after bounce  to allow for a tentative
evaluation of the explosion and remnant properties
in a 3D model of a perturbation-aided neutrino-driven
explosion.

The evolution of the shock radius and mass shell trajectories for the
entire simulation are shown in the bottom panel of
Figure~\ref{fig:expl}.  By the end of the simulation, the minimum
shock radius exceeds $10,000 \, \mathrm{km}$, i.e.\ the shock has
already traversed the entire O/Si shell and reached the C/O shell. The
shock and the ejecta retain a pronounced global asymmetry at this
stage with considerably stronger shock expansion in the $y$-direction
of the computational grid (Figure~\ref{fig:3dexpl}) with a maximum
shock radius of $27,000 \, \mathrm{km}$.  The shock geometry imprinted
by forced shock deformation at the time of shock revival is largely
preserved until these late times aside from some minor adjustments;
the direction of fastest shock propagation corresponds to one of the
Si-rich convective updrafts in the progenitor shown in
Figure~\ref{fig:x_si_ini}.

\begin{figure}
  \includegraphics[width=\linewidth]{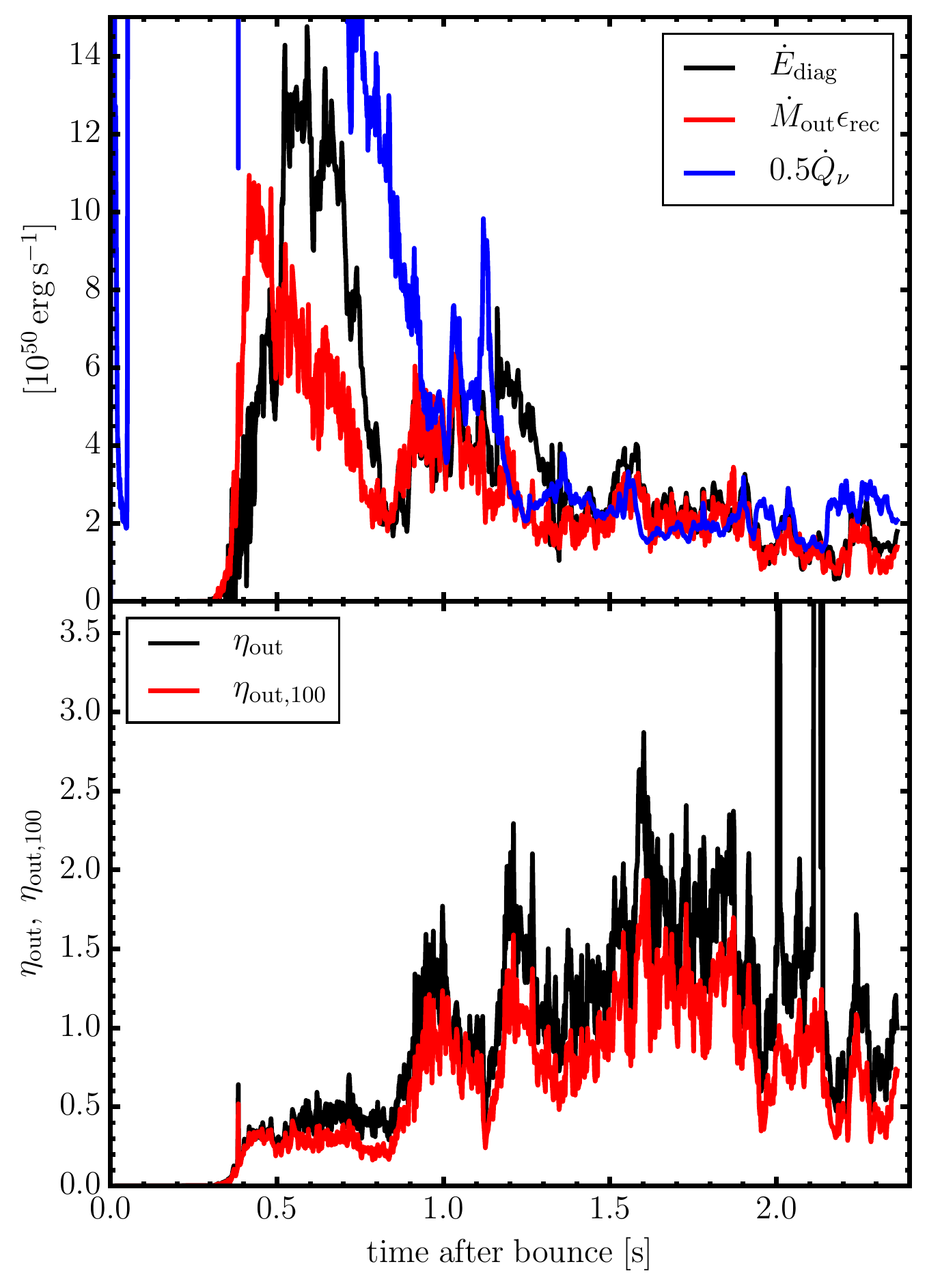}
    \caption{Top panel:
Comparison of the growth
rate $\dot{E}_\mathrm{diag}$ of the diagnostic explosion energy
(black curve)
to the rate of energy released by nucleon recombination (red)
for the measured mass outflow rate $\dot{M}_\mathrm{out}$
and an assumed recombination energy
$\epsilon_\mathrm{rec}=6 \, \mathrm{MeV}$.
The blue curve shows $0.5 \dot{Q}_\nu$ 
(i.e.\ half the volume-integrated neutrino heating rate) to illustrate
that the conversion of neutrino heating 
into explosion energy (via mass outflow and nucleon recombination) carried by the ouflows
is very efficient at late times.
Bottom panel: Outflow efficiency
$\eta_\mathrm{out}$ based on different assumptions
for the typical net binding energy of
ejected matter. For the black curve,
we assume that the ejecta need to be lifted
out of the gravitational potential
from the gain radius
(Equation~\ref{eq:eta_out}), whereas
we use a lower average binding
energy computed from Equation~(\ref{eq:etot_r})
for a typical turnaround radius
of $100 \, \mathrm{km}$ before ejection for
the red curve.
The assumption of a large turnaround
radius results in more reasonable
outflow efficiencies of order unity.
      \label{fig:expl_eff}
  }
\end{figure}
\begin{figure}
  \includegraphics[width=\linewidth]{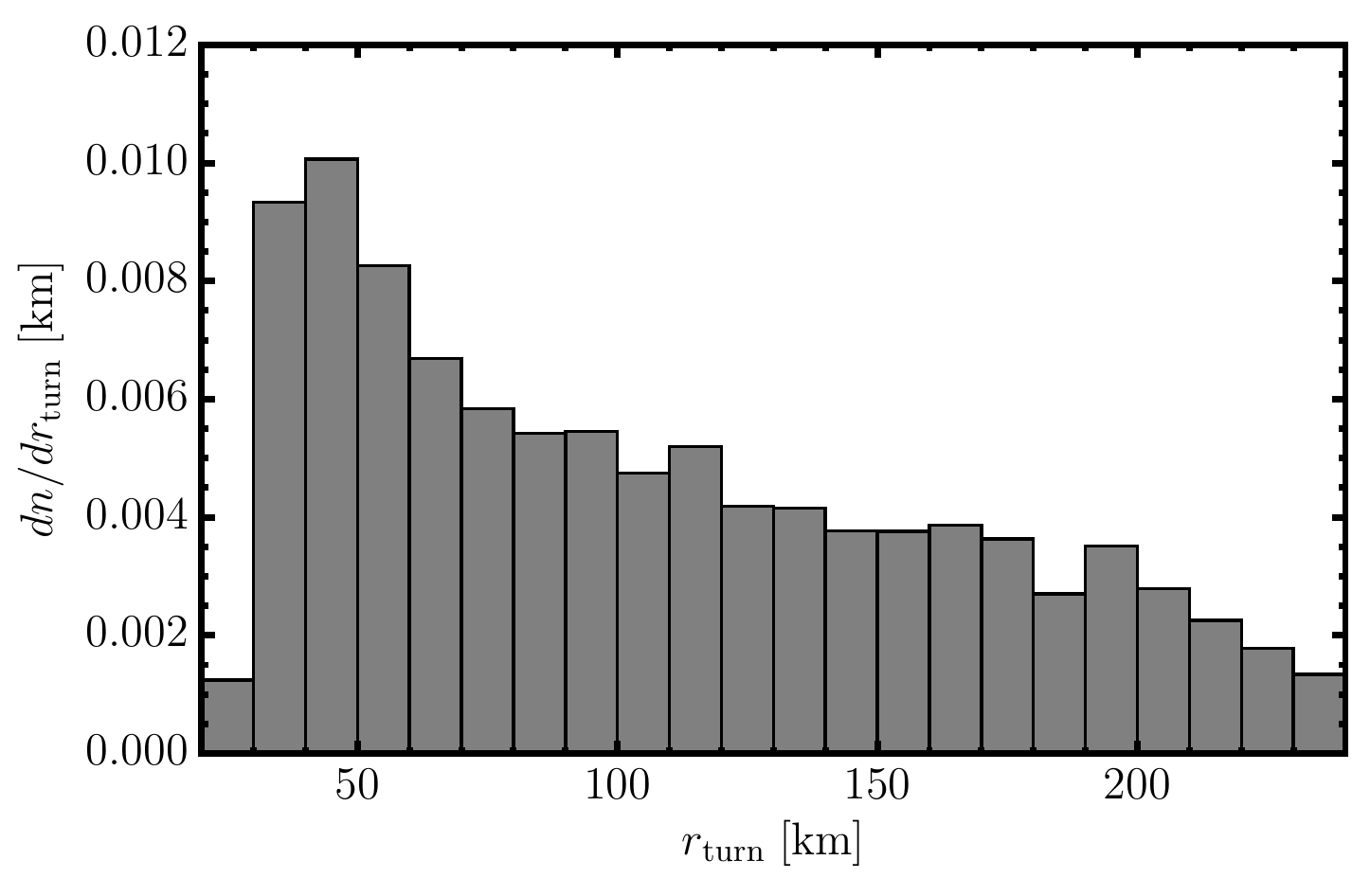}
  \caption{The distribution of turnaround radii for a sample
of 12414 tracer particles
    between $0.6\, \mathrm{s}$ and $2.1 \, \mathrm{s}$ after bounce.
Note that the distribution does not peak at the gain radius and is
strongly right-skewed, implying that the energy (per unit mass)
that needs to be pumped into the gain region
to eject matter is much smaller than
 $|e_\mathrm{gain}|$.
      \label{fig:turnaround}
  }
\end{figure}

\begin{figure}
  \includegraphics[width=\linewidth]{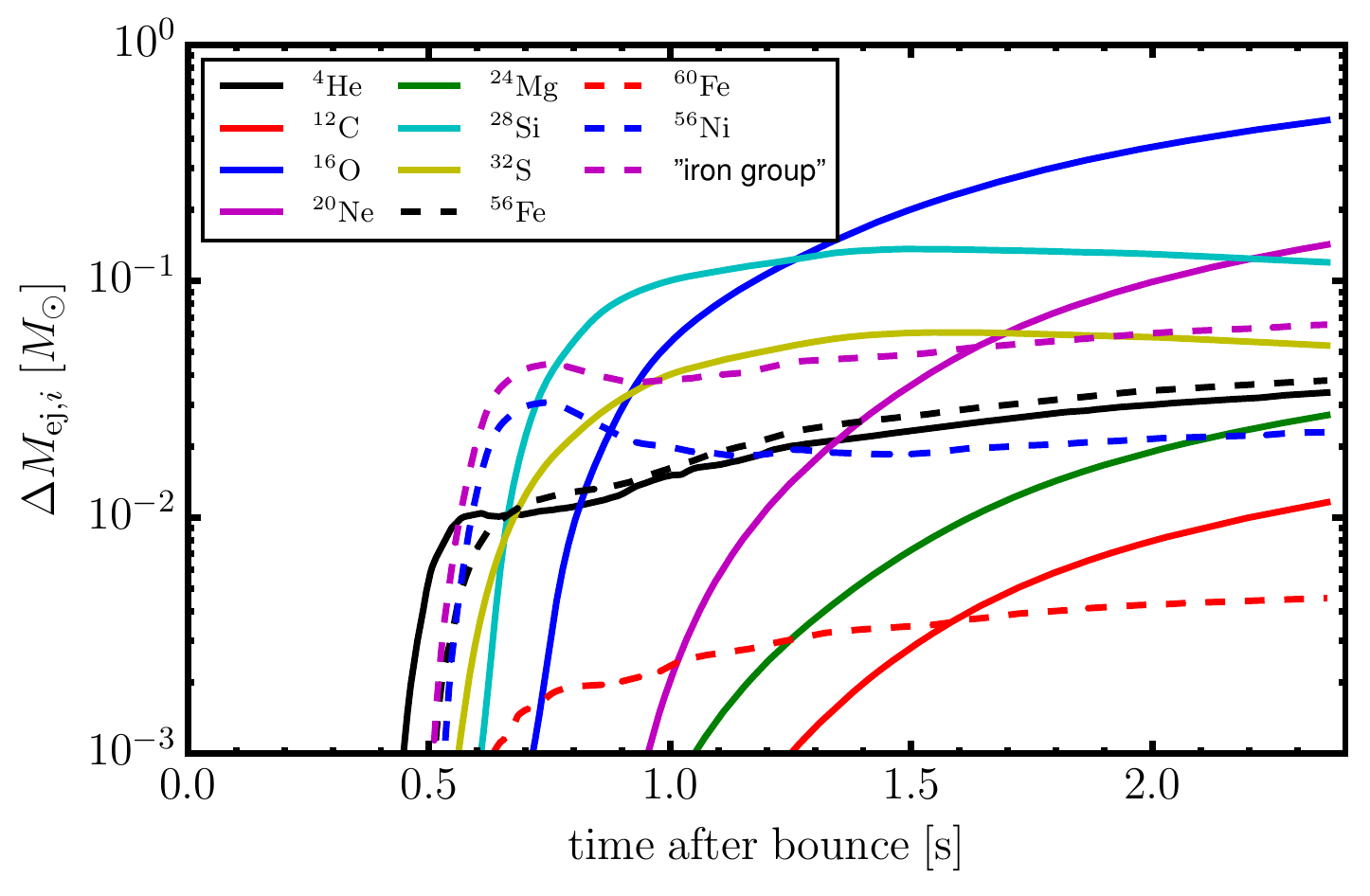}
    \caption{Partial mass of several $\alpha$ and iron-group elements
      in the ejecta as obtained by our simple treatment of nuclear
      burning in model s18-3D (flashing and freeze-out from NSE at a
      temperature of $0.5 \, \mathrm{MeV}$).  We also plot the total
      mass contained in iron-group nuclides (``iron group'', dashed
      magenta curve).  Although this composition is only vaguely
      indicative of the true situation, the Figure suggests that much
      of the iron-group material synthesised by explosive burning at
      early times eventually falls onto the proto-neutron
      star. Instead, most of the iron-group material comes from the
      neutrino-driven outflows. Because of slightly (and artificially)
      neutron-rich conditions in our simulation, most of it is made as
      ${}^{56}\mathrm{Fe}$; under more realistic (i.e.\ slightly
      proton-rich) conditions, most of the iron-group ejecta would be
      ${}^{56}\mathrm{Ni}$ instead. Note that the
      simple treatment of freeze-out from NSE also introduces
      an uncertainty in the overall mass of iron-group ejecta.
      \label{fig:comp}
  }
\end{figure}

\subsection{Explosion Energy}
\label{sec:energy}
Even at the end of the simulation, the cycle of mass accretion and
mass ejection due to neutrino heating is still ongoing, and we can
therefore only obtain tentative estimates for the final explosion
properties as in previous 2D and 3D explosion models with multi-group
neutrino transport. In lieu of the final explosion energy, one
typically considers the ``diagnostic'' energy
$E_\mathrm{diag}$
\citep{buras_06b,suwa_10,mueller_12a,bruenn_16} as an estimator, which
is calculated by integrating the net total
(kinetic+internal+gravitational) energy $e_\mathrm{tot}$ per unit mass over the region where
it is positive (i.e.\ where matter is formally unbound,
\begin{equation}
  \label{eq:ediag}
  E_\mathrm{diag} =\int_{e_\mathrm{tot}>0} \rho e_\mathrm{tot}\,\ud V,
\end{equation}
where the volume element is implicitly taken to include general
relativistic corrections.

The proper definition of $e_\mathrm{tot}$ is not straightforward, especially
in general relativity. \citet{mueller_12a} suggested
computing $e_\mathrm{tot}$
as
\begin{equation}
  \label{eq:etot_gr_old}
  e_\mathrm{tot}= \alpha[ (c^2+\epsilon+P/\rho) W^2-P/\rho]-W c^2,
\end{equation}
in terms of the pressure $P$, density $\rho$, internal
energy density $\epsilon$ (excluding
rest-mass differences between different nuclear species), Lorentz
factor $W$ and lapse function $\alpha$ in the relativistic
case. In the Newtonian limit, this reduces to the familiar form
\begin{equation}
  \label{eq:etot_old}
  e_\mathrm{tot}= \epsilon+v^2/2+ \Phi,
\end{equation}
where $\Phi$ is the gravitational potential. This form, however,
disregards an important subtlety: While
the integral of $\rho(\epsilon+v^2/2+ \Phi)$ is conserved
for a system in an external potential, the conserved
quantity for a self-gravitating system is
\begin{equation}
\int \rho
  (\epsilon+v^2/2+ \Phi/2)\, \ud V.
\end{equation}

Calculating $E_\mathrm{diag}$ based on
Equations~(\ref{eq:etot_gr_old},\ref{eq:etot_old}) effectively
amounts to double-counting the potential energy of the ejecta
that is due to their own self-gravity (and not to
the gravitational field of the proto-neutron star) and therefore overestimates
the energy needed to expel the ejecta to infinity.
In the Newtonian case, \citet{buras_06a,buras_06b} circumvented
this problem by computing $\Phi$ for a given mass shell from
the enclosed mass only, but this is not a suitable solution
in our case because it would imply discarding GR corrections
to the gravitational field of the proto-neutron star. To avoid
the problem of double-counting, we can, however, subtract
the Newtonian potential $\Phi_\mathrm{grav,out}$ generated by the
shells outside a given radius $r$, 
\begin{equation}
  \label{eq:etot_new}
  e_\mathrm{tot}=
  \alpha[ (c^2+\epsilon+P/\rho) W^2-P/\rho]- W c^2-
   \Phi_\mathrm{grav,out}(r),
\end{equation}
where
\begin{equation}
  \label{eq:phiout}
  \Phi_\mathrm{grav,out}=
\int_r^\infty  \frac{4\pi \rho r'^2 G}{r'}\, \ud r'.
\end{equation}
This is justified because relativistic corrections
to the outside potential are negligible in the region
in question ($r>100\, \mathrm{km}$).

The diagnostic explosion energy computed from
Equations~(\ref{eq:ediag}), (\ref{eq:etot_new}), and (\ref{eq:phiout}) is
shown in the top panel of Figure~\ref{fig:expl}.  Model s18-3D
exhibits a steady increase in diagnostic energy, even though 
the rate of increase slows down at about $700\, \mathrm{ms}$. By the end
of the simulation, the model has reached a value of
$E_\mathrm{diag}= 7.7\times 10^{50} \, \mathrm{erg}$, which is still
increasing at a rate of $1\text{-}2  \times 10^{50} \,
\mathrm{erg} \, \mathrm{s}^{-1}$.

$E_\mathrm{diag}$ does not include any correction for the binding
energy of the material outside the shock (for which
\citealt{bruenn_13,bruenn_16} introduced the convenient term
``overburden'').  Taking all the material outside a radius of $10,000
\, \mathrm{km}$ with negative $e_\mathrm{tot}$ into account, we obtain
an overburden of $2.6 \times 10^{50}\, \mathrm{erg}$.  The true
correction is likely somewhat smaller because part of this bound
material will still be channelled onto the proto-neutron through the
downflows, so that its negative total energy will not contribute to
the energy budget of the ejecta.  We can therefore put a relatively
confident lower limit of $5 \times 10^{50} \, \mathrm{erg}$ 
on the explosion energy of s18-3D. The true value is likely higher,
the continuing slow increase of the diagnostic energy will
probably equalise the correction from the overburden within
a second, so that a final explosion energy of 
$7\text{-}8 \times 10^{50} \, \mathrm{erg}$ appears reasonable.

In agreement with the argument of \citet{marek_09} and
earlier long-time simulations of supernova explosions
in 2D and 3D \citep{mueller_15b,bruenn_16}, the growth
of the diagnostic energy at late times is essentially
determined by the mass outflow rate
$\dot{M}_\mathrm{out}$ of neutrino-heated ejecta, which
are first lifted to
$e_\mathrm{tot} \approx 0$ and then obtain most of their
net positive energy from nucleon recombination.
Other contributions to the energy budget of the
ejecta (e.g.\ nuclear burning and the accumulation
of weakly bound material by the shock) are subdominant
at late times.
Figure~\ref{fig:expl_eff} demonstrates that
\begin{equation}
  \frac{\ud E_\mathrm{diag}}{\ud t}
  \approx \epsilon_\mathrm{rec} \dot{M}_\mathrm{out}
\end{equation}
describes the evolution of the diagnostic
energy well at late times with a value
of  $\epsilon_\mathrm{rec}=6 \, \mathrm{MeV}$
that accounts for incomplete recombination
and some turbulent mixing between the outflows and downflows.

What is quite remarkable about the late-time evolution of the
diagnostic energy in model s18-3D is that neutrino heating proves very
efficient at driving outflows.
Based on the assumption that the
volume-integrated neutrino heating rate $\dot{Q}_\nu$ and the specific binding
energy at $|e_\mathrm{gain}|$ at the gain radius are relevant
parameters regulating the outflow rate $\dot{M}_\mathrm{out}$,
\citet{mueller_15b} suggested the ratio
\begin{equation}
\label{eq:eta_out}
  \eta_\mathrm{out}=\frac{\dot{M}_\mathrm{out}
    |e_\mathrm{gain}|}{\dot{Q}_\nu}
\end{equation}
as an appropriate efficiency parameter for mass ejection.
\citet{mueller_15b}, however, already observed
efficiency parameters $\eta_\mathrm{out}>1$ during some
phases of their $11.2 M_\odot$ explosion model. Model s18-3D
is much more extreme in this respect. At late times
($\gtrsim 1.2 \, \mathrm{s}$ after bounce), we find
typical outflow efficiencies
$\eta_\mathrm{out}\mathord{\sim} 2$ and even excursions
to $\eta_\mathrm{out}>3$. Evidently, $|e_\mathrm{gain}|$
is no longer the relevant energy scale regulating
the mass outflow rate at late stages.

\citet{mueller_15b} suggested that values $\eta_\mathrm{out}>1$ can be
explained if a sizeable fraction of the ejecta is not lifted out of the
gravitational potential of the proto-neutron star from the gain
radius, but is channelled from the downflows to the outflows by
turbulent mixing at a larger radius.  To verify this, we compute
trajectories for tracer particles in the gain region and consider the
distribution of turnaround radii $r_\mathrm{turn}$ for tracer
particles that are eventually ejected. Following tracer trajectories
from the onset of collapse for about $2 \, \mathrm{s}$, however, would
be problematic because of the accumulation of integration errors and
would also require an extremely large number of particles to sample
the neutrino-driven ejecta adequately. We therefore opt for an
adaptive placement and deletion of a smaller number of 512 tracer
particles; they are initially distributed randomly in $\theta$,
$\varphi$, and mass coordinate $m$ in the analysis region between the
gain radius and $r=250 \, \mathrm{km}$ at a time of $0.5 \,
\mathrm{s}$ after bounce. Whenever a tracer particle crosses the gain
radius or $r=250 \, \mathrm{km}$, it is removed and a new particle is
again randomly placed within the analysis region.  For particles that
cross $r=250 \, \mathrm{km}$, we record the turnaround point
$r_\mathrm{turn}$ at which their radial velocity changed sign
(and discard particles that are ejected but always
had positive radial velocities since they were
placed in the gain region so that their turnaround
point cannot be determined)
In a state of quasi-stationary accretion, this
procedures allows us to reasonably sample the properties of the
neutrino-driven ejecta.\footnote{As a sanity check, we verified, for
  example, that the rate of ejection $\ud n_\mathrm{out}/\ud t$ of
  tracer particles through the outer boundary of the analysis region
  reproduces the mass outflow rate as $\dot{M}_\mathrm{out} \approx
  dn_\mathrm{out}/dt M_\mathrm{gain}/N$ if we assign each particle a
  mass $M_\mathrm{gain}/N$, where $N$ is the number of tracers.}

The distribution of turnaround points $r_\mathrm{turn}$ for the tracers
is shown in Figure~\ref{fig:turnaround} for the phase between $0.6 \,
\mathrm{s}$ and $2.1 \, \mathrm{s}$ after bounce. Although the
distribution of $r_\mathrm{turn}$ exhibits considerable short-term
variability (which is likely related to the spikes of
$\eta_\mathrm{out}$ in Figure~\ref{fig:expl_eff}), this
distribution is fairly representative for the entire explosion phase;
there is no strong secular trend. The distribution of turnaround
points confirms that most of the neutrino-heated ejecta need not be
unbound from $r_\mathrm{gain}$. Even around at $t=0.6 \, \mathrm{s}$
the gain radius is already smaller than the peak of the distribution
of $r_\mathrm{turn}$ ($40 \text{-}50 \, \mathrm{km}$, and at late
times the gain radius shrinks to $\approx 24 \, \mathrm{km}$.
The distribution of turnaround points is strongly right-skewed,
and the average turnaround radius is rather of the order
of $100 \, \mathrm{km}$, i.e.\ $3\text{-}4$ times larger
than the gain radius.

In the light of the distribution of turnaround points, the
efficient driving of outflows by neutrino heating no longer
seems exorbitant. If we consider instead of
$\eta_\mathrm{out}$ an efficiency parameter
based on the binding
energy $\eta_\mathrm{out,100}$ at a typical turnaround
radius of $\mathord{\sim} 100 \, \mathrm{km}$,
\begin{equation}
  \eta_\mathrm{out,100}
  =
  \frac{\dot{M}_\mathrm{out}  |e_\mathrm{tot}|}{\dot{Q}_\nu},
\end{equation}
we obtain more reasonable values fluctuating around unity at late
times (bottom panel of Figure~\ref{fig:expl_eff}). Here, we
estimate $|e_\mathrm{tot}|$ for material entrained from the downflows
into the outflows as
\begin{equation}
  \label{eq:etot_r}
  |e_\mathrm{tot}|\approx
  \frac{GM}{4r}+\frac{3}{4}\epsilon_\mathrm{diss}
\end{equation}
following \citet{mueller_15b}.

The large average turnaround radius 
implies that
model s18-3D can maintain an appreciable growth
of the explosion energy despite modest neutrino heating
rates of only a few $10^{50}\, \mathrm{erg} \, \mathrm{s}^{-1}$.
Since the typical binding energy at the turnaround point
is not much larger than the recombination
energy $\epsilon_\mathrm{rec}$, 
the effective ``return of investment'' for using
up the neutrino heating to
unbdind the ejected material is large, and
the time derivative of the explosion energy is
\emph{formally} a large fraction of the neutrino heating
rate. At late times, we find (top panel of Figure~\ref{fig:expl_eff}),
\begin{equation}
  \dot{E}_\mathrm{diag}\approx 0.5 \dot{Q}_\nu
\end{equation}
which is even more extreme than in the 3D long-time simulation of
\citet{mueller_15b} and considerably higher than in 2D simulations,
where $\dot{E}_\mathrm{diag}\approx (0.15\text{-}0.2) \dot{Q}_\nu$
\citep{mueller_12a,bruenn_16} is more typical.

Though this diagnosis reveals a plausible cause for the steady growth
of $E_\mathrm{diag}$ in model s18-3D, the crucial role of turbulent
mixing between the downflows and outflows raises further questions
that need to be addressed with more sophisticated analysis methods and
high-resolution simulations: Why do the efficiency parameters
$\eta_\mathrm{out}$ and $\eta_\mathrm{out,100}$ grow after the first
few hundreds of milliseconds of the explosion phase? Could this come
about because turbulent mixing between the downflows and outflows
becomes more effective as the expanding structures in the post-shock
region are resolved by a larger number of grid points and the
effective numerical Reynolds number increases? Is there a simple
physical explanation for the distribution of turnaround points?  It is
clear  that we cannot provide better answers based on a single,
moderately resolved 3D simulation, but the late-time behaviour of
model s18-3D furnishes further evidence that -- contrary to the
problem of shock revival -- the effects of 3D turbulent flow could be
beneficial in the explosion phase
\citep{handy_14,melson_15a,mueller_15b} and help to achieve explosion
energies in line with observational constraints.

\begin{figure}
  \includegraphics[width=\linewidth]{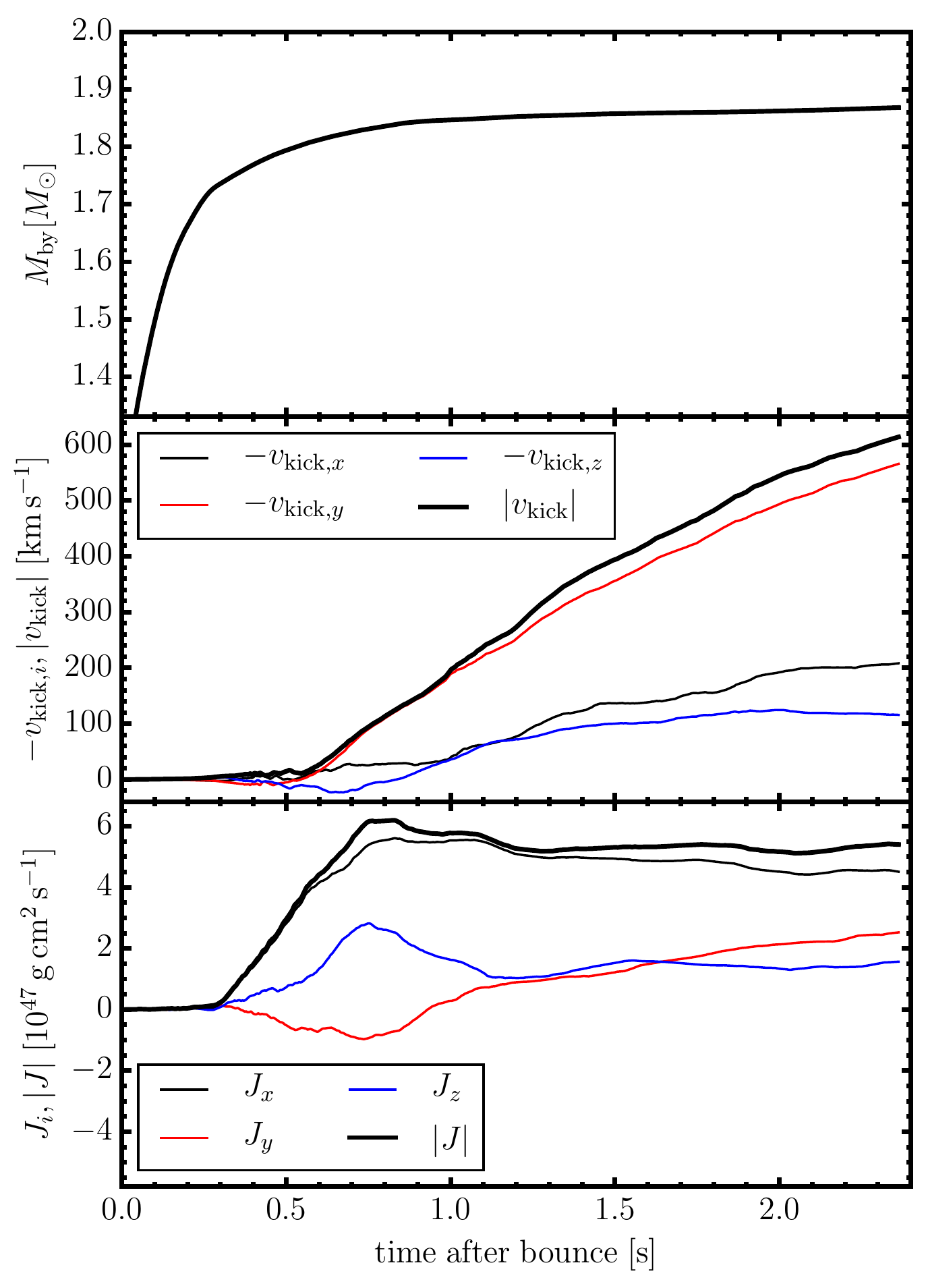}
  \caption{Baryonic mass $M_\mathrm{by}$
of the proto-neutron star (top panel),
  \mbox{$x$-,} \mbox{$y$-,} and $z$-components
(thin black, red, and blue curves)
and absolute value (thick curve)
of the inferred kick velocity $\mathbf{v}_\mathrm{kick}$
computed according to Equation ~(\ref{eq:vkick})
(middle panel) and
the proto-neutron star
angular momentum $\mathbf{J}$ (bottom panel). \label{fig:pns}
  }
\end{figure}

\begin{figure}
  \includegraphics[width=\linewidth]{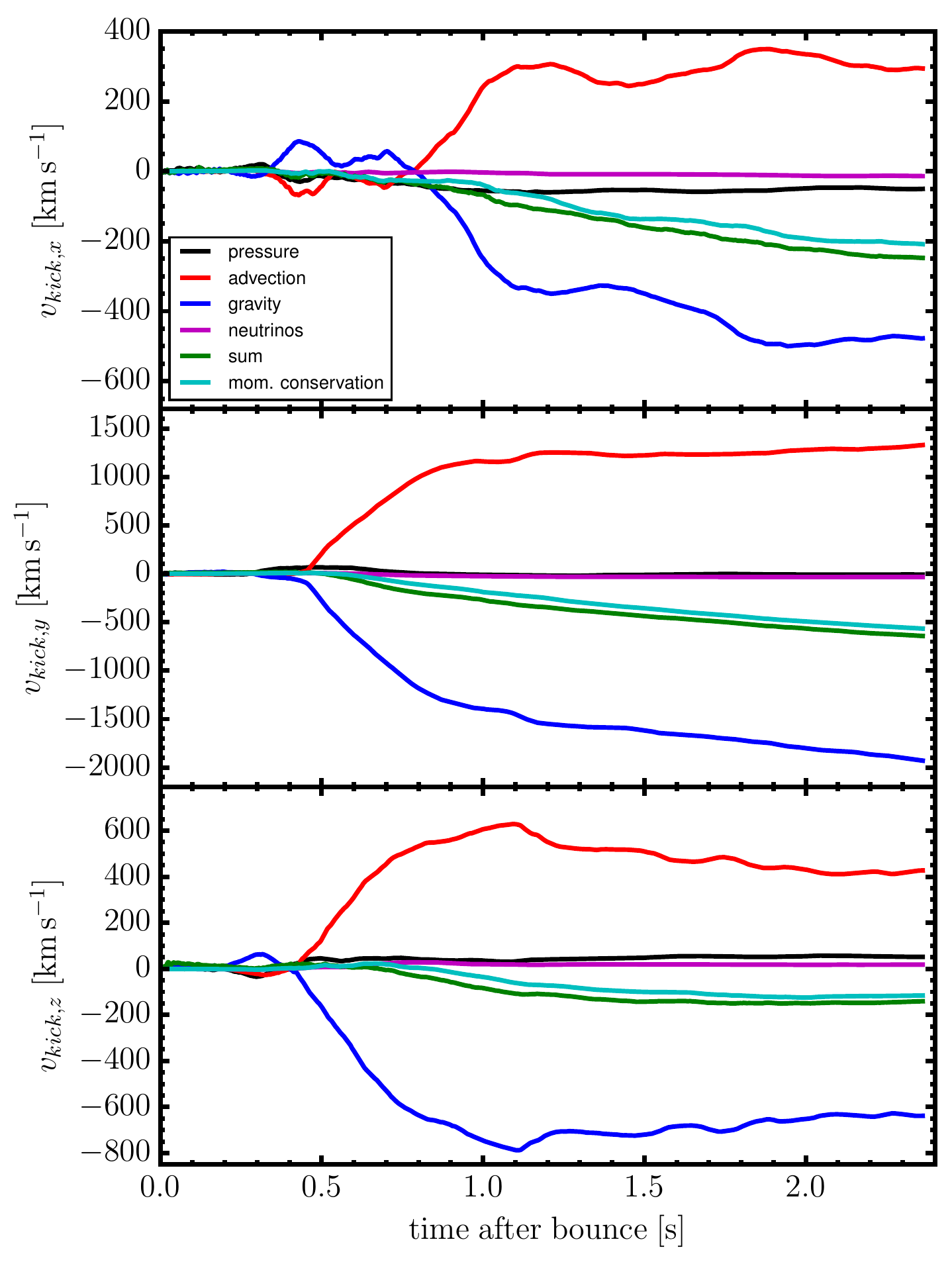}
\caption{Time-integrated contributions of the different flux and 
force terms to the kick velocity
of the proto-neutron star
(pressure anisotropy
$\mathbf{F}_P$: black,
advective momentum flux  $\mathbf{F}_\mathrm{adv}$: red,
asymmetrical gravitational tug $\mathbf{F}_\mathrm{grav}$: blue,
anisotropic neutrino emission: violet.
The sum of these contributions is shown
in green and compared to the estimate
based on momentum conversion
(cyan, Equation~\ref{eq:vkick}).
$x$-, $y$-, and $z$-components are shown in
the top, middle, and bottom panel. Note that the sustained accretion
of matter results in large contributions from the gravitational
tug and the advection of momentum onto the neutron star that almost cancel.
The cancellation is not perfect because of the gravitational tug exerted by
the actual ejecta, which determines the net kick.
     \label{fig:kick}
  }
\end{figure}

\begin{figure}
  \includegraphics[width=\linewidth]{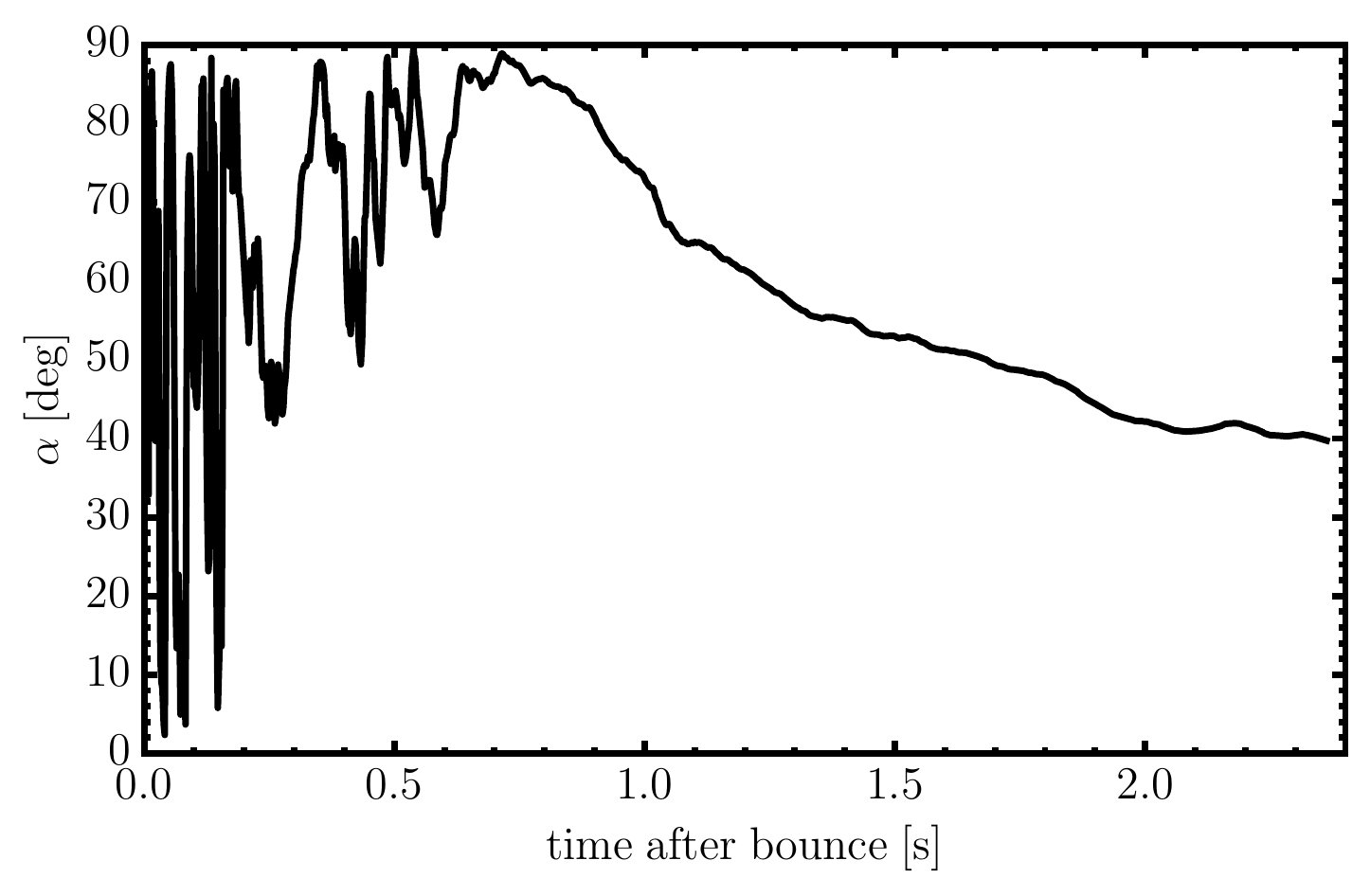}
  \caption{Angle $\alpha$ between the direction of the proto-neutron
star kick and spin. Around the onset of the explosion,
the spin and kick start out as almost perpendicular, but
after a post-bounce time of $0.7 \, \mathrm{s}$,
the continued accretion of angular momentum reduces
$\alpha$ rather steadily to $42^\circ$ at the end of the
simulation.
    \label{fig:alignment}
  }
\end{figure}

\subsection{Ejecta Composition}
With a small set of nuclear species and a very simplified
treatment of nuclear burning, we can only draw limited conclusions
on the composition of the ejecta in model s18-3D; obtaining
the detailed nucleosynthesis would require post-processing
the model and is beyond the scope of the current paper. Moreover,
our use of the FMT method for neutrino transport introduces
uncertainties in the electron fraction of the neutrino-processed
ejecta. Nonetheless, a closer look at the ejecta composition in s18-3D
is useful both as a tentative plausibility check
for our simulation and as an indicator of potential problems
and crucial sensitivities of 3D explosion models.

The contribution $\Delta M_{\mathrm{ej},i}$ of important $\alpha$- and
iron-group nuclei to the mass of the ejecta is shown in
Figure~\ref{fig:comp}. During the early phase of the explosion, the
immediate post-shock temperatures are sufficiently high for burning
the shocked material into NSE, which produces about $0.03 M_\odot$ of
${}^{56} \, \mathrm{Ni}$. A considerable fraction of the synthesised
iron group material does not remain unbound, however, and is
eventually channelled around the expanding neutrino-heated bubbles and
accreted onto the proto-neutron star.  About half of the ${}^{56} \,
\mathrm{Ni}$ synthesised by explosive burning is thus lost again by a
post-bounce time of $1\, \mathrm{s}$. For the same reason, only about
half of the intermediate-mass elements synthesised in the O shell (S,
Si, etc.) in the progenitor are eventually ejected
(see Figures~1 and 14 in \citealt{mueller_16c} for
the initial composition of the O shell, which contains
about $0.2 M\odot$ of ${}^{28}\mathrm{Si}$). By contrast,
shocked material from the almost completely processed C shell mostly
retains positive velocities by the end of the simulation, though the 
simulation would need to be extended to verify that this shell
is ejected completely.

Most of the eventually ejected iron-group material in model s18-3D is
produced by freeze-out from NSE in the neutrino-heated
bubbles. Together with the ${}^{56}\mathrm{Ni}$ produced by explosive
burning, roughly $0.06 M_\odot$ of iron-group material are thus
ejected by the end of the explosion.  The relatively high content of
iron-group nuclei in the neutrino-driven ejecta (compared to only
$0.03 M_\odot$ of ${}^4 \mathrm{He}$ is the result of turbulent mixing
between the bubbles and the colder downflows, which implies
that the freeze-out from NSE occurs at relatively low entropies.

It is important to note that the precise composition of the ejecta
from the iron-group would certainly differ from the one plotted in
Figure~\ref{fig:comp}, where ${}^{56}\mathrm{Fe}$ is the dominant
iron-group nucleus.  This is not only an artefact of the small number
of iron-group nuclei in our NSE table, but in all likelihood related to
the FMT transport solver.  The FMT scheme tends to produce somewhat
higher luminosities for $\bar{\nu}_e$ than for ${\nu}_e$ and a slowly
widening gap between the mean energies of these two neutrino species
at late times (Figure~\ref{fig:neutrino}). As a result, the
neutrino-heated ejecta tend to become slightly neutron-rich, different
from simulations relying on more sophisticated transport solvers
\citep{mueller_12a,bruenn_16,wanajo_17}, where the neutrino-heated
ejecta tend to be proton-rich for massive progenitors.  With an
electron fraction $Y_e$ slightly lower than $Y_e=0.5$ in the
neutrino-heated ejecta, ${}^{54}\mathrm{Fe}$ (which is not included
in our NSE table) and then ${}^{56}\mathrm{Fe}$ replaces
$^{56}\mathrm{Ni}$ as the dominant nucleus in NSE
\citep{hartmann_85}. Since the overall explosion dynamics and the
entropy of the neutrino-heated bubbles is less sensitive to subtle
differences in electron neutrino and antineutrino emission,
we would expect, to first order, that the effect of
a more rigorous transport treatment would be to shift
the ${}^{56}\mathrm{Fe}$ to ${}^{56}\mathrm{Ni}$, which
dominates the nucleosynthesis
in freeze-out from NSE/QSE 
for $Y_e>0.5$. The true mass of ejected ${}^{56}\mathrm{Ni}$
in s18-3D could therefore well be of order $\gtrsim 0.05 M_\odot$.

Considering the modelling uncertainties (small set of 17 nuclei in
NSE, fixed freeze-out temperature of $5.8 \times 10^{9} \,
\mathrm{K}$, flashing prescription below NSE temperatures, uncertainties
in $Y_e$), the
production of a few $0.01 M_\odot$ of iron-group elements does not
present any significant conflict with observational constraints: Based
on observed correlations between explosion energy and nickel mass
\citep{hamuy_03,pejcha_15b}, we expect a few $0.01 M_\odot$ of
${}^{56} \mathrm{Ni}$ to be produced in a core-collapse supernova with
an explosion energy of $7\text{-}8 \times 10^{50} \, \mathrm{erg}$
such as model s18-3D. This is roughly compatible with our
simulations, but it is also obvious that more rigorous simulations
will be required in the future to obtain precise predictions
for the iron-group nucleosynthesis in 3D supernova explosion models.

\subsection{Neutron Star Properties}
The evolution of the baryonic mass, kick, and spin of the
proto-neutron star up to that point is shown in Figure~\ref{fig:pns};
a detailed description of how we evaluate the neutron star kick and
spin is given further below in this section.  As for the explosion
energy and the nickel mass, we can only give tentative limits for the
final neutron star properties since accretion is still ongoing at the
end of the simulation.

\subsubsection{Neutron Star Mass}
The baryonic proto-neutron star mass $M_\mathrm{by},$ however, appears
to be close to its asymptotic limit already. Later than $1.2 \,
\mathrm{s}$ after bounce the accretion rate onto the proto-neutron
star typically fluctuates between $0.01 M_\odot \, \mathrm{s}^{-1}$
and $0.02 M_\odot \, \mathrm{s}^{-1}$. With $M_\mathrm{by}=1.865 M_\odot$
at the end of the simulation, this suggests a final neutron
star of $M_\mathrm{by} \lesssim 1.9 M_\odot$ even if we
allow for another $2\, \mathrm{s}$ of accretion. This is
also in line
with the mass shell trajectories in Figure~\ref{fig:expl},
which show an emerging mass cut at around
$2000 \, \mathrm{km}$; the mass coordinate for
which the spherically averaged velocity vanishes
is $M_\mathrm{cut}=1.887 M_\odot$.
Using the fit formula of \citet{lattimer_89} and \citet{,lattimer_01}
for the neutron star binding energy $E_\mathrm{bind}$,
\begin{equation}
  E_\mathrm{bind} \approx 0.084 M_\odot c^2 (M_\mathrm{grav}/M_\odot)^2,
\end{equation}
in terms of the gravitational mass $M_\mathrm{grav}$, we thus estimate
a final value of $M_\mathrm{grav} \lesssim 1.67 M_\odot$ (assuming
$M_\mathrm{by}=1.9 M_\odot$).  Although higher than the typical masses
in double neutron star binaries \citep{schwab_10,oezel_12,oezel_16},
such a neutron star birth mass is not implausible.  It is well within
the mass distribution of slow pulsars that likely have undergone
little accretion \citep{oezel_16}, and more detailed analyses of the
evolutionary pathways of some binary systems also suggest some neutron
stars are born at least with $M_\mathrm{grav}=1.7 M_\odot$
\citep{tauris_11}.

\subsection{Neutron Star Kick}
Since we model the interior of the proto-neutron star at densities
larger than $10^{11} \, \mathrm{g}\, \mathrm{cm}^{-3}$ in spherical
symmetry, we need to infer the kick and spin indirectly via balance
equations.  For the kick, one can invoke momentum conservation
\citep{scheck_06} and obtain the kick velocity
$\mathbf{v}_\mathrm{kick}$ by assuming that the neutron star momentum,
and the momenta $\mathbf{p}_\mathrm{ej}$ and
$\mathbf{p}_\nu$
 of the ejecta and the emitted neutrinos add up to
zero, i.e.\
\begin{equation}
  \label{eq:vkick}
  \mathbf{v}_\mathrm{kick}=-\frac{\mathbf{p}_\mathrm{ej}+\mathbf{p}_\nu}{M},
\end{equation}
where $M$ is the inertial mass of the neutron star.
In the relativistic case, this is non-trivial because
one needs the inertial mass of the neutron star to obtain the kick
velocity from the momentum. 
By dint of the 
equivalence principle, we can identify the inertial mass
with the gravitational mass $M_\mathrm{grav}$, but, strictly
speaking, $M_\mathrm{grav}$ can only be defined for isolated systems.
For practical purposes, the gravitational field of the ejecta
can still be considered as sufficiently weak to regard the
proto-neutron star as isolated, however. We can thus define
the approximate gravitational and inertial mass   $M_\mathrm{grav}$ enclosed
within a radius $r_0$ (assumed to lie outside the neutron star
surface) in terms of the derivative of the lapse
function in the xCFC metric,
\begin{equation}
  \label{eq:mgrav}
  M_\mathrm{grav}=
  \frac{c^2 r_0^2}{G}\frac{\pd \alpha}{\pd r}(r_0),
\end{equation}
which is analogous to the Newtonian case where
the enclosed mass is related to the surface
integral of the gravitational acceleration $\mathbf{g}$,
\begin{equation}
  \oint \mathbf{g} \cdot \ud \mathbf{A}
  =-4\pi G M.
\end{equation}
For our analysis, we use $r_0=50 \, \mathrm{km}$, which adds
little mass outside the proto-neutron star, but is
sufficiently large to justify
the computation of an effective gravitational mass from
the local gravitational acceleration.

The kick velocity computed from Equation~(\ref{eq:vkick}) (middle
panel of Figure~\ref{fig:pns}) reaches more than $600 \, \mathrm{km}
\, \mathrm{s}^{-1}$ and is oriented opposite to the biggest
neutrino-heated plume (Figure~\ref{fig:3dexpl} and bottom left
  panel of Figure~\ref{fig:entropy_slices}) along the negative
$y$-direction of the grid. Different from long-time simulations with
parameterised gray transport in 2D \citep{scheck_06} and 3D
\citep{wongwathanarat_13}, where accretion onto the neutron star
  had ended after at most $1 \, \mathrm{s}$, the kick velocity shows
little sign of asymptoting to its final value as late as $2\,
\mathrm{s}$ after bounce. At this stage, we can only conclude that the
final kick will lie on the high side of the observed kick distribution
\citep{hobbs_05,faucher_06,ng_07}, but not to the degree that there is
a conflict with observations. Even if the kick were to increase at the
current rate for another $2\, \mathrm{s}$, model s18-3D would still be
within the bounds of the observed distribution that extends to
$\mathord{\gtrsim} 1000 \, \mathrm{km} \, \mathrm{s}^{-1}$
\citep{chatterjee_05}. Neither does the behaviour of model s18-3D
appear too unusual in terms of kick velocity compared to extant
parametric and self-consistent simulations in 2D
\citep{scheck_04,scheck_06,nordhaus_10b,nordhaus_12,bruenn_16} and 3D
\citep{wongwathanarat_10b,wongwathanarat_13} if we leave aside the
slower saturation of the kick.

This slow saturation  is nonetheless the key to an
interesting peculiarity of model s18-3D. It can be explained naturally
by the relatively slow-paced propagation of the shock in our
simulation and the persistence of stronger global accretion asymmetries than in
the 3D models of \citet{wongwathanarat_10b,wongwathanarat_13}: Whereas
the models shown in \citet{wongwathanarat_13} exhibit average shock
radii of $\mathord{\sim}15,000 \, \mathrm{km}$ about $1.3 \,
\mathrm{s}$ after bounce and already show at least the onset of an
isotropic neutrino-driven wind (see their Figure~3), the average shock
radius at this time is smaller than $\mathord{\sim}10,000 \, \mathrm{km}$ in our
models, and
downflows onto the proto-neutron star still persist.  This implies
that the acceleration by the ``gravitational tug'' of the ejecta as
well as impulsive momentum transfer by the accretion downflows can be
maintained longer in our simulations.

Considering the pronounced global asymmetries and the relatively
slow shock propagation in model s18-3D, it is in fact somewhat
surprising that the kick is not too different from
the 3D models of \citet{wongwathanarat_10b,wongwathanarat_13}
in our case. The reason for this emerges from a closer
analysis of the different effects contributing to the
kick: Following \citet{scheck_06}, \citet{nordhaus_10b},
and \citet{wongwathanarat_13}, one can
write the time derivative of the neutron star
momentum $\mathbf{p}_\mathrm{PNS}$ in terms of
the gravitational force $\mathbf{F}_\mathrm{grav}$ exerted
onto the proto-neutron star by material outside
$r_0$, the net pressure force $\mathbf{F}_P$ on the volume enclosed
by $r_o$, the advective
momentum flux $\mathbf{F}_\mathrm{adv}$ into this region,
and the backreaction term $\ud \mathbf{p}_\nu/\ud t$ due to anisotropic
neutrino emission:
\begin{equation}
  \label{eq:dpdt}
  \frac{\ud \mathbf{p}_\mathrm{PNS}}{\ud t}
  =
  \mathbf{F}_\mathrm{grav}+
  \mathbf{F}_P+
  \mathbf{F}_\mathrm{adv}-
\frac{\ud \mathbf{p}_\nu}{\ud t}.
\end{equation}
Here, $\mathbf{F}_P$
and 
  $\mathbf{F}_\mathrm{adv}$ are evaluated as
\begin{eqnarray}
  \mathbf{F}_P&=&
  -\int \alpha \phi^4 r_0^2 P \mathbf{n}\,\ud \Omega,\\
  \mathbf{F}_\mathrm{adv}&=&
  -\int  \alpha \phi^4 r_0^2 \rho v_r \mathbf{v} \,\ud \Omega,
\end{eqnarray}
and $\mathbf{F}_\mathrm{grav}$ is computed using
the monopole approximation for the gravitational
field of the neutron star (but not the ejecta),
\begin{equation}
\label{eq:fgrav}
  \mathbf{F}_\mathrm{grav}=
  \int_{r>r_0} \frac{G M_\mathrm{grav} \rho \mathbf{r}}{r^3} \phi^6 \ud V,
\end{equation}
where $M_\mathrm{grav}$ is calculated
according to Equation~(\ref{eq:mgrav}). 
We compute the backreaction term from the flux of
neutrinos through the outer boundary of the grid,
\begin{equation}
\frac{\ud \mathbf{p}_\nu}{\ud t}=
\oint \mathbf{F}_\nu \mathbf{n}\cdot \ud \mathbf{A},
\end{equation}
where $\mathbf{F}_\nu$ is the sum of
 the frequency-integrated neutrino energy flux
for all species.\footnote{Since
the FMT scheme involves an approximate
solution to the \emph{stationary} transport equation,
it is more appropriate not to include the momentum
nominally carried by the radiation field on
the grid in the total momentum budget.}

Figure~\ref{fig:kick} shows the time-dependent contributions of
$\mathbf{F}_\mathrm{grav}$, $\mathbf{F}_P$, and momentum flux
$\mathbf{F}_\mathrm{adv}$ to the components of the kick velocity (and
incidentally shows that the two different ways evaluating the kick via
Equation~(\ref{eq:dpdt}) or via momentum balance (Equation~\ref{eq:vkick})
are
consistent with each
other). It is noticeable that the contribution of the gravitational
tug and the advective momentum flux almost cancel for our choice of
$r_0=50 \, \mathrm{km}$ with $\mathbf{F}_\mathrm{grav}$ slightly
outweighing $\mathbf{F}_\mathrm{adv}$.  Since the evaluation of the
flux and force terms is sensitive to $r_0$ \citep{nordhaus_10b}, this
can, to some degree, be viewed as accidental, but it nonetheless
enunciates a physical peculiarity of models with sustained accretion
onto the proto-neutron star. The crucial point for understanding the
near-cancellation of these two terms is that a sizeable contribution of
the asymmetric gravitational tug in s18-3D comes from material
that is either not accelerated to positive velocities by the shock at
all or moves so slowly that it eventually falls back onto the
proto-neutron star.  Accretion of this material will then almost
exactly cancel the proto-neutron star momentum generated by
gravitational acceleration by virtue of Newton's third law
(discounting the small effect of momentum transfer from the downflows
to the outflows by hydrodynamic forces).

It must be emphasised that despite this seeming peculiarity,
model s18-3D is fully in line the
established theory of hydrodynamical neutron star kicks in 3D.  The
\emph{net} kick is still due to the slight preponderance of the
gravitational tug over the advective momentum flux, which is
eventually bound to become stronger as accretion ceases. Model s18-3D
merely illustrates the difficulty of peeling out the net gravitational
tug of the \emph{eventual ejecta} by a fixed-volume analysis of the
forces and fluxes onto the proto-neutron star.

In accordance with \citet{nordhaus_10b} and \citet{wongwathanarat_13}
we find the contribution from anisotropic neutrino emission to the kick to be
of minor importance. It contributes only $30 \, \mathrm{km} \,
\mathrm{s}^{-1}$ to the dominant component of the kick in the
$y$-direction. Although one might expect sustained strong global
asymmetry of the accretion flow during the explosion phase to be
favourable for a large kick by anisotropic neutrino emission, this
expectation is thwarted by the lateral redistribution of matter at the
base of the gain region and the decline of the accretion rate.

Finally, a word is in order about the potential implications of
  two approximations made in our model, namely the use of a non-moving
  spherical inner core and a spherical metric. 
The effects of both approximations are very similar, since they imply
that the proto-neutron star becomes a ``momentum sink'' whose
momentum budget \emph{in the simulation} is not captured accurately.
The effects on the momentum budget of the ejecta are less severe since
the gravitational potential of the proto-neutron star does not
deviate strongly from spherical symmetry, and since any higher multipoles
decrease rapidly with higher powers of $r$.
While long-time
simulations without a spherically symmetric core and a full
3D metric remain desirable, previous studies suggest that these
approximation do not constitute severe limitations:
In their 2D simulations with the \textsc{Vulcan} code, \citet{nordhaus_12} found that
evaluating the momentum of the proto-neutron star using
Equation~(\ref{eq:dpdt})
(assuming a spherically symmetric potential of the neutron star)
yields results in very good agreement with the direct evaluation
of the neutron star in their hydrodynamical simulation,
which allowed for free movement of the proto-neutron star across
the grid, included a 2D potential, and maintained
momentum conservation up to truncation error. \citet{scheck_06}
also investigated potential problems with a ``momentum sink''
in the centre by performing neutrino hydrodynamics
simulations in an accelerated reference frame that tracks
the motion of the proto-neutron star, and found no major differences
compared to the naive implementation of the non-moving
neutron star core as a momentum sink.

\subsection{Neutron Star Spin}
The evolution of the angular momentum of the proto-neutron star is
more remarkable against the background of extant 3D
explosion models with parameterised neutrino heating
and cooling or gray  transport
\citep{fryer_07,wongwathanarat_10b,wongwathanarat_13,rantsiou_11}.
Following \citet{wongwathanarat_10b,wongwathanarat_13}, we evaluate
the proto-neutron star angular momentum
$\mathbf{J}_\mathrm{PNS}$ by integrating the flux of angular momentum
across the sphere with radius $r_0$,
\begin{equation}
  \frac{ \ud \mathbf{J}_\mathrm{PNS}}{\ud t}
  =
  \int \alpha \phi^4 r_0^2   \rho v_r \mathbf{v} \times \mathbf{r} \,\ud \Omega.
\end{equation}
Again we do not expect that neglecting higher-order multipoles
of the gravitational field has a major impact on the angular momentum
of the accreted material because the deviations of the gravitational
field of the proto-neutron star from spherical symmetry are very small
(which has prompted the use of the monopole approximation also
in earlier studies such as \citealt{rantsiou_11}). Exchange of
momentum and angular momentum by gravitational torques between
the ejecta and the material accreted after
shock revival (less than $0.1 M_\odot$) is likely also
of minor importance, though simulations with a full 3D metric
 are warranted in the future.

$J_\mathrm{PNS}$ grows quickly immediately after the onset of the
explosion to reach values of order $\mathord{\sim}6 \times 10^{47} \,
\mathrm{g}\, \, \mathrm{cm}^{2}\, \mathrm{s}^{-1}$. Assuming
a final neutron star mass and radius of $M=1.67 M_\odot$ and
$R=12 \, \mathrm{km}$ respectively, we obtain a moment of inertia of
about $I\approx 2 \times 10^{45}\, \mathrm{g}\, \mathrm{cm}^2$
from the fit formula of
\citep{lattimer_05},
\begin{equation}
  I\approx 0.237 M_\mathrm{grav} R^2 \left[1+4.2 \left(\frac{M_\mathrm{grav} \, \mathrm{km}}{M_\odot R}\right)
    +90\left(\frac{M_\mathrm{grav} \, \mathrm{km}}{M_\odot R}\right)^4\right].
\end{equation}
This implies a spin period at birth of around $20 \, \mathrm{ms}$,
i.e.\ close two the lower end of the period distribution of young
pulsars \citep{muslimov_96,marshall_98}. The inferred distribution of
birth periods is somewhat model-dependent, but likely centred at
longer periods of several hundred milliseconds
\citep{faucher_06,perna_08,noutsos_13}, although on the low side
\citet{popov_12} still obtain typical periods of $\mathord{\sim} 100
\, \mathrm{ms}$. Similar to the kick, model s18-3D thus appears
to represent an outlier rather than the norm in terms of the neutron
star spin, but is not in conflict with observational constraints.

Whether or not we classify the neutron star spin in s18-3D as
unusually fast or not, there is a clear contrast to extant 3D
simulations \citep{wongwathanarat_10b,wongwathanarat_13,rantsiou_11}
as the spin period is about 10 times shorter than for the fastest
rotator in \citet{wongwathanarat_10b}.  This is again a consequence of
the long duration of accretion after the onset of shock revival.
Under these conditions, the spin-up mechanism of \citet{spruit_98},
i.e.\ the stochastic transfer of angular momentum to the proto-neutron
star by downflows that strike the proto-neutron star slightly
off-centre, becomes more effective for several reasons: More impulsive
spin-up events occur and the individual events transfer
more angular momentum if late-time accretion is stronger. Furthermore,
the plumes striking the proto-neutron star at late times originate
from large radii, and can have relatively large angular momentum
even if their initial non-radial velocity is small.

It is important to note that an appreciable amount angular momentum is
actually transferred onto the proto-neutron star even at late
times. While the absolute value of $J_\mathrm{PNS}$ remains roughly
constant from about $0.7 \, \mathrm{s}$ after bounce, the direction of
the spin axis of the proto-neutron star still undergoes considerable
re-orientation. The $y$-component of the angular momentum vector
changes most appreciably at late times.  This results in a relatively
stable decrease of the angle $\alpha$ between the proto-neutron star
kick and spin from $80^\circ$ at $0.7 \, \mathrm{s}$ to about
$40^\circ$ at the end of the simulation, i.e.\ a trend towards
increasing spin-kick alignment (Figure~\ref{fig:alignment}).

Model s18-3D would need to be extended longer to determine whether
this trend continues and clear alignment is reached, and even if 
this is the case, a tendency towards spin-kick alignment in a single
simulation may still prove a statistical fluke. Moreover, the
proto-neutron star \emph{initially} acquires angular momentum in
the direction perpendicular to the kick after shock revival, and if there
is a systematic alignment mechanism acting afterwards, it is not clear
whether it would generally be sufficiently effective to ``right''
the proto-neutron star later. A further caveat concerns late-time
fallback, which may alter the direction of the neutron star spin
on much longer time-scales. If fallback occurs predominantly from
directions where the shock is weaker (i.e.\ in the direction
of the kick), this would destroy spin-kick alignment even if
the the fallback mass is  quite small.

Nonetheless, it is worth speculating whether there could be a physical
reason behind this trend: The preferential ejection of mass into the
$y$-direction (Figure~\ref{fig:3dexpl}) suggests that there is a weak
preference for accretion to occur in the $x$-$z$-plane, i.e.\ the
plane perpendicular to the kick direction. Since the angular momentum
imparted onto the proto-neutron star by downflows is perpendicular to
their direction, this could imply that the downflows change $J_y$ more
strongly than $J_x$ and $J_z$ on average. Even after averaging over
many impulsive events, one expects that such an asymmetry between the
angular momentum components parallel and perpendicular to the kick
direction subsists. It should be noted that this suggested mechanism
for spin-kick alignment is distinct from the one initially posited by
\citet{spruit_98}, who claimed that stochastic angular momentum
transfer by downflows is sufficient for spin-kick alignment. Their
suggestion may not work generically \citep{wang_07} and is problematic
because it hinges on the isotropy of the impulsive events in the
rotating frame of the neutron star. By contrast, the mechanism we just
outlined is based on the realisation that angular momentum transfer by
downflows onto the proto-neutron star is \emph{non-isotropic} and not
completely stochastic, but exhibits an asymmetry related to
the explosion geometry (and hence to the direction of the kick).

This hypothesis will need to be investigated further based on a larger
sample of 3D simulations in the future. At present, the evolution of
the proto-neutron star spin only opens up an interesting
perspective. It suggests that sustained accretion may offer an
alternative road towards short neutron star birth periods of a few
tens of milliseconds other than spin-up by the standing accretion
shock instability \citep{blondin_06,fernandez_14b,kazeroni_16} or the
assumption of pre-collapse rotation rates of $\mathord{\sim} 100 \,
\mathrm{s}$ \citep{heger_05,ott_06c}. This is especially
interesting considering that asteroseismic measurements
of core rotation in evolved low-mass stars \citep{cantiello_14}
suggest that current stellar evolution models 
still underestimate the efficiency of angular momentum
transport in stellar interiors, which implies that
pre-collapse rotation rates of supernova progenitors
may be even lower than currently \citep{heger_05} predicted.
Even spin-up mechanisms for strongly braked cores
during late evolutionary stages, such a stochastic
spin-up by internal gravity waves could likely not explain
neutron star spin periods shorter than a few hundred milliseconds
\citep{fuller_15}.
The possibility that late-time accretion may explain
the spin-kick alignment posited by several observational papers
\citep{lai_01,ng_07,noutsos_13,rankin_15} is still more
speculative, but remains noteworthy since many alternative
mechanisms (see \citealt{janka_17} for an overview)
are still somewhat problematic (such as the invocation
of jets, which cannot achieve sufficiently large kicks)
or remain more schematic, such as the putative
spin-kick alignment for explosions triggered
by a spiral SASI mode \citep{janka_17}.

\section{Summary and Conclusions}
\label{sec:conclusions}
In this paper, we presented the first 3D multi-group neutrino
hydrodynamics simulations based on initial conditions from 3D models
of O shell burning. We studied the impact of the perturbations on the
dynamics in the supernova core for an $18 M_\odot$ star (s18-3D,
\citealp{mueller_16c}) and a model with artificially reduced nuclear
burning rates and smaller convective velocities in the O shell
(s18-3Dr). For comparison, we conducted a run based on the
corresponding 1D progenitor with small random seed perturbations.

In the simulations starting from 3D initial conditions a
neutrino-driven explosion develops around $0.3 \, \mathrm{s}$ (s18-3D)
and $0.5 \, \mathrm{s}$ (s18-3Dr) after bounce thanks to the mechanism
of forced shock deformation whereas the shock is not revived in the
control model s18-1D before the end of the simulation $0.645 \,
\mathrm{s}$ after bounce.  Different from the 3D leakage models of
\citet{couch_15}, the initial perturbations from convective shell
burning thus have a significant and qualitative impact on the fate of
the progenitor.  When s18-3D and s18-3Dr explode, the heating conditions in model s18-1D are far below
the runaway threshold  with values of
the critical time-scale ratio $\tau_\mathrm{adv}/\tau_\mathrm{heat}$
are only around $0.5$.  This implies that
at least for some progenitors, the perturbation-aided neutrino-driven
mechanism could be an important part of a solution to the problem of
shock revival.

Our simulations confirm that the perturbation-aided mechanism works in
a similar fashion in 3D as in the earlier parameterised 2D models of
\citet{mueller_15a}. At least in the case of perturbations from O
shell burning, the beneficial role of the perturbations consists in
providing slowly-varying large-scale ``forcing'' of the shock due to
the anisotropic pre-shock density and ram pressure that arise from the
convective velocity field in the progenitor by advective-acoustic
coupling during the collapse.  This facilitates the formation of large
neutrino-heated bubbles whose geometry is dictated by the pre-shock
density perturbations.

At present, some of the links in the chain from convective seed
perturbations in the progenitor to the excitation of more violent
non-spherical flow in the post-shock region and the concomitant
reduction of the critical luminosity for shock revival can still only
be understood in qualitative terms.  The first element of the
perturbation-aided mechanism, i.e.\ the generation of pre-shock
density perturbations during the collapse phase, appears to be
relatively straightforward; the spectrum of pre-shock density
perturbations closely reflects the turbulent velocity spectrum in the
infalling convective shell in our models. 

It is much more difficult to pinpoint how these pre-shock
perturbations affect the violence of non-spherical motions in the gain
region qualitatively. Although strongly perturbed models generally
show higher turbulent kinetic energies and average Mach numbers in the
gain region \citep{couch_13,couch_15,mueller_15a}, this is largely due
to the positive feedback effects that occur for slight changes in the
shock radius and the neutrino heating conditions.  Our models allow us
to better pin down the impact of the infalling perturbations: They
primarily result in a more efficient excitation of turbulent motions
in the radial direction (but not the transverse direction) than would
be expected by neutrino heating alone.  This is consistent with the
idea that the additional turbulent driving is provided by the buoyancy
of the shocked density perturbations \citep{mueller_16c}.  Upon closer
inspection, the comparison of models s18-3D and s18-3Dr to s18-1D
proves more complicated: Part of the differences between the
simulations may be stochastic in nature, as the SASI-dominated models
s18-1D exhibits long-term fluctuations in the heating conditions. At
early times, weak perturbations in s18-3D and s18-3Dr can apparently
either boost or dampen shock oscillations under different
circumstances.  Moreover, some of the differences between the runs may
be related to the disruption of the SASI spiral mode in s18-3D and
s18-3Dr so that one ought to be careful before extrapolating our
results to the convection-dominated regime,
where the interaction between infalling perturbations and
the shock may be somewhat different.
One of the key differences in the convection-dominated regime
could be that the buoyancy-driven modes excited by the infalling
perturbations are already unstable to begin with and could
be triggered by weaker infalling perturbations, whereas
advection stabilises these modes in the SASI-dominated
regime \citep{foglizzo_06} unless the flow is strongly
perturbed \citep{scheck_08}.

For these reasons, it is not yet possible to validate quantitative
models for the additional turbulent driving and the concomitant
reduction of the critical luminosity for shock revival
\citep{mueller_16c,abdikamalov_16}, and it must be emphasised that the
current analytic approaches to the interaction of perturbations with
the shock are undoubtedly too simple to capture the complex nuances
found in models s18-3D and s18-3Dr.  Still, the inferred reduction of
the critical luminosity by about $22 \%$ and $16 \%$
for s18-3D and s18-3Dr compared to s18-1D is roughly compatible with
the analytic model of \citet{mueller_16c}, which may retain some
usefulness as a reasonable fit for extant 2D and 3D simulations with
large-scale initial perturbations.

We studied the explosion dynamics of model s18-3D by extending this
run to more than $2.35 \, \mathrm{s}$ after bounce. Although this makes
s18-3D the longest 3D simulation with multi-group neutrino transport,
this is still no sufficient to obtain final values for the explosion
energy, the nickel mass, and the mass, kick, and spin of the
proto-neutron star.  Different from parameterised long-term
simulations based on grey transport
\citep{wongwathanarat_10b,wongwathanarat_13,handy_14}, the cycle of
accretion and mass ejection is still ongoing in our simulation even at
these late times, and we cannot capture the transition to
the neutrino-driven wind phase yet (cp.\ \citealt{bruenn_16}), which may
still occur on even longer time scales. Nonetheless, model s18-3D exhibits interesting
and encouraging trends that strengthen the paradigm of
neutrino-driven supernova explosions.

The diagnostic explosion energy reaches $7.7 \times 10^{50} \,
\mathrm{erg}$ in model s18-3D at the end of simulations and is still
increasing at an appreciable rate.  Even if we subtract the residual
binding energy of the shells ahead of the shock, the explosion is
clearly sufficiently energetic to unbind the envelope, and we can
place a relatively firm lower limit of $5 \times{10}^{50} \,
\mathrm{erg}$ on the final explosing energy.  3D turbulence plays a crucial
role for the stable growth of the explosion energy, as accreted
material from the downflows is halted and mixed into the outflows
at relatively large radii so that little neutrino heating
is required to unbind it. 

The prospective explosion energy is well within the observed range for
Type~IIP supernovae of $1\text{-}40\times 10^{50} \, \mathrm{erg}$
\citep{kasen_09,pejcha_15b} and already reasonably close to the
``typical'' value of $\mathord{\sim} 9 \times 10^{50} \, \mathrm{erg}$
\citep{kasen_09} (although such ``typical'' values have yet to be
  determined for volume-limited samples).  Because of the simplified
treatment of nuclear burning and neutrino transport, the nickel mass
remains rather uncertain, but is likely of the order of
  $\mathord{\sim} 0.05 M_\odot$, i.e.\ there is no apparent conflict
with observational constraints.

The parameters of the proto-neutron star also appear to lie within the
the observed spectrum, although the kick and spin have not reached
their final values yet. The proto-neutron star mass, however, has
almost converged. Barring the possibility of late-time fallback, it is
not likely to exceed $M_\mathrm{grav}=1.67 M_\odot$. This would
put model s18-3D above the peak of the neutron star
mass distribution \citep{oezel_12,oezel_16}, but is
again not an implausible value. The kick velocity
of $\mathord{\sim} 600 \, \mathrm{km} \, \mathrm{s}^{-1}$ is already
sizable at the end of the simulation, and may still considerably
exceed that value. Interestingly, continued accretion through the
downflows imparts considerable angular momentum onto the proto-neutron
star, resulting in a spin period of only $\mathord{\sim} 20 \,
\mathrm{ms}$. 
With all due caveats about the limitations of
model s18-3D
this suggests that a reasonable spectrum of neutron
star birth spin periods could be obtained even with slower progenitor
core rotation than predicted by current stellar evolution models
\citep{heger_05}. Moreover, we observe a trend towards spin-kick
alignment at late times. It is not completely clear, however, whether
this is accidental or the result of preferred accretion through
the plane perpendicular to the largest neutrino-heated bubble 
and the kick direction. 

In summary, our simulations provide further evidence for the viability
of the neutrino-driven mechanism in 3D beyond the extant successful
explosion models with rigorous \citep{melson_15a,melson_15b,lentz_15}
and more simplified neutrino transport
\citep{takiwaki_12,mueller_15b,roberts_16}: At least for some progenitors,
large-scale initial perturbations can apparently negate the penalty of
slightly less optimistic heating conditions in 3D. With relatively
early shock revival around the infall of the Si/O shell interface,
plausible explosion and remnant parameters can then be reached.

Obviously, however, model s18-3D is only a step towards
a solution of the problem of shock revival and the ``energy
problem'' that has plagued many supernova simulations in the past.
For one thing, the $18 M_\odot$ progenitor considered here
provides very favourable conditions for a perturbation-aided explosion,
as it exhibits high nuclear generation rates in an extended O shell
that admits large-scale convective modes. Such conditions are not
found in all supernova progenitors, and hence only future simulations
will be able to determine whether convective seed perturbations
are generally of major relevance for shock revival.

Model s18-3D also provides some hints that more refined simulations
and additonal physics are still needed for better agreement with
observational constraints: While s18-3D exhibits plausible explosion
and remnant properties, it apparently does not represent a typical
core-collapse supernova yet: Its explosion energy is slightly below
average, the neutron star mass lies in the high-mass tail of the
distribution, and so does the neutron star kick. It is possible that
the $18 M_\odot$ progenitor is simply not a typical case, but there is
some observational evidence that suggests otherwise. Based on the
observed loose correlation between the ejecta mass and explosion
energy of Type~IIP supernovae
\citep{poznanski_13,chugai_14,pejcha_15b}, one would expect
progenitors in this mass range to explode with more than $10^{51} \,
\mathrm{erg}$ like SN~1987A which had $\mathord{\sim} 1.4 \times
10^{51} \, \mathrm{erg}$ \citep{utrobin_05} for a very similar helium
core mass.

It is easy to conceive of effects that could make model s18-3D somewhat
more energetic and perhaps also shift the proto-neutron star mass and
spin towards more typical values considering the limitations of our
current simulation. Both a more rigorous treatment of the neutrino
transport and higher numerical resolution could move the explosion
parameters further in the right direction. In the current
implementation of the FMT scheme, we neglect, for example, the effect
of nucleon correlations at high densities, which lead to faster PNS
cooling \citep{huedepohl_10} and at least slightly more favourable
heating conditions \citep{horowitz_17} (though the effect is likely
less pronounced than claimed by \citealt{burrows_17}). This is
especially noteworthy because correlation effects will become more
important as the neutrinospheric densities increase in models that
accrete as long as s18-3D. Higher resolution could also be
beneficial during the explosion phase if it leads to faster mixing
between the downflows and outflows and larger turnaround radii for the
ejecta. It could also help to entrain more of the $^{56}\mathrm{Ni}$
synthesised by explosive burning in the shock into the neutrino-heated
ejecta and thereby increase the nickel mass.
In the pre-explosion phase, better resolution may be required
to more accurately capture the drag on the high-entropy bubbles
seeded by the infalling perturbations via forced shock deformation.
 Further effects are
conceivable and cannot be extensively discussed here. It is clear that
considerable challenges remain before we can demonstrate that the
neutrino-driven mechanism can explain the bulk of core-collapse
supernova explosions by means of well resolved state-of-the-art
simulations employing rigorous neutrino transport across a wide range
of progenitors. Our current results merely underscore that 3D supernova
modelling is now on a promising track towards this ultimate goal.

\section*{Acknowledgements}
%\acknowledgements
We acknowledge support by the Australian Research
Council through an ARC Future Fellowship FT160100035,
STFC grant ST/P000312/1 (BM), an ARC Future Fellowship FT120100363 (AH), and at Garching by the Deutsche
Forschungsgemeinschaft through the Excellence Cluster Universe EXC 153
and the European Research Council through grant ERC-AdG
No.~341157-COCO2CASA.  This research was undertaken with the
assistance of resources from the National Computational Infrastructure
(NCI), which is supported by the Australian Government and was
supported by resources provided by the Pawsey Supercomputing Centre
with funding from the Australian Government and the Government of
Western Australia.  This work used the DiRAC Data Centric system at
Durham University, operated by the Institute for Computational
Cosmology on behalf of the STFC DiRAC HPC Facility
(\url{www.dirac.ac.uk}); this equipment was funded by a BIS National
E-infrastructure capital grant ST/K00042X/1, STFC capital grant
ST/K00087X/1, DiRAC Operations grant ST/K003267/1 and Durham
University. DiRAC is part of the UK National E-Infrastructure.  This
material is based upon work supported by the National Science
Foundation under Grant No.~PHY-1430152 (JINA Center for the Evolution
of the Elements).

\bibliography{paper}

\end{document}